\documentclass[useAMS,usenatbib]{mn2e}
\usepackage{aecompl}
\usepackage{graphicx}
\usepackage{amssymb}
\usepackage{amsmath}
\usepackage{algpseudocode}
\usepackage{multicol}
\usepackage{longtable}
\usepackage{algorithm}
\usepackage{longtable}
\usepackage{float}

\usepackage{hyperref}

\title[KROME]{KROME - a package to embed chemistry in astrophysical simulations}

\author[T. Grassi et al.]{\parbox{\textwidth}{T. Grassi$^{1}$\thanks{Corresponding author: tommasograssi@gmail.com}, 
S. Bovino$^{2}$, 
D. R. G. Schleicher$^{2}$,
J. Prieto$^{3}$, 
D. Seifried$^{4}$, 
E. Simoncini$^{5}$, 
F. A. Gianturco$^{1,6,7}$}\vspace{0.4cm}\\
\parbox{\textwidth}{$^{1}$Department of Chemistry, Sapienza University of Rome, P.le A. Moro 5, I-00185 Roma, Italy\\
$^{2}$Institut f\"ur Astrophysik Georg-August-Universit\"at, Friedrich-Hund Platz 1, D-37077 G\"ottingen, Germany\\
$^{3}$ICC, University of Barcelona (IEEC-UB), Marti i Franques 1, E-08028 Barcelona, Spain\\
$^{4}$Hamburger Sternwarte, Universit\"at Hamburg, Gojenbergsweg 112, D-21029 Hamburg, Germany\\
$^{5}$INAF-Arcetri Astrophysical Observatory, Largo Enrico Fermi 5, I-50125 Firenze, Italy\\
$^{6}$Institute of Ion Physics and Applied Physics, University of Innsbruck, Technikerstrasse 25, A-6020 Innsbruck, Austria\\
$^{7}$Scuola Normale Superiore, Piazza dei Cavalieri, 7 I-56126 Pisa, Italy}}

\begin{document}

\newcommand{\krome}{\textsc{Krome} }
\newcommand{\kromes}{\textsc{Krome}}
\newcommand{\enzo}{\textsc{Enzo} }
\newcommand{\enzos}{\textsc{Enzo}}
\newcommand{\flash}{\textsc{Flash} }
\newcommand{\flashs}{\textsc{Flash}}
\newcommand{\ramses}{\textsc{Ramses} }
\newcommand{\ramsess}{\textsc{Ramses}}
\newcommand{\fortran}{\textsc{Fortran} }
\newcommand{\python}{\textsc{Python} }
\newcommand{\pythons}{\textsc{Python}}
\newcommand{\dlsodes}{\textsc{DLSODES} }
\newcommand{\dlsodess}{\textsc{DLSODES}}
\newcommand{\ith}{$i$th }
\newcommand{\jth}{$j$th }
\newcommand{\nth}{$n$th }
\newcommand{\kth}{$k$th }

\newcommand{\dd}{\mathrm d}
\newcommand{\mA}{\mathrm A}
\newcommand{\mB}{\mathrm B}
\newcommand{\mC}{\mathrm C}
\newcommand{\mD}{\mathrm D}
\newcommand{\mE}{\mathrm E}
\newcommand{\mH}{\mathrm H}
\newcommand{\mHe}{\mathrm{He}}
\newcommand{\me}{\mathrm e}
\newcommand{\mSi}{\mathrm Si}
\newcommand{\mO}{\mathrm O}
\newcommand{\mX}{\mathrm X}
\newcommand{\cmc}{\mathrm{cm}^{-3}}
\newcommand{\real}{\mathbb R}
\newcommand{\superscript}[1]{\ensuremath{^{\scriptscriptstyle\textrm{#1}\,}}}
\newcommand{\trader}{\superscript{\textregistered}}

\newcommand{\arl}[1]{\url{#1}}
\newcommand{\eqnref}[1]{Eq.(\ref{#1}) }
\newcommand{\eqnrefs}[1]{Eq.(\ref{#1})}
\newcommand\mnras{MNRAS}
\newcommand\apj{ApJ}
\newcommand\apjs{ApJS}
\newcommand\apjl{ApJL}
\newcommand\aap{A\&A}
\newcommand\apss{Ap\&SS}
\newcommand\ssr{Springer}
\newcommand\jgr{JGeophysRes}
\newcommand\icarus{Icarus}

\newcommand{\tgcomment}[1]{{#1}}

\newcommand\osu{osu\_01\_2007}
\def\tm{\leavevmode\hbox{$\rm {}^{TM}$}}

\date{Accepted *****. Received *****; in original form ******}

\pagerange{\pageref{firstpage}--\pageref{lastpage}} \pubyear{2012}

\maketitle

\label{firstpage}

\begin{abstract}
\tgcomment{Chemistry plays a key role in many astrophysical situations regulating the cooling and the thermal properties of the gas, which are relevant during gravitational collapse, the evolution of disks and the fragmentation process.}

In order to simplify the usage of chemical networks in large numerical simulations, we present the chemistry package \kromes, consisting of a \python pre-processor which generates a subroutine for the solution of chemical networks which can be embedded in any numerical code. For the solution of the rate equations, we make use of the high-order solver \dlsodess, which was shown to be both accurate and efficient for sparse networks, which are typical in astrophysical applications. \krome also provides a large set of physical processes connected to chemistry, including photochemistry, cooling, heating, dust treatment, and reverse kinetics.

The package presented here already contains a network for primordial chemistry, a small metal network appropriate for the modelling of low metallicities environments, a detailed network for the modelling of molecular clouds, a network for planetary atmospheres, as well as a framework for the modelling of the dust grain population. In this paper, we present an extended test suite ranging from one-zone and 1D-models to first applications including cosmological simulations with \enzo and \ramses and 3D collapse simulations with the \flash code. The package presented here is publicly available at \mbox{\arl{http://kromepackage.org/}} and \mbox{\arl{https://bitbucket.org/krome/krome_stable}}.

\end{abstract}


\section{Introduction}\label{introduction}

Chemistry plays a central role in many astrophysical environments, including the formation of stars at low and high metallicities \citep{Omukai2005, Glover2007}, the interstellar medium \citep{Hollenbach1979, Wakelam2010}, starbursts and active galactic nuclei \citep{Maloney1996, Meijerink2005}, protoplanetary disks \citep{Semenov2004, Woitke2009}, planetary atmospheres \citep{Burrows1999, Lodders2002} and the early Universe \citep{DalgarnoLepp87,Stancil1998,Galli1998, Schleicher2008}. It is important as it regulates the cooling, therefore influencing gravitational collapse \citep{Yahil1983, Peters2012}, disk stability \citep{Lodato2007} and fragmentation \citep{Li2003}. At the same time, the chemical abundances regulate the appearance of astrophysical objects by influencing the line emission from atoms, ions and molecules. Both for an accurate modelling, but also to pursue a comparison with observations, it is thus necessary to include chemical models in numerical simulations.

The required machinery is however complex for at least two reasons: (i) chemical kinetics has a non-negligible computational cost, which can be prohibitive for large numerical simulations even after a complexity reduction, as described by \citet{Grassi2012}. It therefore requires a sophisticated framework where the rate equations are efficiently solved. Moreover, (ii) building the set of ordinary differential equations (ODEs) associated with a given chemical network (including the Jacobian and its sparsity structure for efficient evaluation) can be difficult, especially when the number of reactions involved is large ($\gtrsim150$, up to more than 6000).

As the latter presents an important problem, there are a number of attempts to make the incorporation of chemical networks in astrophysical simulations more feasible. For instance, \textsc{XDELOAD} performs a pre-processing task required for the modelling of astrochemical kinetics, including the associated derivatives and the Jacobian \citep{Nejad2005}. The code is available on request, but currently not maintained. \textsc{ASTROCHEM} is a network pre-processor  that is based 
on the Kinetic PreProcessor (KPP)\footnote{\url{http://people.cs.vt.edu/~asandu/Software/Kpp/}}, and is a proprietary code \citep{Kumar2013}. Another code with a similar name is \textsc{astrochem} by S. Maret and aims at studying the evolution of the ISM. It allows to build the associated equations which can be employed within its own framework. This code can be forked on GitHub\footnote{\url{http://smaret.github.com/astrochem/}}, but a newer version has been announced on \citet{Maret2013}. The same approach is followed by \textsc{Nahoon}\footnote{\url{http://kida.obs.u-bordeaux1.fr/models}}, calculating the chemical evolution of a molecular cloud with a pseudo time-dependent approach \citep{Wakelam2012}. The chemical network is created from on a datafile provided by the user. Also this code can build the ODEs 
and the Jacobian only within its framework. Finally,  \textsc{ALCHEMIC} \citep{Semenov2010}  is a code optimized for the chemistry in protoplanetary disks employing the \textsc{DVODPK} (Differential Variable-coefficient Ordinary Differential equation solver with the Preconditioned Krylov method) solver; this code is available on request by the authors. A recent tentative to provide a general package has been made by extracting the \enzo chemistry module to create a sort of primordial chemistry library for astrophysical simulations called \textsc{Grackle}\footnote{\url{https://bitbucket.org/brittonsmith/grackle}}. \tgcomment{Finally, the \textsc{MeudonPDR} code\footnote{\url{http://pdr.obspm.fr/PDRcode.html}} \citep{LePetit2006} allows to study the chemical evolution for a photon-dominated region (PDR) with an incident radiation field. This code includes some thermal processes and a default chemical network, in line with the PDR models.}

However, not all of these codes are publicly available, and often they are restricted to special purposes. An additional restriction is due to the computational efficiency that can be achieved with a given framework. While one-zone and even 1D calculations can still be pursued at a moderate efficiency, including even moderate-sized networks in 3D hydrodynamical simulations presents an additional challenge and requires the usage of optimized numerical techniques that are both accurate and efficient.

For this reason, we have developed the chemistry package \kromes, which includes a pre-processor generating the subroutines for solving the chemical rate equations for any given network that is provided by the user. By default, it provides a set of chemical networks including primordial chemistry, low metallicity gas, molecular clouds and planetary atmospheres. \krome is developed not only to study the evolution of a given chemical network, but also to include many other processes that are tightly connected to astrochemistry (e.g. see Tab.\ref{tab:proc_list}). \tgcomment{In particular \krome permits to add thermal processes, as cooling from endothermic reactions and from several molecules and atoms, but also heating from photochemistry and exothermic chemical reactions. We also include a dust grain model with destruction and formation processes and catalysis of molecular hydrogen on the grain surface. To increase the applicability of \krome the current version allows one to use the rate equation approximation for grain surface chemistry (e.g. see \citealt{Semenov2010}), as well as cosmic-rays ionization which follows a similar scheme.} The \krome pre-processor builds the \fortran subroutines that can be directly included in other simulation codes. The rate equations are solved with the high-order solver \dlsodess, which was recently shown to be both accurate and efficient, as it makes ideal usage of the sparsity in astrochemical networks \citep{Grassi2013, Bovino2013b}. As an additional option we also include the \textsc{DVODE} solver in its \fortran 90 version\footnote{\url{http://www.radford.edu/~thompson/vodef90web/}}.

\begin{table*}
\caption{List of the main processes for different environments. Please note that this list is indicative and may depend on parameters other than temperature and density such as metallicity.}\label{tab:proc_list}
\begin{tabular}{llll}
\hline
&$T$(K)& $n$(cm$^{-3}$)&processes\\
\hline
HIM & $\gtrsim3\times10^5$ 	& $\sim0.004$	& atomic cooling\\
HII & $10^4$			& $0.3-10^4$	& atomic and metal cooling, photoheating\\
WNM & $\sim5\times10^3$ 	& 0.6		& atomic and metal cooling, CR, photoheating, dust sputtering\\
CNM & $\sim100$ 		& $30$		& metal and H$_2$ cooling, dust growth\\
diff H$_2$ 	& $\sim50$ 	& $\sim100$	& H$_2$ and metal cooling\\
dense H$_2$ 	& 10-50 	& $1-10^6$	& CR and photoionisation/dissociation, H$_2$ and HD cooling, dust growth\\
cold dense	& $\lesssim10^2$	& $10^6-10^{14}$	& dust cooling, photoheating, chemical heating\\
warm collapsing	& $\sim10^3$	& $>10^{14}$	& chemical cooling, H$_2$ cooling, chemical heating, CIE\\
\hline
\end{tabular}
\end{table*}

The overall structure of this paper is as follows: in Section 2, we describe the physical modelling including the rate equations, photoionisation reactions, the thermal evolution and the treatment of dust. The computational framework of \krome is described in Section 3, including the code structure,  the evaluation of the ODEs, the solver, the parsing of chemical species and the calculation of inverse reactions (if not provided by the user). In Section 4, we present a test suite including the chemistry in molecular clouds, gravitational collapse in one-zone models with varying metallicity and radiation backgrounds, 1D-shocks, tests for the dust implementation, 1D planetary atmospheres and slow-manifold kinetics. To demonstrate the applicability of the package in 3D simulations, we employ \krome in Section 5 in the hydrodynamical codes \enzos, \ramsess, and \flash to follow the evolution of primordial chemistry and during gravitational collapse. A summary and outlook is finally provided in Section 6.


\section{The physical framework}
\subsection{Rate equations}\label{sect:network}
The evolution of a set of initial species that react and form new species via a given set of reactions is described by a system of ordinary differential equations (ODEs), and it is 
mathematically represented by a Cauchy problem. The ODE associated to the variation of the number density of the \ith species is
\begin{equation}\label{eqn:ODEs}
 \frac{\dd n_i}{\dd t} = \sum_{j\in F_i} \left(k_j\, \prod_{r\in R_j} n_{r(j)}\right) - \sum_{j\in D_i} \left(k_j\, \prod_{r\in R_j} n_{r(j)}\right)\,,
\end{equation}
where the first sum represents the contribution to the differential by the reactions that form the \ith species (belonging to the set $F_i$), while the second part is the analogous for the reactions that 
destroy the \ith species (set $D_i$). The \jth reaction has a set of reactants ($R_j$), and the number density of each reactants at time $t$ of the \jth reaction is $n_{r(j)}$, while the corresponding reaction rate coefficient is $k_j$, which 
is often a function of the gas temperature, but it can be a function of any parameter, including the number densities. The rate coefficient $k_j$ has units of cm$^{3(n-1)}$~s$^{-1}$, where $n$ is the number of reactants of the \jth reaction.
Each species has an initial number density $n_i(t=0)$, and our aim is to find a solution after a given $\Delta t$ to obtain the updated set of number densities $n_i(t=\Delta t)$.

In most cases of astrophysical interest, the system of Eq.(\ref{eqn:ODEs}) is a stiff system, i.e. it has two or more very different scales of the independent variable on which the dependent variables are 
changing \citep{Press1992}, which requires a solver that is tailored for this class of problems (see Section \ref{sect:solver}).

Finally, it is worth noting that also other physical quantities may have their own differential equations often coupled with Eq.(\ref{eqn:ODEs}), like the temperature which will be discussed in Sect.\ref{sect:thermal} below.

\subsection{Photoionisation and photodissociation}\label{sect:phioniz}
To complete the set of the chemical reactions involved, we also consider the reactions of photoionisation and photodissociation in the forms 
\begin{eqnarray}\label{eqn:photo_proc}
	\mathrm A + h\nu &\rightarrow& \mathrm A^+ + \mathrm e^-\nonumber\\
	\mathrm A + h\nu &\rightarrow& \mathrm B + \mathrm C\,,
\end{eqnarray}
where in the first reaction A is an atom or an atomic ion, while in the second case A is a molecule or a molecular ion, and B and C are two generic products.
Following the scheme of \citet{Glover2008} and \citet{Grassi2012} the ionisation reaction is controlled by the rate
\begin{equation}\label{eqn:ph_integral}
	R_\mathrm{ph} = 4\pi\,\int_{E_t}^\infty \frac{I(E)\sigma(E)}{E} e^{-\tau(E)} \dd E\,,
\end{equation}
in units of s$^{-1}$, where $E_t$ is the ionisation potential of the ionized species, $I(E)$ is the energy distribution of the impinging photon flux, $\sigma(E)$ is the cross section of the given process, $\tau(E)$ is the optical depth of the gas, and $E$ is the energy. Standard units are $E$ and $E_t$ in eV, $\sigma$ in cm$^2$, hence $I$ is in eV s$^{-1}$ cm$^{-2}$ Hz$^{-1}$.

The default cross sections for atoms (and ions) are provided by the fit of \citet{Verner1996}, and for molecules we let the user select his own functions, for example following \citet{Glover2008}. The flux is described based on \citet{Efstathiu1992,Vedel1994,Navarro1997}, namely
\begin{equation}\label{eq:flux}
	I(E) = 10^{-21} J_{21} \left(\frac{E_0}{E}\right)^\alpha\,,
\end{equation}
in erg s$^{-1}$ cm$^{-2}$ Hz$^{-1}$, where $E_0=13.6$ eV is the ionisation potential of the hydrogen atom, $\alpha=1$ is a coefficient that controls the slope of the distribution, and $J_{21}$ = $J/(10^{-21}$~erg s$^{-1}$ Hz$^{-1}$ str$^{-1}$).

One of the main problems for this kind of reactions is that the integral of Eq.(\ref{eqn:ph_integral}) has a non-negligible computational cost, since in principle it must be evaluated 
whenever the impinging flux, or the number density of the ionized species are changing, since the optical depth depends on it. To avoid time-consuming procedures for the tests presented with this release of 
\krome, we assume that $I(E)$ is constant with time, and neglect the frequency dependence of $\tau(E)$ by considering only the dominant spectral contribution. 
It is then only a function of the number density of the ionizing species. 
In this way, it is sufficient to integrate Eq.(\ref{eqn:ph_integral}) only at the beginning of a given simulation and directly update the $e^{-\tau}$, which is now a function of the number density only. 
This approximation allows a faster integration of the ODE system associated to the chemical network, but it can be modified by the user to obtain more accurate results in small systems. 

In the current version of \krome we do not include any self-shielding model.

\subsubsection{Cosmic-rays processes}
\tgcomment{Analogously, we include cosmic-ray (CR) ionization and dissociation processes as
\begin{eqnarray}\label{eqn:CR_proc}
	\mathrm A + \mathrm{CR} &\rightarrow& \mathrm A^+ + \mathrm e^-\nonumber + \mathrm{CR}\\
	\mathrm A + \mathrm{CR} &\rightarrow& \mathrm B + \mathrm C +\mathrm{CR}\,,
\end{eqnarray}
modelled by using the following rate approximation
\begin{equation}
	k_\mathrm{CR}=\alpha\zeta\,,
\end{equation}
in units of s$^{-1}$ where $\zeta$ is the molecular hydrogen ionisation rate.}


\subsection{Thermal evolution}\label{sect:thermal}
Many astrophysical problems require to consider the evolution of the temperature along with the evolution of the chemical species. This means that we introduce a new ODE to the system of Eqs.(\ref{eqn:ODEs}), namely
\begin{equation}\label{eqn:cooling}
 \frac{\dd T}{\dd t} = (\gamma-1)\frac{\Gamma(T,\bar n)-\Lambda(T,\bar n)}{k_b\sum_i n_i}\,,
\end{equation}
where $\Gamma$ is the heating term in units of erg~cm$^{-3}$~s$^{-1}$ and is a function of the temperature ($T$) and of the vector that contains the abundances of all the species ($\bar n$), $\Lambda$ is the analogous cooling term, $k_b$ is the Boltzmann constant in erg~K$^{-1}$, and the sum determines the total gas number density in cm$^{-3}$. The dimensionless adiabatic index $\gamma$ is defined as in \citet{Grassi11}
\begin{equation}\label{eq:adiabatic}
	\gamma = \frac{5n_\mathrm{H} + 5n_\mathrm{He} + 5n_\mathrm{e} + 7n_\mathrm{H_2}}{3n_\mathrm{H} + 3n_\mathrm{He} + 3n_\mathrm{e} + 5n_\mathrm{H_2}}
\end{equation}
where $n_X$ is the number density of the element indicated by the subscript. This equation gives $\gamma$ = 5/3 for an ideal atomic gas composed by hydrogen, and $\gamma$ = 7/3 if the gas is made only of molecular hydrogen (see option\footnote{Note that all the options are discussed with more details 
the online guide \mbox{\arl{http://kromepackage.org/}}.} \verb+-gamma=FULL+).
The arguments of both the cooling and the heating function suggest that Eq.(\ref{eqn:ODEs}) and Eq.(\ref{eqn:cooling})  are tightly coupled because of the rate coefficients $k_j(T)$ that control the behaviour of the differential equations associated with the species number densities are temperature dependent, and because the abundances of the species control the behaviour of $\Gamma$ and $\Lambda$, which determine the variation of the temperature of the system. In some cases, the two sets of equations can be decoupled and one can decide to calculate the temperature by using Eq.(\ref{eqn:cooling}) independently at the beginning of the integration time-step. With this assumption the time-step must be chosen small enough to avoid large changes of the temperature. In the present scheme we solve both equations simultaneously which is more stable and accurate, but the user can also decide to solve the thermal evolution independently by introducing $\dd T/\dd t = 0$ and determining the temperature out of the system of ODEs with an appropriate cooling and heating functions (see option \verb+-skipODEthermo+). 

\subsubsection{Cooling}\label{sect:cooling}

\krome already incorporates a variety of cooling functions that can be employed for different applications. A short summary is given in Tab.\ref{tab:cooling_function}.\\

\begin{table*}
	\begin{center}
		\caption{Cooling physical processes included in the \krome thermal model.} \label{tab:cooling_function}
		\begin{tabular}{l r}
		  \hline
		  Process & Ref. \\
		  \hline
		  Atomic cooling\\
		  \hline
		  H, He, He$^+$ collisional ionisation & \citet{Cen1992}\\
		  H$^+$, He$^+$, He$^{2+}$ recombination & \citet{Cen1992}\\
		  He dielectric recombination & \citet{Cen1992}\\
		  H (all levels) collisional excitation & \citet{Cen1992}\\
		  He (2,3,4 triplets) collisional excitation & \citet{Cen1992}\\
		  He$^+$ ($n=2$) collisional excitation & \citet{Cen1992} \\
		  bremsstrahlung all ions & \citet{Cen1992}\\
                  \hline
                  Molecular cooling\\
                  \hline
                  H$_2$ roto-vibrational lines & \citet{Galli1998} \\ 
                  H$_2$ roto-vibrational lines & \citet{Glover2008}\\ 
		  Collisionally induced emission (CIE) & \citet{Ripamonti2004}\\
                  HD roto-vibrational lines & \citet{Lipovka2005}\\
		  H$_2$ collisional dissociation & \citet{Martin1998,Glover2007} \\
		  \hline
		  Other processes\\
                  \hline
	 	  Metals (C, O, Si, Fe and ions) & \citet{Maio2007,Grassi2012}\\
		  Compton cooling & \citet{Cen1992,Glover2007} \\
		  Dust cooling & \citet{Hollenbach1979} \\ 
                  Continuum & \citet{Omukai2000,Lenzuni1991}\\
                  \hline
		\end{tabular}
	\end{center}
\end{table*}

\begin{figure}
	\includegraphics[width=0.45\textwidth]{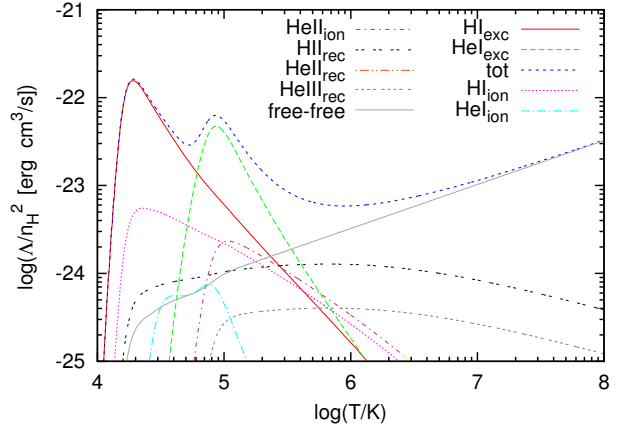}
        \caption{Cooling rates as a function of the temperature for a primordial gas in collisinal equilibrium.}\label{fig:katz96}
\end{figure}

\textsc{Atomic cooling:} the cooling from \citet{Cen1992} considers the collisional ionisation of H, He, and He$^+$ by electrons, the recombination of H$^+$, He$^+$, and He$^{++}$, the dielectronic recombination of He, the collisional excitation of H (all levels), He (2, 3, 4 triplets), and He$^+$ (level $n=2$), and finally bremsstrahlung for all the ions. The rates are included in \krome as they are listed in \citet{Cen1992}. 
In Fig.~\ref{fig:katz96} we report the cooling rates obtained by following the approach described in \citet{Katz96}, i.e. considering collisional equilibrium abundances at any given temperature: the collisional excitation contributions coming from H and He dominate the cooling until the free-free transitions (bremsstrahlung) become important (see option \verb+-cooling=ATOMIC+).\\

\textsc{Molecular hydrogen:} the cooling by H$_2$ has been included following two models, namely \citet{Galli1998} and \citet{Glover2008}. In both cases the final functional form of the total cooling in units of erg~cm$^{-3}$~s$^{-1}$ is given as 
\begin{equation}\label{eq:H2_cooling}
	\Lambda_\mathrm{H_2} = \frac{n_\mathrm{H_2}\Lambda_\mathrm{H_2,LTE}}{1+\Lambda_\mathrm{H_2,LTE}/\Lambda_\mathrm{H_2,n \rightarrow 0}}\,,
\end{equation}
and the high-density limit is the same, expressed by $\Lambda_\mathrm{H_2,LTE}=H_R+H_V$ as the sum of the vibrational and the rotational cooling at high densities 
\begin{eqnarray}
 H_R &=& (9.5\times10^{-22}T_3^{3.76})/(1+0.12 T_3^{2.1})\nonumber\\
	&\cdot&\exp\left[-(0.13/T_3)^3\right]+3\times10^{-24}\exp\left[-0.51/T_3\right]\nonumber\\
 H_V &=& 6.7\times10^{-19}\exp(-5.86/T_3)\nonumber\\ 
	&+& 1.6\times10^{18}\exp(-11.7/T_3)\,,
\end{eqnarray}
with $T_3=T/10^3$ \citep{Hollenbach1979}.

The two cooling functions have however different low density limits: (i) in the \citet{Galli1998} the following approximation is used 
\begin{eqnarray}\label{eq:hd_pall}
\log(\Lambda_\mathrm{H_2,n \rightarrow 0}) = [-103 +97.59\log(T)-48.05\log(T)^2\nonumber\\ 
10.8\log(T)^3 -0.9032\log(T)^4]\,n_\mathrm{H}\,,
\end{eqnarray}
and the final cooling is valid in the range $13<T<10^5$ K; (ii) the low-density limit cooling by \citet{Glover2008} considers H and He as the colliding partners in the temperature range $10<T<6\times10^3$ K, and H$^+$ and  e$^-$ in the range $10<T<10^4$ K, and is expressed as 
\begin{equation}
	\Lambda_\mathrm{H_2,n \rightarrow 0} = \sum_k{\Lambda_{\mathrm{H_2},k} n_k}\,,
\end{equation}
with $k={ \mH, \mH_2, \mHe, \mH^+, \me^-}$. Each term of the sum is reported in Table \ref{tab:total-coeffs} considering an ortho to para ratio of 3:1.

\begin{figure}
	\includegraphics[width=0.45\textwidth]{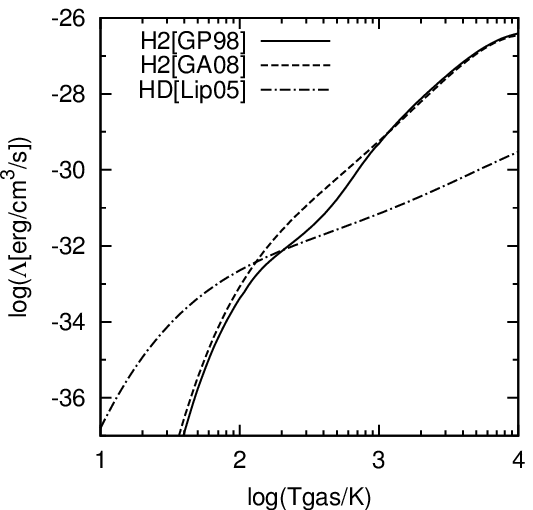}
        \caption{Cooling functions for H$_2$ as in \citet{Glover2008} and \citet{Galli1998}, and HD as in \citet{Lipovka2005}.}\label{fig:H2HDcool}
\end{figure}

Both cooling functions are reported in Fig.\ref{fig:H2HDcool} using $n_{\mH}=~1$ cm$^{-3}$, $n_{\mH_2}=10^{-5}$ cm$^{-3}$, $n_{\me^-}=10^{-4}$ cm$^{-3}$, and $n_{\mH\mD}=10^{-8}$ cm$^{-3}$ as in \citet{Maio2007}.

It should be noted that the above approach is valid only in the optically thin limit, while once the gas becomes optically thick ($n \gtrsim 10^{10}$ cm$^{-3}$) an opacity term should be included. We follow the model by \citet{Ripamonti2004} and define the H$_2$ cooling in the optical thick regime as

\begin{equation}
	\Lambda_\mathrm{H_2,thick} = \Lambda_\mathrm{H_2,thin}\times \mathrm{min}\left[1,\left(\frac{n}{8\times10^9 \mathrm{cm^{-3}}}\right)^{-0.45}\right]\,,
\end{equation}
where $\Lambda_\mathrm{H_2,thin}$ = $\Lambda_\mathrm{H_2}$ as in Eq.(\ref{eq:H2_cooling}). See options \verb+-cooling=H2+ and \verb+-cooling=H2GP98+.
\\

\textsc{HD:} we follow \citet{Lipovka2005}, who provide a fit that is a function of the gas temperature and the density. In \kromes, it is computed by using two nested loops to give 
\begin{equation}\label{eqn:fitHD}
	\Lambda_\mathrm{HD} = \left[\sum_i\sum_j c_{ij} \log(T)^i\log(n_\mathrm{tot})^j\right]\,n_\mathrm{HD}\,,
\end{equation}
where $c_{ij}$ are the fit data provided by Tab.\ref{tab:lipHD}.
The cooling function is reported in Fig.\ref{fig:H2HDcool} as discussed in the previous paragraph. Note that the HD cooling has been extended under 100 K following the implementation of \citet{Maio2007}. See option \verb+-cooling=HD+.\\

\begin{table*}
	\begin{center}
		\caption{Coefficients $c_{ij}$ used  in the Eq. (\ref{eqn:fitHD}) and
		taken from \citet{Lipovka2005}.} \label{tab:lipHD}
		\begin{tabular}{r r r r r r}
		  \hline
		   & $i=0$ & $i=1$ & $i=2$ & $i=3$ & $i=4$\\
		  \hline
		  $j=0$ & $-42.56788$ &$0.92433$ &$0.54962$ &$-0.07676$&$0.00275$\\
		  $j=1$ & $21.93385$ &$0.77952$ &$-1.06447$ &$0.11864$&$-0.00366$\\
		  $j=2$ & $-10.19097$ &$-0.54263$ &$0.62343$ &$-0.07366$&$0.002514$\\
		  $j=3$ & $2.19906$ &$0.11711$ &$-0.13768$ &$0.01759$&$-0.00066631$\\
		  $j=4$ & $-0.17334$ &$-0.00835$ &$0.0106$ &$-0.001482$&$0.000061926$\\
		  \hline
		\end{tabular}
	\end{center}
\end{table*}

\textsc{Collisionally induced emission:} the collisionally induced emission (CIE) cooling is very important at high densities and represents the continuum emission of a photon due to the formation of a ``supermolecule'' with a non-zero electric dipole induced by collisions between pairs of atoms/molecules (H$_2$-H$_2$, H$_2$-He, H$_2$-H). We follow the fit given by \citet{Ripamonti2004} and also discussed in \citet{Hirano2013}:
\begin{equation}\label{eq:cie}
	\Lambda_\mathrm{CIE,thick} = \Lambda_\mathrm{CIE,thin}\times \mathrm{min}\left[1,\frac{1-\mathrm{e^{-\tau_{CIE}}}}{\mathrm{\tau_{CIE}}}\right]
\end{equation}
where
\begin{equation}
       \mathrm{\tau_{CIE}} = \left(\frac{n_\mathrm{H_2}}{7\times10^{15} \mathrm{cm}^{-3}}\right)^{2.8}.
\end{equation}

The optically thin term ($\Lambda_{\mathrm{thin}}$) has been fitted based on the values of \citet{Borysow2002,Borysow2001} which include collisions between H$_2$-H$_2$ and H$_2$-He. The original data, defined in the temperature range of 400-7000 K, have been extended and fitted by using three different functions and are valid between 100-10$^6$ K. The expressions are:
\begin{eqnarray}\label{eqn:CIE}
\log(\Lambda_\mathrm{CIE,thin}) =
\left\{
        \begin{aligned}
	& \sum_{i=0}^5 a_i(\log T)^i \qquad 100<T<891\,\mathrm{K}\\
	&  \sum_{i=0}^5 b_i(\log T)^i \qquad 891\le T<10^5\,\mathrm{K}\\
	& c\log(T)-d \qquad \qquad T\ge10^5\,\mathrm{K}
         \end{aligned}
         \right. 
\end{eqnarray}
and the fitting coefficient are reported in Table \ref{tab:cie_fitting}.

\begin{table}
	\begin{center}
		\caption{Collisionally induced emission (CIE) fitting coefficients for the fit provided in Eq.\ref{eqn:CIE}.} \label{tab:cie_fitting}
		\begin{tabular}{r r r}
		  \hline
		   & $a_i$ & $b_i$\\
		  \hline
		   $i=0$ & $-30.3314216559651$ & $-180.992524120965$\\ 
		   $i=1$ & $19.0004016698518$  & $168.471004362887$ \\
                   $i=2$ & $-17.1507937874082$ & $-67.499549702687$  \\
                   $i=3$ & $9.49499574218739$  &$13.5075841245848$   \\
                   $i=4$ & $-2.54768404538229$ & $-1.31983368963974$ \\
                   $i=5$ &$0.265382965410969$ & $0.0500087685129987$\\
                  \hline
		   & $c$ & $d$  \\
	           \hline
                   &  $3.0$ & $21.2968837223113$\\
                  \hline
		\end{tabular}
	\end{center}
\end{table}

The fit provided by \citet{Ripamonti2004} has been tested for environments where the H$_2$ fraction is still important ($\sim$0.5) and might not work for extremely dissociated media. In addition \citet{Hirano2013} have shown a substantial difference between the approximation by \citet{Ripamonti2004} and their more detailed treatment. They concluded that to accurately follow the thermal and chemical evolution of a primordial gas to high densities well posed physical assumptions should be made. We suggest the \krome users to carefully use the current CIE implementation and to check if it is suitable for their specific problem. See option \verb+-cooling=CIE+.\\

\textsc{Continuum:} the CIE cooling can be extended by including other processes as bound-free absorption by H$^0$ and H$^-$, free-free absorption by H$^0$, H$^-$, H$_2$, H$_2^-$, H$_2^+$, H$_3^+$, He$^0$, and He$^-$, photodissociation of H$_2$, and H$_2^+$ by thermal radiation, Rayleigh scattering by H$^0$, H$_2$ and He$^0$, and Thomson scattering by e$^-$. The gas has a continuum emission due to its temperature, but, in order to take into account the different processes that influence the opacity we use the opacity calculations found in \citet{Lenzuni1991}, which includes CIE and the other processes listed above. We follow the formulation of \citet{Omukai2000} that employs the Planck opacity as the process opacity into the cooling function
\begin{equation}
	\Lambda_\mathrm{cont} = 4\,\sigma_\mathrm{sb} T^4 \kappa_\mathrm{p} \rho_\mathrm{g} \beta\,,
\end{equation}
where $\sigma_\mathrm{sb}$ is the Stefan-Boltzmann constant, $T$ is the gas temperature, $\kappa_\mathrm{P}$ is the \citet{Lenzuni1991} opacity (see discussion below), $\rho$ is the gas density, $\beta$ is an opacity-dependant term given by
\begin{equation}\label{eqn:beta}
	\beta = \min\left(1,\tau^{-2}\right)\,,
\end{equation}
with
\begin{equation}
	\tau = \lambda_\mathrm{J} \kappa_\mathrm{p} \rho_\mathrm{g} + \epsilon \,,
\end{equation}
where $\epsilon=10^{-40}$ is intended to avoid a division by zero in Eq.(\ref{eqn:beta}), and $\lambda_\mathrm{J}$ is the Jeans length defined by
\begin{equation}
	\lambda_\mathrm{J} = \sqrt{\frac{\pi k_b T}{\rho_\mathrm{g} m_\mathrm{p} G \mu}}\,,
\end{equation}
being $k_b$ the Boltzmann constant, $m_\mathrm{p}$ the proton's mass, and $\mu$ the mean molecular weight (in our case we use a constant value of 1.22, but in principle it can be evaluated from the specific gas composition).

It is worth noting that the density- and temperature-dependent fit of $\kappa_\mathrm{p}$ proposed by \citet{Lenzuni1991} has a wrong formulation (probably a typo in their Tab.14) which leads to non-physical results. For this reason we propose here a simpler density-dependent fit which can be safely employed for $T\le3\times10^4$ K regime. We found for $\kappa_\mathrm{p}$ in units of cm$^{2}$~g$^{-1}$
\begin{equation}
	\log\left(\kappa_\mathrm{p}\right) = a_0 \log\left(\rho_\mathrm{g}\right) + a_1\,,
\end{equation}
with $a_0=1.000042$ and $a_1=2.14989$, while we assume that the fit is valid from $\rho_\mathrm{g}>10^{-12}$ g~cm$^{-3}$ (otherwise $\kappa_\mathrm{p}=0$) and can be extrapolated over $0.5$ g~cm$^{-3}$ by defining $\rho_\mathrm{g} = \min\left(\rho_\mathrm{g},0.5\,\mathrm{g\,cm}^{-3}\right)$. See option \verb+-cooling=CONT+.\\

\textsc{Chemical cooling:} according to \citet{Omukai2000} the processes related to the reactions listed in Tab.\ref{tab:chem} remove energy from the gas. The \jth reaction removes an energy of $E_j$ giving a cooling of
\begin{equation}\label{eqn:cool_chem}
	\Lambda_j = E_j\,k_{j}\,n(R_{j1})\,n(R_{j2})\,,	
\end{equation}
where $k_j$ is the reaction rate coefficient, while $n(R_{j1})$ and $n(R_{j2})$ are the abundances of the two reactants.
All these reactions listed in Tab.\ref{tab:chem} are recognized by \krome from the user-defined chemical network and added automatically to the cooling calculation.
The total amount of cooling is then
\begin{equation}
	\Lambda_\mathrm{CHEM} = \sum_{j}\Lambda_j\,,
\end{equation}
with $j$ that runs over the reactions of Tab.\ref{tab:chem} found by \krome in the given reaction network. The cooling is in unit of eV~cm$^{-3}$~s$^{-1}$ if $E_j$ in eV. See the analogous heating in Sect.\ref{sect:heating}. See option \verb+-cooling=CHEM+.

\begin{table}
	\caption{Reactions associated to the cooling processes recognized by \krome including the energy of the process in eV.\label{tab:chem}}
	\centering
	\begin{tabular}{lll}
		\hline
		&Reaction & Energy (eV)\\	
		\hline
		1 & $\mH + \me^- \to \mH^+ + 2\me^-$ & $13.6$\\
		2 & $\mHe + \me^- \to \mHe^+ + 2\me^-$ & $24.6$\\
		3 & $\mHe^+ + \me^- \to \mHe^{++} + 2\me^-$ & $24.6$\\
		4 & $\mH_2 + \mH \to 3\mH$ & $4.48$\\
		5 & $\mH_2 + \me^- \to 2\mH + \me^-$ & $4.48$\\
		6 & $\mH_2 + \mH_2 \to \mH_2 + 2\mH$ & $4.48$\\
		\hline
	\end{tabular}
\end{table}

\textsc{Metals:} each metal has a different number of energy levels, transitions, and colliding partners listed in Tab. (\ref{tab:info_zcool}), and their values are reported in several works \citep{Hollenbach1989,Maio2007,Glover2007,Grassi2012}. It is important to remark that metal cooling requires a complex implementation, due to the fact that we must know the distribution of the metal population in the fine-structure levels to determine the exact amount of cooling. These calculations have a non-negligible computational cost since we need to solve a linear system of equations in which the number densities of the metals and their collision partners are involved.

Each metal has a matrix $M$ that contains the transition probabilities between the levels and it has the following components
\begin{eqnarray}
	M_{ij}&=&\sum_k n_k \gamma_{ij}^{(k)}\\
	M_{ji}&=&\sum_k n_k \gamma_{ji}^{(k)} + A_{ji}\,,\\
\end{eqnarray}
where $k$ is the index of the \kth collider, $\gamma_{ji}$ is the reaction rate for the $j\to i$ transition (de-excitation), $A_{ji}$ is the Einstein coefficient for the spontaneous transitions. Note that the rates for the de-excitation and the excitation are related by using
\begin{equation}
	\gamma_{ij}=\frac{g_j}{g_i}\gamma_{ji}\exp{\left[\frac{-\Delta E_{ji}}{k_bT}\right]}\,,
\end{equation}
where $g_i$ and $g_j$ are the statistical weights, $k_b$ is the Boltzmann constant, $T$ is the gas temperature, and $\Delta E_{ij}$ is the energy separation between the \ith and the \jth levels.

To find the population distribution between $N$ levels we must solve an $N\times N$ linear system consisting of (i) a conservation equation
\begin{equation}
	\sum_i n_i = n_\mathrm{tot}\,,
\end{equation}
for the excited levels of the given metal, and (ii) $N-1$ equations as 
\begin{equation}\label{eqn:linearsys}
	n_i \sum_j M_{ij} = \sum_j n_j M_{ji}\,,
\end{equation}
where $M$ is the matrix with the transition probabilities and $n_i$ are the number densities of the \ith level which represent the unknowns of this linear system \citep{Santoro2006,Maio2007}. In \krome this linear system is solved using the \textsc{LAPACK} function \textsc{dgesv} \citep{LAPACK} which is robust and can be found in several optimised libraries.
Once the number densities of each level $n$ are found, the total metal cooling is $\Lambda_Z=\sum_{ij}n_i\Delta E_{ij}A_{ij}$, which is the sum of the energy losses in each level decay $i\to j$ of each metal \citep{Maio2007,Grassi2012}.

\begin{figure}
	\includegraphics[width=0.45\textwidth]{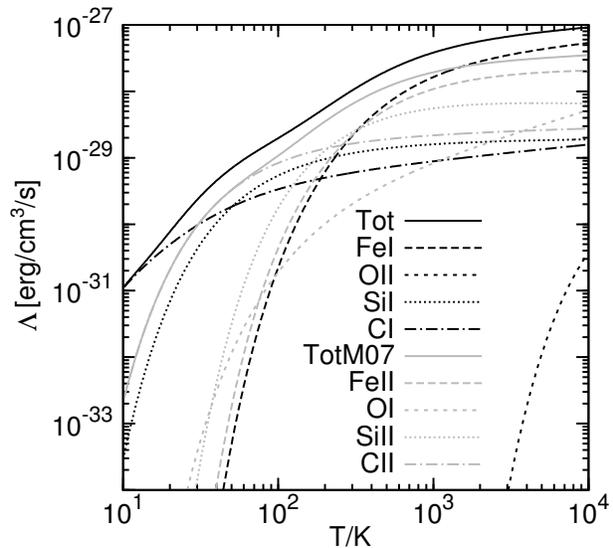}
        \caption{Cooling functions for CII, OI, SiII, FeII, and their total (grey) as in \citet{Maio2007}, and for CI, OII, SiI, and FeI as in the text. Note that the solid black line represents the total of all the contributions. See text and appendix for details.}\label{fig:zcool}
\end{figure}

We report in Fig.\ref{fig:zcool} a comparison of the cooling rates for the C, O, Si, Fe, and their first ions with $n_{\mH}=1$ cm$^{-3}$, $n_{\me^-}=10^{-4}$ cm$^{-3}$, and each metal $n_{\mX}=10^{-6}$ cm$^{-3}$ as in \citet{Maio2007}. See option \verb+-cooling=Z+ and the analogous for individual metals.\\

\begin{table}
	\caption{Characteristics of the atomic systems adopted for metal cooling, and colliding partners. Note that H$_2$ is split in para and ortho with a standard default ratio of 1:3, but in principle it can be modified in the code.\label{tab:info_zcool}}
	\centering
	\begin{tabular}{l|l|l|l}
		\hline
		Coolant & Levels & Transitions & Partners\\	
		\hline
		C & 3 & 3 & H$_2$, H, H$^+$, e$^-$\\
		O & 3 & 3 & H$_2$, H, H$^+$, e$^-$\\
		Si & 3 & 3 & H, H$^+$\\
		Fe & 5 & 6 & H, e$^-$\\
		C$^+$ & 2 & 1 & H, e$^-$\\
		O$^+$ & 3 & 3 & e$^-$\\
		Si$^+$ & 2 & 1 & H, e$^-$\\
		Fe$^+$ & 5 & 5 & H, e$^-$\\
		\hline
	\end{tabular}
\end{table}

\textsc{Dust:} The dust cooling is included following \citet{Hollenbach1979,Omukai2000,Schneider2006}, who employ the expression
\begin{equation}\label{eqn:dust_cool}
	\Lambda_\mathrm{dust} = 2 \pi a^2 n_g n_d v_g k_b(T - T_d)\,,
\end{equation}
where $n_g$ and $n_d$ are the number densities of the gas and the dust respectively, $v_g$ is the gas thermal velocity, $k_b$ is the Boltzmann constant, while $T_d$ and $T$ represent the temperature of the dust and the gas, and $\pi a^2$ is the grain cross section. Note that when the dust is hotter than the gas (i.e. $T_d>T$) Eq.(\ref{eqn:dust_cool}) acts as a heating term for the gas. See option \verb+-cooling=DUST+.\\

\textsc{Compton cooling}: According to \citet{Cen1992}, for cosmological simulations of ionized media it is important to include the Compton scattering of CMB photons by free electrons, the so called Compton cooling. We employ the formula reported by \citet{Glover2007} who refer to the original work of \citet{Cen1992}. Then the cooling function in units of erg~cm$^{-3}$~s$^{-1}$ is expressed by: 
\begin{equation}\label{eq:compton}
	\Lambda_\mathrm{compt} = 1.017\times10^{-37}T_{CMB}^4(T-T_{CMB})n_e
\end{equation} 
with $T_{CMB} = 2.73 (1 + z)$ in K and $z$ being the redshift. See option \verb+-cooling=COMPTON+.

\subsubsection{Heating}\label{sect:heating}
The heating implemented in \krome is divided in three components: chemical ($\Gamma_\mathrm{chem}$), compressional ($\Gamma_\mathrm{compr}$), and photoheating ($\Gamma_\mathrm{ph}$). The total heating in Eq.(\ref{eqn:cooling}) is the sum of these three heating terms in units of erg~cm$^{-3}$~s$^{-1}$. Each heating term can be switched on or off depending on the type of the environment chosen. The compressional heating is included in order to simulate cloud collapse processes in one-zone tests.

\textsc{Chemical heating:} The main contribution that we currently consider is the heating due to the H$_2$ formation. We choose to employ the heating functions from \citet{Omukai2000} that describe the formation of H$_2$ via H$^-$, H$_2^+$, and three-body reactions. According to \citet{Hollenbach1979} the heat deposited per formed molecular hydrogen is weighted by a critical density factor
\begin{equation}
	f = \left(1 + \frac{n_{cr}}{n_\mathrm{tot}}\right)^{-1}
\end{equation}
with $n_{cr}$ in cm$^{-3}$ defined as
\begin{eqnarray}
	n_{cr} &=& 10^6 T^{-1/2} \left\{1.6\,n_\mH \exp\left[-\left(\frac{400}{T}\right)^2\right]\right.\\
		&+&\left.1.4\,n_{\mH_2} \exp\left[-\frac{12000}{T + 1200}\right]\right\}\,.
\end{eqnarray}
The total chemical heating is then given as:
\begin{equation}
	\Gamma_\mathrm{chem} = \Gamma_\mathrm{H_2,3b} + \Gamma_\mathrm{H^-} + \Gamma_\mathrm{H_2^+}\,,
\end{equation}
where the individual terms are based on the amount of energy deposited and the reaction rate of the process (see Tab.\ref{tab:rates}), which leads to the formation of an H$_2$ molecule:
\begin{equation}\label{eqn:heat_3b}
	\Gamma_\mathrm{H_2,3b} = 4.48 f k_\mathrm{H_2,3b} n_{\mH}^3\qquad \mathrm{eV\,cm^{-3} s^{-1}}
\end{equation}

\begin{equation}\label{eqn:heat_hm}
	\Gamma_\mathrm{H^-} = 3.53 f k_\mathrm{H^-} n_{\mH}n_{\mH^-}\qquad \mathrm{eV\,cm^{-3} s^{-1}}
\end{equation}

\begin{equation}\label{eqn:heat_h2p}
	\Gamma_\mathrm{H_2^+} = 1.83 f k_\mathrm{H_2^+} n_{\mH} n_{\mH_2^+}\qquad \mathrm{eV\,cm^{-3} s^{-1}}\,,
\end{equation}
where $k_i$ represents the rate coefficient of the relative chemical process that \krome automatically chooses from the reactions network. The  template reactions recognized by \krome are $3\mH\to\mH_2+\mH$ and $2\mH+\mH_2\to2\mH_2$ for processes in Eq.(\ref{eqn:heat_3b}), $\mH^-+\mH\to\mH_2+\me^-$ for Eq.(\ref{eqn:heat_hm}), and $\mH_2^++\mH\to\mH_2+\mH^+$ for Eq.(\ref{eqn:heat_h2p}). See option \verb+-heating=CHEM+.\\

\textsc{Molecular hydrogen formation on dust:} analogously, the formation of H$_2$ on dust grains (see Sect.\ref{sect:H2dust}) provides a contribution to the total heating 
\begin{equation}
	\Gamma_\mathrm{H_2dust} = k_\mathrm{d} (0.2 + 4.2 f) n_{\mH} n_\mathrm{d}\qquad \mathrm{eV\,cm^{-3} s^{-1}}\,,
\end{equation}
where $k_\mathrm{d}$ is the rate coefficient for the formation of the dust on the grain surface and $n_\mathrm{d}$ is the dust number density. In \krome this heating term is included in the chemical heating routine described above. See option \verb+-heating=CHEM+.\\

\textsc{Compressional:} the compressional heating is defined as in \citet{Glover2008} and \citet{Omukai2000}, 
from who we derive the following expression
\begin{equation}
	\Gamma_\mathrm{compr} = \frac{n_\mathrm{tot}k_bT}{t_{ff}}
\end{equation}
where $n$ the total number density, $k_b$ the Boltzmann constant, $T$ the gas temperature, and  $t_{ff}$  the free fall time defined as
\begin{equation}\label{eqn:ff_time}
	t_{ff} = \sqrt{\frac{3\pi}{32G\rho_\mathrm{g}}}
\end{equation}
and $G$ and $\rho_\mathrm{g}$ the gravitational constant and the mass particle density in g~cm$^{-3}$, respectively. See option \verb+-heating=COMPRESS+.\\

\textsc{Photoheating:} this heating term originates from a photoionisation process that releases energy in the gas transferring the photon energy $h\nu$ to a free electron. It is modelled following the same approach as in Sect.\ref{sect:phioniz}, and hence we employ the equation \citep{Glover2008,Grassi2012},
\begin{equation}\label{eqn:heat_integral}
	H_\mathrm{ph} = 4\pi\,\int_{E_t}^\infty \frac{I(E)\sigma(E)}{E} e^{-\tau(E)} (E-E_t) \eta(E)\dd E\,,
\end{equation}
where the terms have the same meaning as in Eq.(\ref{eqn:ph_integral}), with $E_t$ the ionisation energy of the given species and $\eta(E)$ an efficiency factor that determines the amount of energy released into the gas. The heating $H_\mathrm{ph}$ is in units of erg~s$^{-1}$ while $E$ and $E_t$ are in erg. To determine the total energy that heats the gas the latter quantity must be weighted by the number density of the ionizing species as 
\begin{equation}
 \Gamma_\mathrm{ph} = H_\mathrm{ph}n_\mathrm{X}
\end{equation}
when the reaction has the form $\mathrm X + h\nu \to \mathrm X^+ + \mathrm e^-$, and in this case we have that $\Gamma_\mathrm{ph}$ is in units of erg~cm$^{-3}$~s$^{-1}$. See option \verb+-heating=PHOTO+.

\subsection{Dust}\label{sect:dust}
The scheme proposed here for the dust is a simplified version of the one in \citet{Grassi2011} that includes dust formation via aggregation and destruction via sputtering. In this first release, we do not include any shuttering process since its implementation depends on the type of model one wants to study. The dust is modelled with a standard distribution of bins of different sizes following an MRN profile \citep{Mathis1977,Draine1984} as $\dd n(a)/\dd a \propto a^{-3.5}$ where $a$ is the grain size and $n(a)$ is the number density of the grain with the size $a$ in the default range $5<a<2500$ \AA. For small values of $a$, the distribution includes also the PAHs (e.g. pentacene is $\approx6$\AA) as suggested by \citet{LiDraine2001b}: this is a crude approximation, and more complex models are suggested by \citet{Weingartner2001}, but for the aims of this code the simplified version is detailed enough. However, both the distribution limits and the MNR index are customisable in \krome as well as the number of bins.

We divide the dust into two components: carbon-based (carbonaceous) and silicon-based (silicate) grains that have different physical properties. In \krome the user can improve this simple two-components model by including more types of dust and hence building up a more realistic model. Note that this release of \krome is not intended to deal more advanced dust models, and adding specific type of dust requires a good knowledge of \kromes. However, in the future releases of \krome we plan to improve the dust modelling increasing the dust features and adding a user-friendly dedicated module.

The dust population changes with time, since (i) the free atoms in the gas can stick to the grains and then they grow (e.g. carbon atoms with carbon grains), and (ii) the dust is sputtered by the hot gas atoms that collide with the grains (thermal sputtering) as already stated. At the present time we do not include any detailed grain-grain shuttering processes \citep{Hirashita2009} and vaporisation \citep{Tielens1994}, but we consider this process only to slow down the grain growth via a delay factor $c_d$ (see below). See options \verb+-dust=+ and \verb+-dustOptions=+.

\subsubsection{Growth}\label{sect:growth}
We have modelled the dust formation following \citet{Dwek1998} and \citet{Grassi2011} who calculated the grain accretion as a function of time as
\begin{equation}\label{eqn:growth}
	\frac{\dd n(a)}{\dd t} = c_d \alpha\left[T,T_d(a)\right]\pi a^2 n_g \left[n_d(a)+n_\mathrm{seed}\right] v_g\,,
\end{equation}
where $c_d$ is a delay factor that will be discussed further, $T$ and $T_d(a)$ are the gas and the dust temperatures in K, respectively, $\pi a^2$ is the grain cross section in cm$^2$, $n_g$ is the number density of the gas partner (e.g. for carbon-based is $n_\mathrm{C}$, while $n_d(a)$ is the dust number density (cm$^{-3}$), and $v_g=\sqrt{8k_b T/\pi/m_p}$ is the thermal speed of the gas in cm~s$^{-1}$ with $k_b$ the Boltzmann constant (erg~K$^{-1}$) and $m_p$ the mass of the proton in grams. Finally, $n_\mathrm{seed}=10^{-12}$ cm$^{-3}$ is a dust seed that allows the grain formation when $n_d=0$.

\begin{figure}
 \includegraphics[width=.47\textwidth]{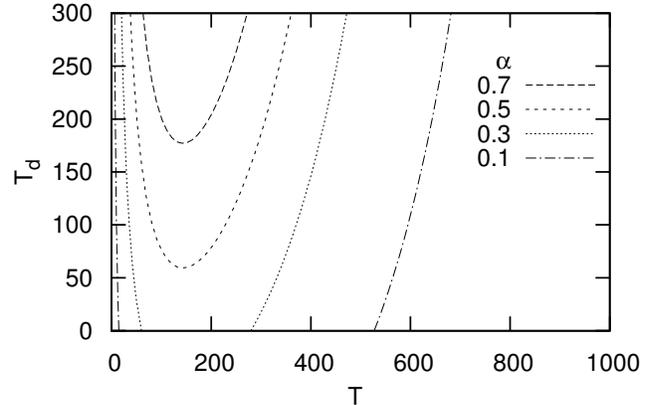}
 \caption{Plot contour of the sticking coefficient $\alpha$ as a function of the gas ($T$) and the dust ($T_d$) temperature. \label{fig:alpha}}
\end{figure}

The dimensionless delay factor $c_d\approx10^{-3}$ takes into account phenomena such as the evaporation caused by cosmic rays or UV heating, and the grain-grain collisions that can reduce the efficiency of the dust growth and may depends on the type of environment studied \citep{Dwek1998,Grassi2011}.
The whole process is controlled by a sticking coefficient $\alpha\left[T,T_d(a)\right]$ that is a fit to the data of \citet{LeitchDevlin1985} of the form
\begin{eqnarray}\label{eqn:alpha}
	 \alpha &=& 1.9\times10^{-2}\, T (1.7\times10^{-3}\, T_d + 0.4) \nonumber\\ 
		&\times& \exp(-7\times10^{-3}\, T)\,,
\end{eqnarray}
which is valid in the range $10<T<1000$ K and $10<T_d<300$ K and is plotted in Fig. \ref{fig:alpha}. This function has been developed for sticking carbon atoms over a carbon lattice, but in our model we assume that this can be employed also for other types of dust such as silicates. However, to save CPU time, Eq.(\ref{eqn:alpha}) can be approximated for low gas temperature environments by $\alpha=0.3$, keeping a good accuracy as suggested by \citet{Hirashita2011}. See option \verb+-dustOptions=GROWTH+.

\subsubsection{Thermal sputtering}\label{sect:sputter}
The thermal sputtering is due to the collisions between the gas and the grains, and to model this in \krome we follow the scheme proposed by \citet{Grassi2011} adapted with the yields from \citet{Nozawa2006}. This can be modelled considering the gas-dust impact probability given by
\begin{equation}\label{eqn:probsput}
	P(T,a) = n_g n_d(a) a^2 v_g\,,
\end{equation}
where $n_g$ is the total gas number density, $n_d(a)$ the amount of dust of the given size $a$, $a^2$ the grain cross-section, and $v_g$ the thermal speed of the gas. Each impact has a yield factor from \citet{Nozawa2006} fitted together with the thermal speed as
\begin{equation}\label{eqn:yield}
	\log\left[Y(T)\,v_g\right] = \frac{\exp\left[-a_0 \log(T)\right]}{a_1 + a_2 \log(T)} + a_3\,,
\end{equation}
with  $a_0 = -0.392807$, $a_1 = 19.746828$, $a_2 = -4.003187$, and $a_3 = 7.808167$, and the fit valid for $T>10^5$ K. The yield in Eq.(\ref{eqn:yield}) represents the number of atoms removed per impact, and hence we obtain
\begin{equation}\label{eqn:sputter}
	\frac{\dd n(a)}{\dd t} =  \frac{Y(T) n_g n_d(a) a^2}{\eta}\,,
\end{equation}
where $\eta=\rho_da^3/m_p$, being $\rho_d$ the intrinsic grain density, $a^3$ its volume, and $m_p$ the mass of the proton. We have employed this model for all the types of dust, and for any colliding species ($n_g$ is the total gas number density): we note that the full model by \citet{Nozawa2006} is more complex, but goes beyond the aims of this release of \kromes. See option \verb+-dustOptions=SPUTTER+.

\subsubsection{Dust temperature}\label{sect:dust_temperature}
The dust temperature is mainly controlled by the radiation in the environment \citep{Hollenbach1979}, following Kirchhoff's law that for a given dust size $a$ allows to compute its temperature $T_d(a)$ from the equation
\begin{equation}\label{eqn:kirckhoff}
	\Gamma_\mathrm{em}(a) = \Gamma_\mathrm{abs}(a)+\Gamma_\mathrm{CMB}(a)\,,
\end{equation}
where $\Gamma_\mathrm{em}$ is the radiance (erg$\,$s$^{-1}\,$cm$^{-2}\,$sr$^{-1}$) of the dust grain, $\Gamma_\mathrm{abs}$ is the absorbed radiance by a generic photon flux (e.g. $I(\nu)$ employed in Sect.\ref{sect:phioniz}), and $\Gamma_\mathrm{CMB}(z)$ is the same quantity, but for the CMB background radiation, which is a black-body with temperature $T_\mathrm{CMB}(z)=T_0(1+z)$, $z$ is the redshift, and $T_0=2.73$ K.
These quantities are given as 
\begin{eqnarray}
	\Gamma_\mathrm{em}(a) &=& \int Q_\mathrm{abs}(a,\nu) B\left[T_d(a),\nu\right] \dd\nu\\
	\Gamma_\mathrm{abs}(a) &=& \int Q_\mathrm{abs}(a,\nu) I(\nu)\, \dd\nu\\
	\Gamma_\mathrm{CMB}(a) &=& \int Q_\mathrm{abs}(a,\nu) B\left[T_\mathrm{CMB},\nu\right] \dd\nu
\end{eqnarray}
where $Q_\mathrm{abs}(a,\nu)$ is the dimensionless absorption coefficient\footnote{Downloadable for graphite and silicates at \url{http://www.astro.princeton.edu/~draine/dust/dust.diel.html}} that depends on the size of the grain and on the photon frequency $\nu$ \citep{Draine1984,Laor1993}, $B\left[T,\nu\right]$ is the spectral radiance of a black-body with temperature $T$, and $I(\nu)$ is a generic spectral radiance.

By using Eq.(\ref{eqn:kirckhoff}) and a root-finding algorithm it is possible to find the dust temperature $T_d(a)$.
Note that when $I(\nu)=0$ we obtain $T_d(a)=T_\mathrm{CMB}$.

It is important to include in this balance also the gas-dust thermal exchange as discussed by \citet{Hollenbach1979,Omukai2000,Schneider2006}, but we note that at high density ($n > 10^{10}$~cm$^{-3}$) this process increases the stiffness of the ODEs system, influencing the stability of the solver. We plan to tackle this problem in a future release of \kromes.

\subsubsection{Molecular hydrogen formation on dust}\label{sect:H2dust}
Dust grains catalyse the formation of molecular hydrogen on their surface (e.g. \citealt{Gould1963}) through the reaction 
\begin{equation}
	\rm H + H + dust \to H_2 + dust\,.
\end{equation}
In our model we employ the rate given in \citet{Cazaux2009} for carbon and silicon-based grains who gives the total amount of H$_2$ catalysed on the dust surface as
\begin{equation}
 \frac{\dd n_\mathrm{H_2}}{\dd t} = \frac{\pi}{2} n_\mathrm{H} v_g \sum_{j\in[\mathrm{C,Si}]}\sum_i  n_{ij} a_{ij}^2 \epsilon_j(T, T_i)  \alpha(T, T_i)\,,
\end{equation}
where each \ith bin of the two grain species (i.e. C and Si) contributes to the total amount of molecular hydrogen. In the above equation $n_\mathrm{H}$ is the number density of atomic hydrogen in the gas-phase, $v_g$ is the gas thermal velocity, $n_{ij}$ is the number density of the \jth dust type in the \ith bin, $a_{ij}$ is its size, and $\epsilon_j$ and $\alpha$ are two functions where $T$ and $T_i$ are the temperature of the gas and of the dust in the \ith bin.  
The function $\epsilon_j$ has two expressions depending on the type of grain employed: for carbon-based dust it is
\begin{equation}
 \epsilon_\mC = \frac{1-T_H}{1+0.25\left(1+\sqrt{\frac{E_c-E_s}{E_p-E_s}}\right)^2} \exp\left(-\frac{E_s}{T_d}\right)\,,
\end{equation}
with  
\begin{equation}
 T_H = 4\left(1+\sqrt{\frac{E_c-E_s}{E_p-E_s}}\right)^2 \exp\left(-\frac{E_p-E_s}{E_p+T}\right)\,,
\end{equation}
where $E_p=800$ K, $E_c=7000$ K, and $E_s=200$ K.
Analogously, silicon grains have
\begin{equation}
 \epsilon_\mathrm{Si} = \left[1+16\frac{T_d}{E_c-E_s}\,\exp\left(-\frac{E_p}{T_d}-\beta a_{pc}\sqrt{E_p-E_s}\right)\right]^{-1}+ \mathcal{F}\,,
\end{equation}
where $E_p=700$ K, $E_c=1.5\times10^4$ K, $E_s=-1000$ K, $\beta=4\times10^9$,  $a_{pc} = 1.7\times10^{-10}$ m (Cazaux, priv. comm. 2012), and
\begin{equation}
 \mathcal{F} = 2\frac{\exp\left(-\frac{E_p-E_s}{E_p+T}\right)}{\left(1+\sqrt{\frac{E_c-E_s}{E_p-E_s}}\right)^2}\,.
\end{equation}
Finally, the sticking coefficient (not to be confused with the one in Sect.\ref{sect:growth}) is given as
\begin{equation}
	\alpha =\left[{1+0.4\,\sqrt{T_2 + \frac{T_d}{100}}+0.2\,T_2 +0.08\left(T_2\right)^2}\right]^{-1}
\end{equation}
with $T_2 = T/(100\,\mathrm{K})$. See option \verb+-dustOptions=H2+

\section{The computational framework}
\subsection{Code structure}\label{sect:code_structure}
\begin{figure*}
 \includegraphics[width=.98\textwidth]{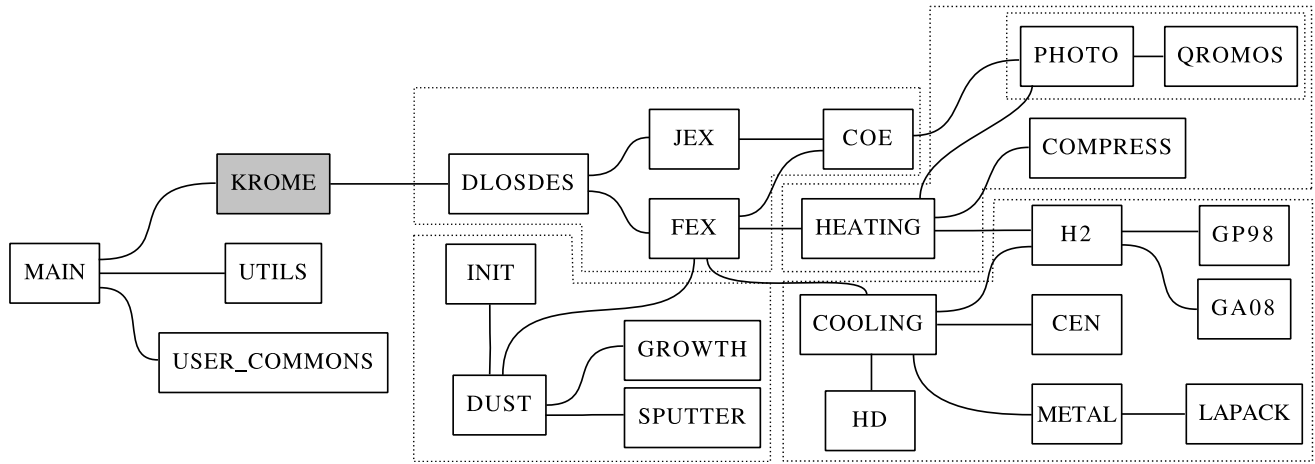}
 \caption{Pictorial view of the graph representing subroutine and modules in the \krome package. \emph{MAIN} is the framework program, \emph{KROME} (greyed) is the main module of \kromes, while the other modules are explained in detail within the text.  \label{fig:fwork}}
\end{figure*}

The code \krome is based on a \python pre-processor that creates the necessary \fortran files to compute the chemical evolution, the cooling, the heating, and all the physical processes one wants to include in the model based on a user-defined list of reaction rates.
The typical scenario is represented by an external code (called hereafter \emph{framework} code) that needs a call to a function that computes the chemical evolution, e.g. a hydrodynamical model. The user can choose one of the networks provided with the package or provide a list of reactions that contains for each reaction the list of the reactants and the products, the temperature limits, and the rate coefficient as a numerical constant or a \fortran expression. Finally, depending on the physics involved, the various command-line options of \krome allow to add or remove modules like cooling, heating, and dust.

\krome has a main module that calls the solver and actually represents an interface between the framework program and the selected solver. There are other modules that provide utilities, physical constants, problem parameters, and common variables.
The scheme in Fig. \ref{fig:fwork} provides a pictorial view of the \krome package with its modules and functions: \emph{MAIN} is the framework program which needs to call \krome for computing the chemical evolution. There are several modules (dotted contours in Fig.\ref{fig:fwork}) that identify different features of the package:
\begin{itemize}
	\item the part devoted to solving the ODE system employing the \emph{DLSODES} solver (or \textsc{DVODE}), all the differential equations (\emph{FEX}), the Jacobian (\emph{JEX}), and the rate coefficients (\emph{COE}). Note that \emph{FEX} is connected to \emph{COE} to determine the values of the rates at a given time and thus retrieving $\dd n_i/\dd t$ for each species, \emph{HEATING} and \emph{COOLING} to obtain the $\dd T/\dd t$, and finally, \emph{DUST} to include $\dd n_j/\dd t$ in the ODE system for each bin of dust. Also \emph{JEX} has the same links, but we omit them in Fig.\ref{fig:fwork} for the sake of clarity.

	\item The cooling that contains the \emph{COOLING} driver which calls the individual cooling functions listed in Sect.\ref{sect:cooling}: \emph{ATOMIC}, both \emph{H2} models, \emph{HD}, and \emph{METAL} that needs a call to the \textsc{LAPACK} libraries to solve the system of Eq.(\ref{eqn:linearsys}) via the \textsc{dgesv} function.
	
	\item The \emph{HEATING} has a similar structure, and it includes the chemical heating (\emph{CHEM}), the compressional (\emph{COMPRESS}), and the photoheating (\emph{PHOTO}), which is a module itself and contains a call to the routine \emph{QROMOS} to compute the integrals in Eqs.(\ref{eqn:ph_integral}) and (\ref{eqn:heat_integral}) by using a Ronberg integration methods from \citet{Press1992}.

	\item The code has a module for dust modelling, including accretion (\emph{GROWTH}) and destruction (\emph{SPUTTER}), and a subroutine to initialise the grain distribution (\emph{INIT}).

	\item Finally, the package provides some utility functions (\emph{UTILS}) to retrieve information on species and reactions, and a set of common variables (\emph{USER\_COMMONS}) that allows the communication between the framework program and the package modules (for the sake of clarity in Fig.\ref{fig:fwork} the \emph{USER\_COMMONS} block is not connected to the other blocks).
\end{itemize}

The different parts of \krome resemble the physics of the processes that are included in the package, in order to help the user to modify the source code or to add new parts to study other phenomena than the ones embedded in the public version of our code. The \fortran language allows a simple logical distribution through the module units that collect the variables, the functions, and the subroutines that match a specific criteria, in our case that belong to the same physical process. In this sense the dotted contours in Fig.\ref{fig:fwork} give a pictorial representation of these modules.

\subsection{The system of ODEs}\label{sect:ODE}
\begin{figure}
 \includegraphics[width=.48\textwidth]{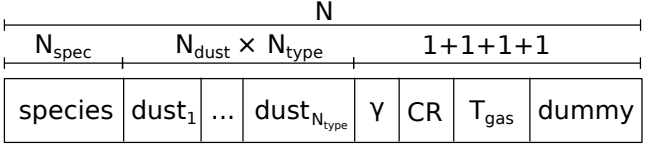}
 \caption{The scheme of the differential equations employed in the \krome package. The label \emph{species} indicates all the molecular and atomic species included their ions, \emph{dust$_i$} are the ODEs of the dust bins: the number of these blocks depends on the $N_{type}$ of grain type employed, and each block has $N_{dust}$ ODE (one for each bin). The last four ODEs are for photons ($\gamma$), cosmic rays (\emph{CR}), temperature $T_{gas}$, and the \emph{dummy} species. The total number of ODEs is $N$.  \label{fig:ODE_array}}
\end{figure}

The system of ODEs is the core of the package (see Fig.\ref{fig:ODE_array}), since the chemistry in the ISM is determined by the differential equations of each species (including the dust bins), the temperature differential, and some species that by default do not participate directly to the solution of the systems since their differential is always equal to zero. In fact, these species are included only for compatibility purposes with networks that include cosmic rays (CR) and photons ($\gamma$) as regular species, even if these species have always unitary abundances during the whole simulation. The same considerations apply to \emph{dummy} which is employed to cope with the implicit ODE representation (see Sect.\ref{sect:implicitODE} below).

The ODEs are stored in the \emph{FEX} module (see Fig.\ref{fig:fwork} in Sect.\ref{sect:code_structure}) which is called by the solver each time $\dd n_i/\dd t$ must be evaluated.
The total number of differentials is $N$, which is divided in the following groups:
\begin{eqnarray}\label{eqn:ODEs2}
 \frac{\dd n_i}{\dd t} &=& \sum_{j\in F_i} R_j(\bar n,T) - \sum_{j\in D_i} R_j(\bar n,T)\,\,\forall\,i\in species\nonumber\\
 \frac{\dd n_{ij}}{\dd t} &=& G_{ij}(n_j, T) - S_{ij}(n_j, T)\,\,\forall\,i\in dust\,,j\in types\nonumber\\
 \frac{\dd n_\gamma}{\dd t} &=& \frac{\dd n_\mathrm{dummy}}{\dd t} = \frac{\dd n_\mathrm{CR}}{\dd t} = 0\nonumber\\
 \frac{\dd T}{\dd t} &=& (\gamma-1)\frac{\Gamma(\bar n,T)-\Lambda(\bar n,T)}{k_b\sum_i n_i}\,,
\end{eqnarray}
where $\bar n$ is the array that contains the number densities of all the species.
More in detail: the first set of equations represents the ODEs of the chemical species (molecules and atoms, including ions and anions), as already discussed in Sect.\ref{sect:network}, Eq.(\ref{eqn:ODEs}). The next set contains the differential equations that represent the evolution of the dust bins, where $G_{ij}$ and $S_{ij}$ are the rates of growth and destruction (via thermal sputtering) as indicated in Sect.\ref{sect:dust}, Eqs.(\ref{eqn:growth}) and (\ref{eqn:sputter}). The subscript $i$ runs over the $N_\mathrm{dust}$ bins, while $j$ runs over $N_\mathrm{type}$
types (e.g. carbonaceous and silicates). The total number of the ODEs in this second set is then $N_\mathrm{dust}\times N_\mathrm{type}$.
The third set includes the ``utility'' species (photons, cosmic rays, and dummy) which, unless some user-defined exception is applied, are set to zero, since their value has no need to change during the evolution of the system. In particular they are $n_\mathrm{\gamma}=n_\mathrm{CR}=n_\mathrm{dummy}=1$ 
when they participate in the implicit ODE scheme (see Sect.\ref{sect:implicitODE}). Finally, the last set contains the differential equation that controls the evolution of the gas temperature which is determined by the cooling ($\Lambda$) and the heating ($\Gamma$) as discussed in Sect.\ref{sect:thermal}. It is worth noting that the first two sets and the last one contain functions on the RHS that depend on the abundances of the species, the abundance of the grains in the dust bins, and the temperature: this suggests that these $N$ equations are tightly connected and for this reason a proper solver with an high order of integration is needed (\dlsodes or \textsc{dvode} in our case).

\subsubsection{Implicit vs explicit evaluation of the ODEs}\label{sect:implicitODE}
In \krome we adopt two approaches to implement the ODE in the \emph{FEX} module, namely \emph{implicit} and \emph{explicit}. These two terms are not to be confused with the same terms employed for the solving method: in our case the solver \dlsodes is \emph{always} an implicit BDF. The most efficient way to write the RHS term in the ODE is the \emph{explicit}: for example we report a two-reactions network with
\begin{eqnarray}\label{eqn:reactions}
	\mathrm H+ \mathrm H &{\overset{k_{1}}{\to}}& \mathrm H_2\nonumber\\
	\mathrm H_2+ \mathrm O &{\overset{k_{2}}{\to}}& \mathrm{OH} + \mathrm H\nonumber\\
	\mathrm{OH} &{\overset{k_{3}}{\to}}& \mathrm{H} + \mathrm O\,,\nonumber
\end{eqnarray}
where the first reaction is controlled by the rate coefficient $k_1$ and the second by $k_2$ and so on, and following that the abundance of the  atomic hydrogen changes as
\begin{equation}\label{eqn:Hdiff}
	\frac{\dd n_\mathrm{H}}{\dd t} = -k_1\,n_\mathrm{H}\,n_\mathrm{H}+k_2\,n_\mathrm{H_2}\,n_\mathrm{O}+k_3n_\mathrm{OH}\,,
\end{equation}
and analogue differential can be written for the other species, for a total of five differential equations. These equations in the \emph{explicit} scheme are translated directly in the \fortran code in the form shown by Eq.(\ref{eqn:Hdiff}), while for the \emph{implicit} evaluation we employ a loop cycle (see Algortithm \ref{alg:ODE})
\begin{algorithm}\label{alg:ODE}
\caption{Pseudo-algorithm of the \emph{implicit} scheme for the \emph{ODE} module. Note that this example is generated for the set of reactions listed in Eqs.(\ref{eqn:reactions}), and it can be different for a different set of reactions as discussed within the text.}
\begin{algorithmic}
\For{$i \in reactions$}
	\State $F=k_i n(r_1) n(r_2)$
	\State $\dd n(r_1) = \dd n(r_1) - F$
	\State $\dd n(r_2) = \dd n(r_2) - F$
	\State $\dd n(p_1) = \dd n(p_1) + F$ 
	\State $\dd n(p_2) = \dd n(p_2) + F$ 
\EndFor
\end{algorithmic}
\end{algorithm}
where the loop is on the three reactions listed in Eq.(\ref{eqn:reactions}), while $r_j$ and $p_j$ are the indexes of the \jth reactants and products of the \ith reaction respectively. Note that \krome can handle reactions with an arbitrary number of reactants and products, and it modifies automatically the implementation of the Algorithm \ref{alg:ODE} according to the chemical network employed. Finally, it is worth noting that in the last reaction of the network in Eq.(\ref{eqn:reactions}), only one reactant is present, and for this reason the evaluation of $F$ will include the \emph{dummy} element, which in fact is always equal to one.

The \emph{implicit} scheme is useful for large networks ($\gtrsim500$ reactions), since it increases the efficiency during the compilation, it allows a more compact implementation of the \emph{ODE} module, and it can be employed for applying the reduction methods as discussed in \citet{Grassi2012,Grassi2013}. Conversely, the \emph{explicit} scheme is faster at run-time, especially when compiled with Intel \fortran compiler. We always suggest to adopt an \emph{explicit} evaluation when the network is small.

\subsection{Extra ODEs}\label{sect:test_lotkav}
\tgcomment{\krome has the capability of handling extra ODE equations to be included in the system of Eq.(\ref{eqn:ODEs2}). These extra equations can be included simply indicating their \fortran expression in a user-defined file, and can be employed alongside the standard ``chemical'' ODE equations that are built from the chemical network. This feature allows to add ODEs that not arise directly from the chemical network, but that need to be coupled to the main ODE system.
To provide an example we include in \krome a simple predator-prey Lotka-Volterra model.
The results of the test provided with the package follow the expected evolution both for predators and for preys.}

\subsection{Solver}\label{sect:solver}
The ODE systems that represent astrochemical networks are often stiff, and for this reason the solver employed must be suitable for such kind of equations. There are several schemes proposed in literature, from simple first order BDF as in \enzo \citep{Anninos97,Enzo2004}, to more complicated ones such as Gears, Runge-Kutta with different orders of integration (even if the most employed is the fourth order), from implicit schemes to semi-implicit as in \flash which employs a multi-order Bader-Deuflhard solver \citep{Bader1983} or the third-order Rosenbrock method \citep{Rentrop1979}.

For \krome we have chosen the widely used \dlsodes \citep{Hindmarsh1983} which takes advantage of compressing a sparse Jacobian matrix. Very often, in fact, astrochemical networks present a sparse or a very sparse Jacobian matrix associated with their ODE system, and hence this solver can give a very large speed-up compared for example with \textsc{DVODE/CVODE} in the SUNDIALS package\footnote{\arl{http://computation.llnl.gov/casc/sundials/main.html}} \citep{Hindmarsh2005}, which is used for example in \textsc{AREPO}, that employs the same integration scheme but without any capability of handling the Jacobian sparsity \citep{Nejad2005,Grassi2013}. For this reason \dlsodes is employed in all the tests presented with the \krome package (see Sect.\ref{sect:test_suite}).

The behaviour of \dlsodes is controlled via the $MF$ flag (see solver's documentation) which is defined as
\begin{equation}
	MF = 100\cdot MOSS + 10\cdot METH + MITER\,,
\end{equation}
where $MOSS=0$ if \krome supplies the sparsity structure,  $MOSS=1$ when \krome supplies the Jacobian and the sparsity is derived from the Jacobian, and $MOSS=2$ if both the Jacobian and the sparsity structure are derived by the solver (see Sect.\ref{sect:sparsity}). The term $METH=1$ is to use an Adams method, while $METH=2$ is for backward differential formula (BDF) as in our case. Finally,
$MITER=1$ describes the case scheme when \krome provides the Jacobian via the \emph{JEX} module, and $MITER=2$ when the Jacobian is evaluated by the solver. The flag $MF=222$ implies that the solver generates the sparsity structure and the Jacobian, while with $MF=21$ our package must provide both the sparsity structure and the Jacobian. The latter case can achieve better results since otherwise the solver must call the \emph{FEX} function for $NEQ+1$ times to evaluate the sparsity structure, and an unpredictable number of calls to evaluate the Jacobian. However, the performances are often problem-dependent: in fact, in case of varying temperature \krome can't provide an analytic Jacobian (see discussion in Sect.\ref{sect:jacobian}).

\begin{figure}
 \includegraphics[width=.48\textwidth]{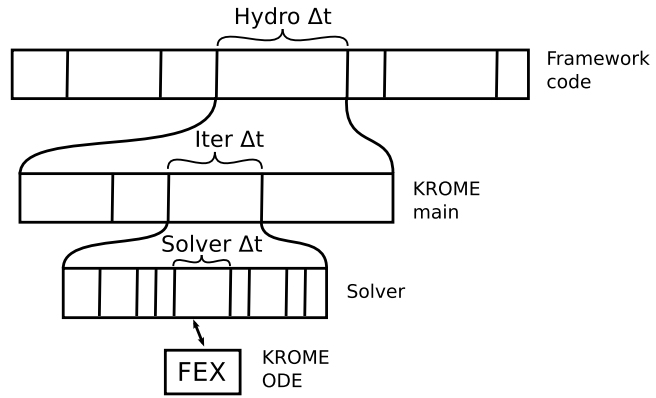}
 \caption{Pictorial view of the time-step hierarchy. Top to bottom: the framework code is divided into hydrodynamical time-steps, the main module of KROME iterates calling the \dlsodes solver that during its internal time-steps call the \emph{FEX} routine inside the \emph{ODE} module of the \krome package.  \label{fig:hierarchy}}
\end{figure}

In Fig. \ref{fig:hierarchy} we show the time-step hierarchy, where the framework code at each hydrodynamical time-step calls the \krome main module to determine the chemical and thermal evolution. \krome then call the solver: if the solver finds a solution before reaching \emph{MAXSTEP} internal steps or without any warnings (see \dlsodes manual for further details), \krome calls the solver only one-time (i.e. a single iteration time-step); otherwise \krome calls the solver again until the end of the hydrodynamical time-step has reached. At the bottom of the hierarchy the solver calls the \emph{FEX} module (and \emph{JEX} if $MF=21$) at each internal time-step to advance the solutions of the ODE system.

\krome provides also the DVODE solver in its \fortran 90 implementation\footnote{\arl{http://www.radford.edu/~thompson/vodef90web/}} for people who are more confident with it. Note that this version contains options to handle with sparse systems, but in the present release of \krome this capability is not enabled.

\subsubsection{Tolerance}
The accuracy and the computational efficiency of the \dlsodes solver is controlled as usual by two tolerances, namely relative (\emph{RTOL}) and absolute (\emph{ATOL}), that are employed by the solver to compute the error as \mbox{$\varepsilon_i=RTOL_i\cdot n_i+ATOL_i$} for the \ith species. The first represents the error relative to the size of each solution, while the latter is the threshold below which the value of the solution becomes unimportant for the purpose of the ODE system discussed. \krome has the default values of \emph{RTOL=$10^{-4}$} and \emph{ATOL=$10^{-20}$}, but they can be modified by the user according to the simulated environment. Note that lower values give more accurate solutions, but conversely the computational time increases, and in some cases this cost becomes prohibitive. In the 3D runs (Sect.\ref{sect:3D}) presented here we set \emph{RTOL=$10^{-4}$} and \emph{ATOL=$10^{-10}$}. 

\subsubsection{Jacobian}\label{sect:jacobian}
For an ODE system with $N$ equations, the Jacobian is a $N\times N$ matrix defined by
\begin{equation}\label{eqn:defjac}
	J_{ij} = \frac{\partial^2 n_i}{\partial t\, \partial n_j}\,,
\end{equation}
where $n_i$ and $n_j$ are the abundances of the \ith and the \jth species, respectively. 
The Jacobian in the module \emph{JEX} can be evaluated by the solver itself ($MITER=2$) or explicitly written by \krome and hence provided to the solver. This latter method is only available when the \emph{explicit} ODEs paradigm has been chosen, since it allows to write algebraically the Jacobian via the definition of Eq.(\ref{eqn:defjac}).

When the temperature is not constant during the evolution, \krome must include in the $(N+1)\times (N+1)$ Jacobian matrix the following rows and columns:
\begin{eqnarray}\label{eqn:jactemp}
	J_{Tj} &=& \frac{\partial^2 T}{\partial t\, \partial n_j}\nonumber\\
	J_{iT} &=& \frac{\partial^2 n_i}{\partial t\, \partial T}\nonumber\\
	J_{TT} &=& \frac{\partial^2 T}{\partial t\, \partial T}\,,
\end{eqnarray}
where $T$ is the gas temperature. However, the ODE for the temperature depends on the cooling and heating functions, thus determining algebraically their derivatives is not as straightforward as for the species ODEs. For this reason the terms in Eqs.(\ref{eqn:jactemp}) must be evaluated numerically with a linear approximation: this is possible since the solver accepts a rough estimate of the Jacobian elements, as suggested by \citet{Hindmarsh1983}.
In particular, we evaluate the Jacobian elements at a given time $t$ as
\begin{equation}
	J_{Tj}(t) = \frac{\Phi(n_j+\delta n) - \Phi(n_j)}{\delta n}\,,
\end{equation}
where $\Phi(n)$ is the function to compute $\dd T/\dd t$, i.e. Eq.(\ref{eqn:cooling}), and $\delta n = \epsilon n_j$ is the amount of change of the \jth species, with $\epsilon=10^{-3}$ (the other species keep the same values). Unfortunately, $\epsilon$ seems to be problem-dependent and one should tune it.
The other terms (i.e. $J_{iT}$ and $J_{TT}$) are evaluated by calling the $FEX$ with $T+\delta T$ giving
\begin{equation}
	J_{Tj}(t) = \frac{FEX_j(T+\delta T) - FEX_j(T)}{\delta T}\,,
\end{equation}
and analogously
\begin{equation}
	J_{TT}(t) = \frac{FEX_T(T+\delta T) - FEX_T(T)}{\delta T}\,,
\end{equation}
by using $\delta T = \epsilon T$, where $FEX_j$ is the \jth term of the function that returns the array of the RHSs of the differential equations.

This evaluation is not necessary when the Jacobian is internally generated by  the solver by using the option $MF=222$ as indicated above.
Note that in many cases providing the Jacobian may help the calculation, that otherwise would be impossible with the internally-generated Jacobian, but in case of varying temperature the user may need to tune the $\epsilon$ parameter above.
\subsubsection{Sparsity structure}\label{sect:sparsity}
The \dlsodes solver takes advantage if the sparsity structure is provided by the user instead of being calculated by the solver itself. This structure consists of a matrix that indicates the positions of the non-zero elements inside the Jacobian matrix. This sparsity matrix is represented by the Yale Sparse Matrix Format \citep{Eisenstat1977,Eisenstat1982}, which consists of two arrays (named IA and JA in the \dlsodes documentation): for a $m\times n$ matrix IA has size $m+1$ and contains the index of the first non-zero element of the \ith row in the array $A$ that represents the sequence of the $N_{NZ}$ non-zero elements of the original matrix. JA has size $N_{NZ}$ and contains the column index of each element of $A$. \krome automatically builds and provides IA and JA to the \fortran routines of the solver. 

Note that \krome builds the sparsity structure during the pre-processor stage calculating IA and JA from the chemical network, thus determining the non-zero elements algebraically instead of numerically. For example in a network with a single reaction $\mA+\mB\to\mC$ the differential 
\begin{equation}
	\frac{\dd n_\mA}{\dd t} = -k\,n_\mA\,n_\mB\,,
\end{equation}
has two associate Jacobian elements:
\begin{equation}\label{eqn:alg_nnz}
	\frac{\partial^2 n_\mA}{\partial t\,\partial n_\mA} = -k\,n_\mB\,,
\end{equation}
and
\begin{equation}\label{eqn:alg_nnz2}
	\frac{\partial^2 n_\mA}{\partial t\,\partial n_\mB} = -k\,n_\mA\,,
\end{equation}
which are algebraically non-zero elements, and 
\begin{equation}\label{eqn:alg_z}
	\frac{\partial^2 n_\mA}{\partial t\,\partial n_\mC} = 0\,,
\end{equation}
which is algebraically a zero element. \krome identifies Eqs.(\ref{eqn:alg_nnz}) and (\ref{eqn:alg_nnz2}) as non-zero elements, while Eq.(\ref{eqn:alg_z}) is a zero element. This is not true if during the temporal evolution $k$, $n_\mA$, and $n_\mB$ become zero, but in this case the pre-calculated sparsity structure suggests that Eq.(\ref{eqn:alg_nnz}) is a non-zero element. This problem does not affect the results of the ODE system, but reduces the maximum speed-up. Depending on the chemical system this option can generate an overhead if the amount of time spent in including false non-zero elements in the ODE is larger than the time spent in evaluating the sparsity of the Jacobian by the solver: evaluating this overhead in the pre-processor stage is non-trivial without some tuning.

\subsection{Parsing chemical species}
\krome automatically determines the chemical species and their properties from the provided network file. It computes the mass, the charge, and the atoms that form any molecule. It can also evaluate the isotopes by using the $[n\mX]$ notation where $n$ is the 
atomic number and $\mX$ is the atom (e.g. $[14\mC]$ for ${}^{14}\mC$). \krome allows to include test species with unitary mass to test non-chemical networks by using the notation FK$i$ where $i=0,9$ (e.g. FK3). Any other non-recognized species is intended with mass and charge equal to zero.

The parser can recognize the excited level of selected molecules as CH$_2$ and SO$_2$ using the notation CH$2\_i$, where $i$ is the excited level, and also O(1D) and O(3P) by including them directly in the network file (see Sect.\ref{sect:test_planet}).

\krome also recognizes special species as GRAIN0, GRAIN$+$, and GRAIN$-$ that represent a general neutral grain with the same mass of a carbon-based aggregate of 100 carbon atoms, hence having $m_{g0}=6\times100\times (m_p+m_n+m_e)$, while the ionized grain has $m_{g+}=m_g-m_e$, and the anion  $m_{g-}=m_g+m_e$, where $m_p$ is the mass of the proton, $m_n$ the mass of the neutron, and $m_e$ the mass of the electron. Another class of species are the PAHs (PAH, PAH+, PAH-) that are similar to grains but their mass is $m_{PAH0}=6\times30\times (m_p+m_n+m_e)$, $m_{PAH+}=m_{PAH0}-m_e$, and $m_{PAH-}=m_{PAH0}+m_e$, respectively. \krome can also handle general colliders (M), cosmic rays (CR), and photons (g), although the last two species can be omitted in the network file. Note that the masses of these species are treated as zero.

Finally, \krome checks mass and charge conservation during the reaction parsing and warns the user with an error message in the pre-processor stage to avoid critical errors at runtime. It also checks the correct bracket balancing in the \fortran expression employed for the rate coefficient, but note here that \krome does not control the provided \fortran syntax.

\subsection{Reverse kinetics}\label{sect:reverse}
\krome offers an utility feature to compute the reverse kinetics of the reactions provided in the network file. It creates a new reverse reaction for each forward reaction with a rate constant computed by using a standard thermodynamical approach:
\begin{equation}
	k_r = \frac{k_f}{K_{eq}}\,, 
\end{equation}
where $k_f$ is the forward kinetic constant, $k_r$ is the reverse kinetic constant, and 
\begin{equation}
  K_{eq} = \exp\left(-\frac{\sum_{i\in P}G_i - \sum_{i\in R}G_i}{RT}\right)\, = \frac{\prod_{i \in P} [i]^{\nu_i}}{\prod_{i \in R} [i]^{\nu_i}}.
 \label{eq:keq}
\end{equation}
is the equilibrium constant of the reaction, $[i]$ the concentration of reactants $R$ and products $P$.
The term $K_{eq}$ is related to the quotient of partial pressures, $K_p$, as
\begin{equation}
	K_p = K_{eq} \cdot (k_b \cdot T)^{\Delta n}
\end{equation}
and then
\begin{equation}
	k_r = \frac{k_f}{K_{eq}} \cdot (k_b \cdot T)^{\Delta n}\,,
\label{eq:deltan}
\end{equation}
where $k_b$ (J K$^{-1}$) is the Boltzmann constant, $\Delta n = n_P - n_R$ is the difference between the number of products and reactants, $R$ (J mol$^{-1}$ K$^{-1}$) is the universal gas constant. The $\Delta n$ term is needed when there is different number of reactants and products \citep{Chase1998, lewis_book_1997, visschermoses2011}. The units correction factor for gas phase reactions $(k_b \cdot T)^{\Delta n}$ must be employed when using cm$^{-3}$ for species concentration. When using molarity, mol L$^{-1}$, the correction factor becomes $(R \cdot T )^{\Delta n}$. 

In the Eq.(\ref{eq:keq}) $G_i$ represents the Gibbs energy of the \ith species evaluated at temperature $T$, while $R$ is the gas constant (i.e. 8.3144621 J mol$^{-1}$ K$^{-1}$). These terms are computed as
\begin{equation}
	G_i = f_i(T) - g_i(T) + R \cdot T \ln \left( \frac{p_i}{p_0} \right) \,,
\end{equation}
 where $p_i$ and $p_0$ are the partial pressure of the gas and the standard pressure, respectively, while $f_i$ and $g_i$ are the NASA polynomials\footnote{\arl{http://www.me.berkeley.edu/gri_mech/data/nasa_plnm.html}} being $f_i(T)$ = $H_i(T)$ (enthalphy) and $g_i(T)$ = $T\,S_i(T)$ (entropy) as
\begin{equation}
	\frac{H_i(T)}{RT} = a_1 + a_2 \frac{T}{2} + a_3 \frac{T^2}{3} + a_4 \frac{T^3}{4} + a_5 \frac{T^4}{5} + \frac{a_6}{T}\,,
\end{equation}
and
\begin{equation}
	\frac{S_i(T)}{T} = a_1 \ln(T) + a_2 T + a_3 \frac{T^2}{2} + a_4 \frac{T^3}{3} + a_5 \frac{T^4}{4} + a_7\,,
\end{equation}
with $a_j$ the \jth polynomial coefficient for the \ith species given by \citep{Burcat1984} available on his website\footnote{\arl{http://garfield.chem.elte.hu/Burcat/THERM.DAT}}. Note that since the Burcat's thermochemical data are intended for many different purposes, we do not provide his complete file, but a much smaller version. If the user needs to include some species that are not present there, he/she can simply add them manually to the file \verb+data/thermo30.dat+ in the package: if the species is not present \krome writes a warning message.

\tgcomment{A test for this method is included in the package: it consists in deriving the inverse reactions for the Zel'dovich thermal nitric oxide (NO) mechanism as discussed in \citet{Ashraf2009}.
This model is based on two forward reactions and the corresponding reverse reactions with the rate coefficients taken from \citet{Baulch1994}. As expected \krome reproduces the results of the NO evolution, and correctly derives the reverse reactions.}

\section{The test suite}\label{sect:test_suite}
In the following, we present a range of astrophysically motivated tests to illustrate the capabilities of the chemistry package \kromes.
\subsection{Chemistry in molecular clouds}
Molecular clouds (e.g. TMC-1 and L134N) represent a standard environment for testing extended chemical networks \citep{Leung1984,Wakelam2008,Walsh2009,Wakelam2010}, since they include a large number of complex species like PAH and carbon linear chains that require several thousands of reactions. Hence, these models represent an interesting computational challenge, as the evolution of the species must be accurate and they need to handle hundreds of ODEs with a RHS that is formed of many terms.
On the other hand these models, in their one-zone or 1D versions, do not require any parallel calculation, as they are simple one-zone models and can run on a standard laptop in a few seconds. These environments are largely studied by the astrophysical community and several networks are available for downloading\footnote{See for example:\\ \arl{http://www.physics.ohio-state.edu/~eric/research.html}\\ \arl{http://kida.obs.u-bordeaux1.fr/}\\ \arl{http://www.udfa.net/}}. For these reasons, they represent valuable benchmarks for any code that is employed for determining the chemical evolution of the ISM.

For this molecular cloud test we choose the \verb+osu_01_2007+ network and the initial conditions proposed by \citet{Wakelam2008}: a constant temperature of $T=10$ K, H$_2$ density of $10^4$ cm$^{-3}$, cosmic rays ionization rate of $1.3\times10^{-17}$ s$^{-1}$, and a visual extinction of $10$. The initial conditions of the species are listed in Table \ref{tab:WHinit} and correspond to the EA2 model of \citet{Wakelam2008}, an high-metal environment observed in the diffuse cloud $\zeta$ Ophiuchi, while the electron abundance is computed summing the number densities of all the ions together, i.e. $n_\mathrm{e^-}=\sum_{i\in ions}n_i$. The density and the temperature of the model remains the same during the evolution, while the abundances of the species are computed following the non-equilibrium evolution accordingly to the ODEs system generated by the set of reactions indicated above. Note that this model does not include PAHs nor dust.

The command line to generate the test is
\begin{verbatim}
 > ./krome -test=WH2008
\end{verbatim}
which is a short-cut\footnote{Note that the test option also copies the Makefile and the test.f90 file from the given test directory (tests folder) to the build directory: in this sense the test option is not properly an alias or a short-cut. In general \kromes, when the option -test=[name] is enabled, copies all the files from test/[name] to the build/ folder.} for 
\begin{verbatim}
 > ./krome -n network/react_cloud
  -iRHS 
  -useN
\end{verbatim}
where the path is the path of the file of the chemical network, \verb+-iRHS+ is the option for forcing the implicit ODE scheme (see Sect.\ref{sect:implicitODE}), and \verb+-useN+ allows to use the numerical densities (cm$^{-3}$) as argument in the \krome call instead of the default fractions of mass.

\begin{figure}
 \includegraphics[width=.23\textwidth]{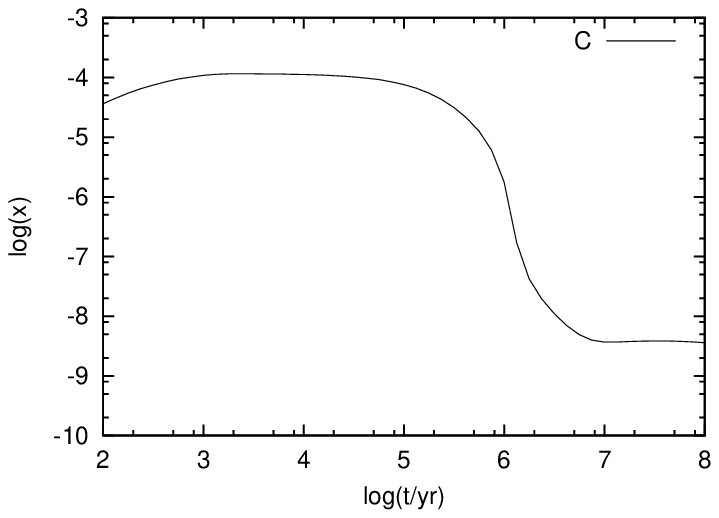}
 \includegraphics[width=.23\textwidth]{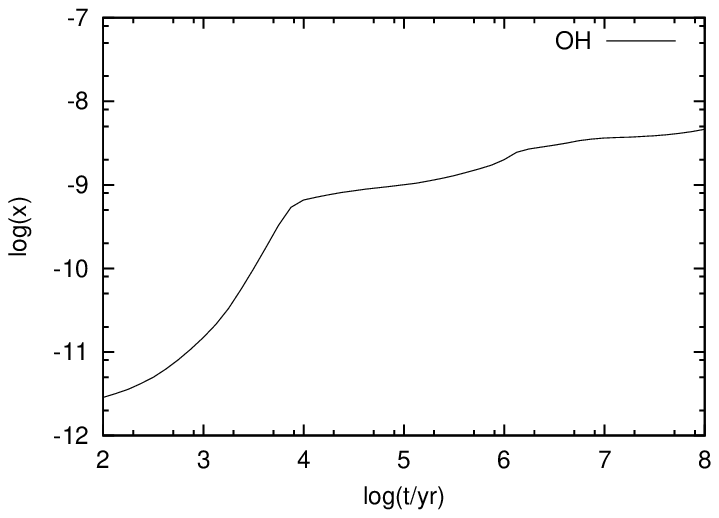}\\
 \includegraphics[width=.23\textwidth]{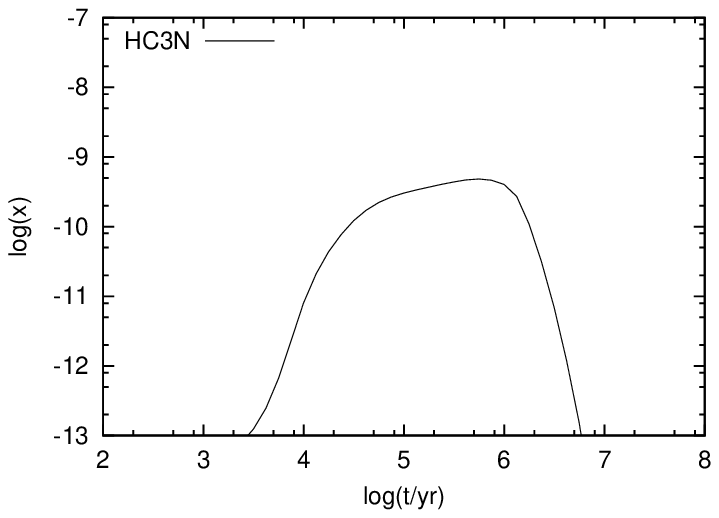}
 \includegraphics[width=.23\textwidth]{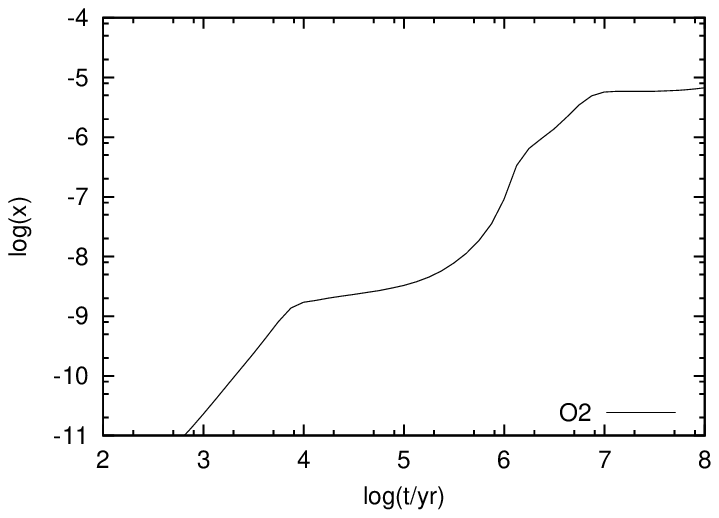}
 \caption{Evolution of carbon (top left), OH (top right), HC$_3$N (bottom left), and O$_2$ (bottom right). \label{fig:testWH2008}}
\end{figure}

The results of the $10^8$ yr time evolution are shown in Fig. \ref{fig:testWH2008} for carbon, OH, HC$_3$N and O$_2$, reproducing the same evolution as shown in Figs.3, 4, and 7 of \citet{Wakelam2008}.

\begin{table}
	\caption{Initial conditions for the molecular cloud test. Fraction of total hydrogen, where $a(b)=a\times10^b$.\label{tab:WHinit}}
	\centering
	\begin{tabular}{l|l|l|l}
		\hline
		Species & Abundance & Species & Abundance\\	
		\hline
		He & 9.00(-2) & Fe$^+$ & 2.00(-7)\\
		N & 7.60(-5) & Na$^+$ & 2.00(-7)\\
		O & 2.56(-4) & Mg$^+$ & 2.40(-6)\\
		C$^+$ & 1.20(-4) & Cl$^+$ & 1.8(-7)\\
		S$^+$ & 1.50(-5) & P$^+$ & 1.17(-7)\\
		Si$^+$ & 1.70(-6) & F$^+$ & 1.8(-8)\\
		e$^-$ & see text &&\\
		\hline
	\end{tabular}
\end{table}

\subsection{Cloud collapse (one-zone)}\label{sect:onezone}
\begin{figure}
	\includegraphics[width=0.48\textwidth]{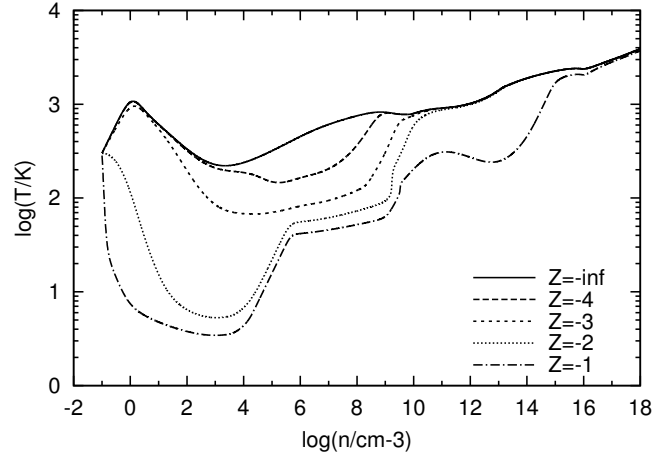}
        \caption{Temperature profile in function of the total number density for the collapse of the primordial cloud irradiated by different UV background.}
\label{fig:collapsez}
\end{figure}

We consider here a simple one-zone spherical cloud collapse including a primordial network listed in Tab.\ref{tab:rates} based on \citet{Omukai2000}  with the one in Tab.1 of \citet{Omukai2005}, labelled from Z1 to Z40. The initial conditions are set to $T=100$ K, $n=n_\mH=1$ cm$^{-3}$, $n_{\me^-}=10^{-4}$ cm$^{-3}$ and $n_{\mH_2}=10^{-6}$ cm$^{-3}$, following \citet{Omukai2000}. We assume also that all the metals are ionized, except for the oxygen as in e.g \citet{Omukai2005,Santoro2006,Maio2007}. The abundances of the metals are rescaled to the metallicity using the solar abundances of \citet{Anders1989}.
The time evolution of the collapsing core is calculated assuming a free-fall collapse,
\begin{equation}
	\frac{\dd\rho}{\dd t} = \frac{\rho}{t_{ff}}\,.
\end{equation}
The thermal evolution as a function of density is reported in Fig. \ref{fig:collapsez}. We can distinguish in the non-metal profile ($Z=-\infty$) the typical features of the temperature evolution as consequence of the processes involved. The cloud is heated up by compression ($n\sim10$~cm$^{-3}$), then starts to cool ($n\sim10^2$~cm$^{-3}$) for the effect of the H$_2$ cooling and the gas temperature drops to $\sim$200 K ($n\sim10^4$~cm$^{-3}$). Once the H$_2$ approaches the local thermal equilibrium (LTE) the cooling is less efficient and the temperature starts to increase again until densities of 10$^8$-10$^{10}$ cm$^{-3}$, when the three-body reactions come into play producing first a slight cooling dip ($n\sim10^{10}$~cm$^{-3}$). At higher densities the gas becomes optically thick and the cooling is less efficient producing a net heating in the cloud until the collisional-induced emission ($n\sim10^{16}$~cm$^{-3}$) cooling acts ($n\sim 10^{16}$~cm$^{-3}$). Finally the gas evolves adiabatically since it becomes also opaque to the continuum (CIE) emission: the evolution is a power-law with a slope that depends on the adiabatic index, that in this case is $\gamma=7/5$ since the gas at this stage is fully molecular. 

\begin{figure}
	\includegraphics[width=0.48\textwidth]{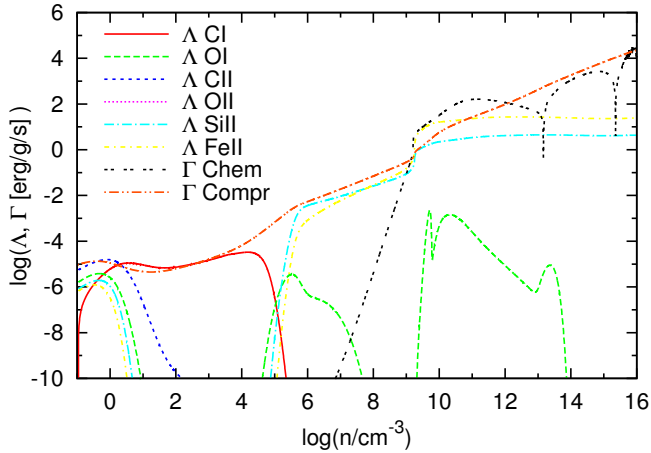}
        \caption{Individual thermal contributions including the main metal coolants ($\Lambda$ CI, $\Lambda$ CII, $\Lambda$ SiII, $\Lambda$ FeII), the compressional and the molecular hydrogen heating ($\Gamma_\mathrm{Compr}$, $\Gamma_\mathrm{Chem}$), for a collapse with $Z=-2$. Colours online.}
\label{fig:coolz}
\end{figure}

\begin{figure}
	\includegraphics[width=0.48\textwidth]{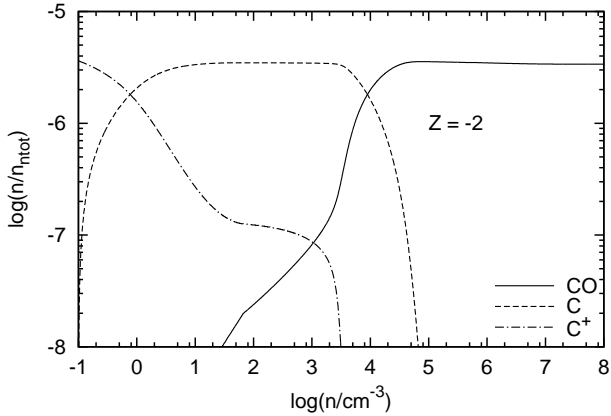}
        \caption{Fractional abundance evolution for three carbon-based species, i.e. C, C$^+$, and CO, for $Z=-2$ cloud collapse. See text for details.}
\label{fig:evolC}
\end{figure}

In Fig.\ref{fig:collapsez} we show the thermal evolution for metallicities ranging from $Z=-\infty$ to $Z=-1$, as indicated in the labels. The metals increase the gas cooling in the low-density regime, where the effects on the evolution are stronger. We also note that the metal cooling effect extends above densities of 10$^6$ cm$^{-3}$ due to the presence of SiII and FeII, as shown in Fig.\ref{fig:coolz}, where also the compressional and the H$_2$ heating are reported for comparison. In the very first stage of the evolution the thermal history is dominated by CI and subsequentially by CII, while other coolants such as OI, OII, are less important. In this test the chemical network does not include any Si or Fe chemistry, and for this reason the initial amounts of Si$^+$ and Fe$^+$ remain unchanged during the whole evolution, and hence the cooling from neutral Si and Fe is not included.  At the beginning of the density range $10^{4}\lesssim n\lesssim 10^{6}$ cm$^{-3}$ the CI cooling drops because C is depleted into CO (Fig.\ref{fig:evolC}), determining an interval without metal cooling that ends when the SiII-FeII cooling starts to act: in this region the thermal evolution is then controlled by the compressional heating determining an adiabatic temperature increase as shown in Fig.\ref{fig:collapsez}. However, different initial abundances of the metal species (i.e. with non-solar ratios), can lead to different thermal history and different interplay by the single metal contributions.

The command line to generate the test is
\begin{verbatim}
 > ./krome -test=collapseZ
\end{verbatim}
which is a short-cut for 
\begin{verbatim}
 > ./krome -n network/react_primordialZ2
   -useN 
   -cooling=H2,COMPTON,CI,CII,OI,OII,SiII,
	FeII,CONT,CHEM
   -heating=COMPRESS,CHEM
   -useH2opacity
   -ATOL=1d-40
   -gamma=FULL
\end{verbatim}
These options are the same as the previous tests, except for the

where \verb+COMPTON+ and \verb+CONT+ are the flags for the Compton and the continuum cooling (Sect.\ref{sect:cooling}), \verb+COMPRESS+ activates the compressional heating (Sect.\ref{sect:heating}), while \verb+useH2opacity+ allows to include the opacity term for the molecular hydrogen (also Sect.\ref{sect:cooling}). Finally, \verb+gamma=FULL+ is for the species-dependent calculation of the adiabatic index as in Eq.(\ref{eq:adiabatic}).
Individual metal cooling functions are from \verb+CI+ to \verb+FeII+, and \verb+-ATOL=1d-40+ forces the absolute tolerance to $10^{-40}$.

\subsection{Cloud collapse under UV radiation}
In this Section we present results from a simple one-zone collapse as in the previous Section, but under the presence of a ultraviolet (UV) background and employing a different reaction network (see Table \ref{tab:rates}). The initial conditions are the same as discussed in \ref{sect:onezone}, we only add the H$_2$ photodissociation rate as in \citet{Shang2010} who provided the following rate parametrization,
\begin{equation}
	k = 10^{-12}\beta J_{21}\\,
\end{equation}
with $J_{21}$ being defined as in Eq. (\ref{eq:flux}) and $\beta$ = 0.9 assumes a T5 spectrum (T$_\mathrm{rad} = 10^5$ K). No self-shielding nor other processes as Rayleigh scattering or H$^-$ cooling are included. This run only provides a simple test to check the behaviour of the gas under UV background. For a complete analysis we refer the reader to \citet{Omukai2001} and \citet{Schleicher2010}.

\begin{figure}
	\includegraphics[width=0.48\textwidth]{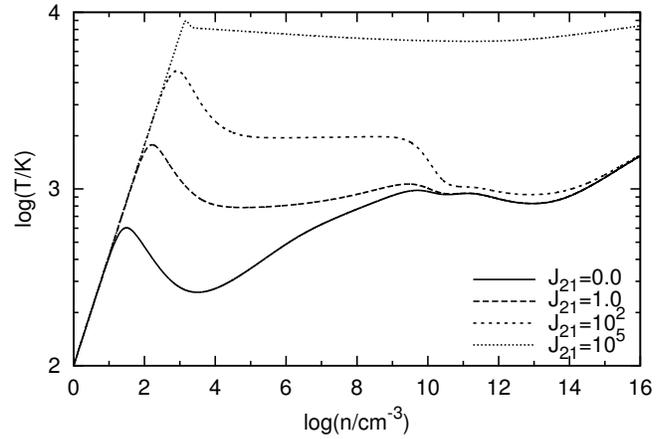}
        \caption{Temperature profile in function of the total number density for the collapse of the primordial cloud irradiated by different UV background.}
\label{fig:uvradiation}
\end{figure}

\begin{figure}
	\includegraphics[width=0.48\textwidth]{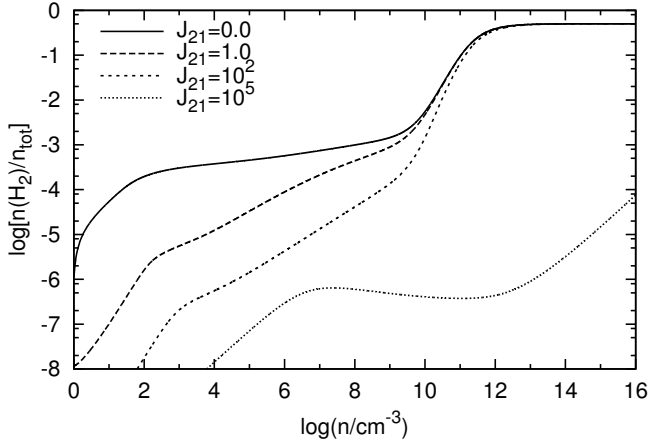}
        \caption{Molecular hydrogen fraction as a function of the total number density for different UV background.}
\label{fig:uvabunda}
\end{figure}

The results for four different $J_{21}$ are reported in Fig.~\ref{fig:uvradiation}. The stronger the UV flux, the larger the amount of H$_2$ dissociation, and the higher the temperature which is reached at low density. For $J_{21} < 10^5$ there is still enough H$_2$ to allow the cooling to act, while for $J_{21} = 10^5$ the temperature heats up to $\sim$8000 K very quickly and the H$_2$ formation and its cooling are strongly reduced, as can be seen from Fig. \ref{fig:uvabunda}. At this stage the hydrogen line emission of atomic hydrogen is the only contribution which cools down the collapsing cloud.

The command line to generate the test is
\begin{verbatim}
 > ./krome -test=collapseUV
\end{verbatim}
alias for
\begin{verbatim}
 > ./krome -n network/react_primordial_UV
   -cooling=H2,COMPTON,CIE,ATOMIC
   -heating=COMPRESS,CHEM
   -useN
   -gamma=FULL
\end{verbatim}
with the same meaning as in the previous tests.

\subsection{Dust}\label{sect:test_dust}
As described in Sect.\ref{sect:dust} \krome has the capability of handling the dust evolution and its interaction with the gas phase. The test proposed here consists of a gas with an initial population of dust, both silicon and carbon-based, with 10 bins each (i.e. a total of $n_d=10\times2=20$ bins). The initial condition for the gas phase are the same as for the primordial shock (Tab. \ref{tab:cooling_init}) extended with carbon and silicon with a dust-to-gas ratio of $10^{-5}$, and the distribution follows a MRN power-law as described in Sect.\ref{sect:dust}.

\begin{figure}
 \includegraphics[width=.47\textwidth]{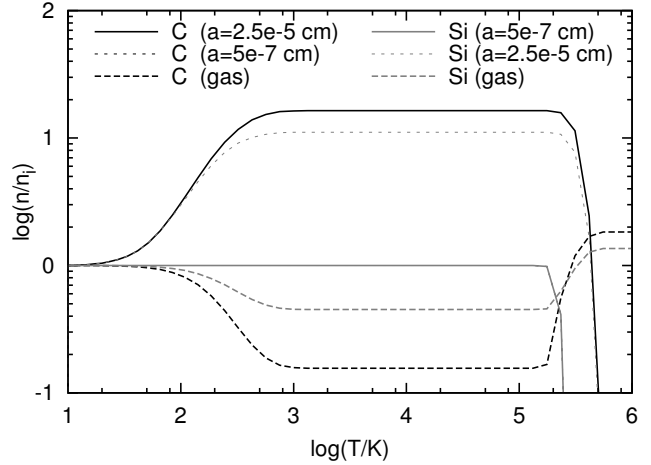}
 \caption{Evolution of the ratio between the number density $n$ of the given species at time $t$ and the its initial value $n_i$ as a function of the temperature (see text). Solid and dotted lines represent the largest and the smallest bins respectively, while the dashed indicates the species in the gas-phase. Black lines are for carbon and grey for silicon.  Note that black and grey dotted lines overlap. \label{fig:dust_test}}
\end{figure}

The aim of the test is to analyse the influence of the dust growth and the dust destruction by changing the gas temperature artificially during the simulation as
\begin{equation}\label{eqn:Tgas_evol_dust}
	T(t) = \left(10^6-10\right)\frac{t}{t_{end}}+10\,\, \mathrm{K}\,,
\end{equation}
where $t$ is the time reached by the simulation and $t_{end}=10^8$ yr is the ending time. Eq.(\ref{eqn:Tgas_evol_dust}) represents a gas that linearly increases its temperature with time from 10 K to $10^6$ K: we expect that at the beginning of the simulation when the temperature is higher the effect of the growth will dominate the dust processes, while when the temperature increases the dust will be sputtered by the hot gas. The results of this calculation is plotted in Fig.\ref{fig:dust_test}, for carbon and silicon-based dust. We represent the evolution with time of the number density of dust for each bin size. \tgcomment{As expected the amount of dust increases at the beginning of the evolution (lower temperatures), subtracting carbon and silicon from the gas-phase, due to the interaction between the two phases that are coupled within the same ODE system. In the later stages (higher temperatures) the sputtering destroys the grains, therefore the species in the gas-phase increase again, since the sum of the mass of C (and Si) in the dust phase and in the gas phase is constant.} Note that small grains are not affected by dust growth, since their smaller size reduces the probability of a gas-dust encounter.

The command to run this test is
\begin{verbatim}
 > ./krome -test=dust
\end{verbatim}
alias for
\begin{verbatim}
 > ./krome -n networks/react_primordial
     -dust=10,C,Si 
     -useN
     -dustOptions=GROWTH,SPUTTER
\end{verbatim}
with \verb+-dust=10,C,Si+ to employ 10 bins of carbon-based and 10 bins of silicon-based dust, and \verb+-dustOptions=GROWTH,SPUTTER+ enables the growth and thermal sputtering processes as discussed in Sects.\ref{sect:growth} and \ref{sect:sputter}, respectively.

\subsection{1D shock}
To test \krome in a hydrodynamical context we propose a one-dimensional spherically symmetric shock as described by \citet{Bodenheimer2006}, which provides a simple implementation of a high temperature and density region that expands into a lower density and cold environment.
In this test we can see the effect of the chemical solver and the thermal processes on the evolution of the shock from a computational point of view, but also from a physical perspective when the cooling is turned on.

The numerical set-up consists in $100$ shells, 20 in the inner region (the so-called fireball), with a temperature of $10^6$ K, a density of 1.24$\times10^{-23}$ g cm$^{-3}$, and a mass of $3\times10^{-4}$ M$_\odot$ per shell. Conversely, the outer region made of 80 shelss is colder and less dense with a temperature of $10$ K, a density of $10^{-26}$ g cm$^{-3}$, and a mass of approximately $8\times10^{-4}$ M$_\odot$ per shell, since they have larger volumes for the same thickness when compared to the inner shells. All the shells are at rest at the beginning of the calculation.

\begin{table}
	\caption{Number densities for the cooling test, where $a(b)=a\times10^b$. All the abundances are mass fractions. \label{tab:cooling_init}}
	\centering
	\begin{tabular}{l|l|l|l}
		\hline
		Species & Abundance & Species & Abundance\\	
		\hline
		H & 0.9225 		& H$_2^+$ & 1.0(-20)\\
		e$^-$ & 1.0(-4) 	& HD & 1.0(-8)\\
		H$^+$ & 1.0(-4) 	& H$^-$ & 1.0(-20)\\
		D & 1.0(-20)	 	& He$^{++}$ & 1.0(-20)\\
		D$^+$ & 1.0(-20) 	& C$^+$ & 1.0(-6)\\
		He & 0.0972 		& Si$^+$ & 1.0(-7)\\
		He$+$ & 1.0(-20) 	& O & 1.0(-4)\\
		H$_2$ & 1.0(-5)		& Fe$^+$ & 1.0(-8)\\
		\hline
	\end{tabular}
\end{table}

This test is similar to the one proposed by \citet{Grassi2013}: the shock develops following only the hydrodynamics, while the evolution of the chemical species follows the density and the temperature of the shell where the ODE system is resolved. At each hydrodynamical time-step each shell calls \krome to solve the chemical evolution using the initial abundances and the temperature provided by the given shell, and the time-step chosen is determined by the hydrodynamics. Note that the internal time-step of \krome is chosen according to the ODE system and the stiffness of the set of equations is evaluated and the user has no need to provide it, since it is automatically determined by the solver using the values of relative and absolute tolerance employed.

We add the thermal evolution of the gas to understand how cooling can influence the evolution of the shock, since the abundances of the species (and the temperature) controls the amount of energy lost by the gas, i.e. the cooling.

The shock test with cooling can be obtained by typing
\begin{verbatim}
 > ./krome -test=shock1Dcool
\end{verbatim}
an alias of
\begin{verbatim}
 > ./krome -n network/react_primordial 
     -cooling=H2,HD,Z,DH
\end{verbatim}
that employs the primordial network, and includes the cooling functions listed above.

\begin{figure}
 \includegraphics[width=.23\textwidth]{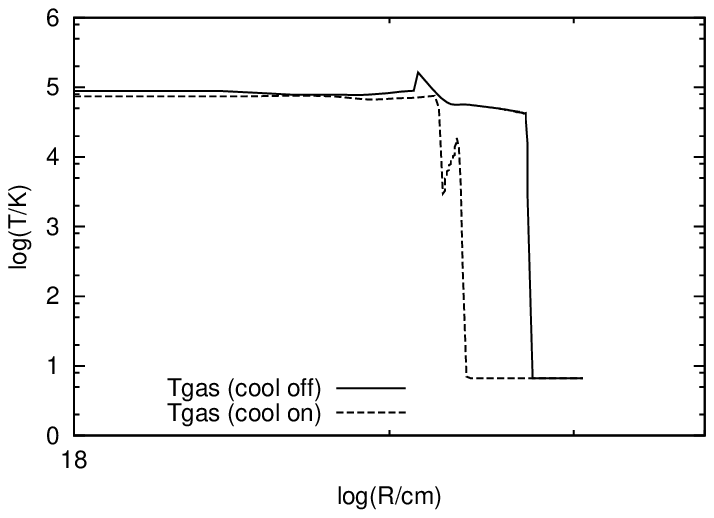}
 \includegraphics[width=.23\textwidth]{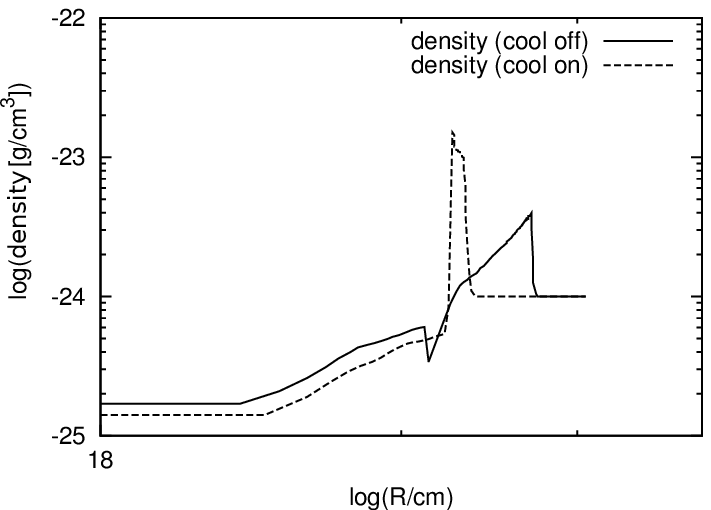}
 \caption{Radial distribution of temperature (left) and density (right) at $10^4$ yr for a shock with (solid)  and without cooling (dashed).\label{fig:test_shock1Dcool}}
\end{figure}

\begin{figure}
 \includegraphics[width=.5\textwidth]{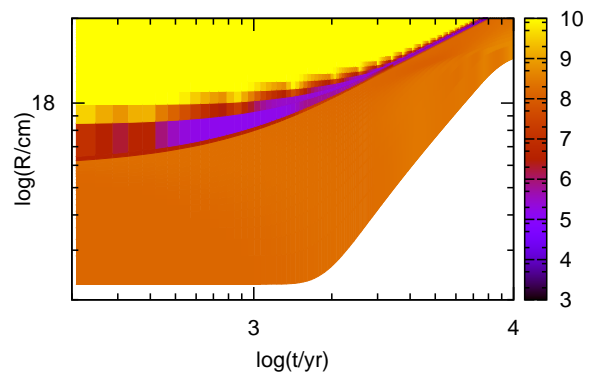}
 \caption{Map for the cooling time during the evolution of the shock as defined by Eqn.(\ref{eqn:coolingtime}). The logarithm of the cooling time in year is plotted as a function of the evolution time $t$ and the radius $R$. The label $>10$ indicates the values that exceed a cooling time of $10^{10}$ yr.\label{fig:test_coolingtime}}
\end{figure}

The results are plotted in Fig.\ref{fig:test_shock1Dcool} for a shock with the cooling functions enabled and disabled. We also plot a map of the cooling time as a function of the time of the simulation and the radius of the shells, where the cooling time is defined as
\begin{equation}\label{eqn:coolingtime}
 \tau_c=\frac{T}{\left|\dd T/ \dd t\right|}\,,
\end{equation}
where the lower term of the fraction is defined in Eq.(\ref{eqn:cooling}). The cooling time $\tau_c$ is defined as the time needed to reduce the temperature of a shell from $T$ to zero with the instantaneous cooling efficiency $\dd T/ \dd t$ evaluated with the cooling function $\Lambda(\bar n, T)$ for the array of abundances $\bar n$ and the temperature $T$ of the shell. Where the  quantity $\tau_c$ has the lower values its influence on the evolution of the shock is higher: in Fig.\ref{fig:test_coolingtime} it is clear that the shock front is the region most affected by the cooling, while in the outer region the cooling has no effects. The efficiency of the cooling on the shocking region is caused by the higher density that increases the number of exciting collisions of the chemical species. The final effect on the shock in Fig.\ref{fig:test_shock1Dcool} is that the cooling slows the advancing front (top), and reduce the temperature immediately before the outer region.

\subsection{Planetary atmosphere with diffusion and chemistry}\label{sect:test_planet}
The aim of the present test is to verify the behaviour of \krome when coupled to a standard one-dimensional diffusion model.
We represent the atmosphere of a planet with 64 cells at different height $z$, where each cell corresponds to a layer of the atmosphere.
The diffusion of the chemical species $n_i$ between each layer follows a standard diffusion equation, namely
\begin{equation}\label{eqn:diffusion}
\frac{\partial n_i}{\partial t} = -k(z)\frac{\partial^2 n_i}{\partial z^2}\,,
\end{equation}
where $k(z)$ is a diffusion coefficient that is a function of the height $z$.
This equation is coupled with the chemical model of \citet{Kasting1980} that contains 77 reactions and 30 species, and we solve the chemical differential equations according to Eqs.(\ref{eqn:ODEs}). It is important to note here that the so-called photochemical models, employed for the study of the planetary atmospheres, need to couple the chemical ODE system directly with Eqs.(\ref{eqn:diffusion}), since the chemistry of the long-living species have the same time-scale of the vertical diffusion. The simplification made in this example is intended only for illustrative purposes and is not intended for reproducing a realistic atmospheric model. The interested reader can get more details in the thorough paper by \citet{Hu2012} and the references therein.

The initial conditions of the model are taken from \citet{Segura2003} and the diffusion coefficients are from Jim Kasting's models website\footnote{\arl{http://vpl.astro.washington.edu/sci/AntiModels/models09.html}}, but, in order to keep the chemical time-scale similar to the dynamical time determined by Eq.(\ref{eqn:diffusion}), we normalize the chemical abundances to the species with the maximum abundances considering all the layers, and the diffusion coefficients are normalized to 0.5 in arbitrary units.

\begin{figure}
 \includegraphics[width=.48\textwidth]{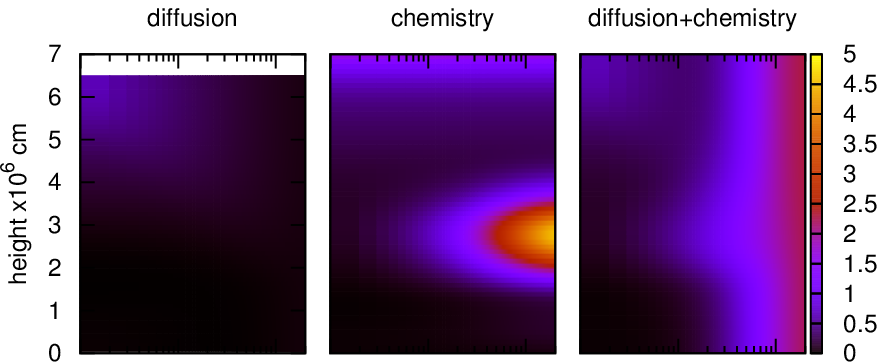}
 \includegraphics[width=.48\textwidth]{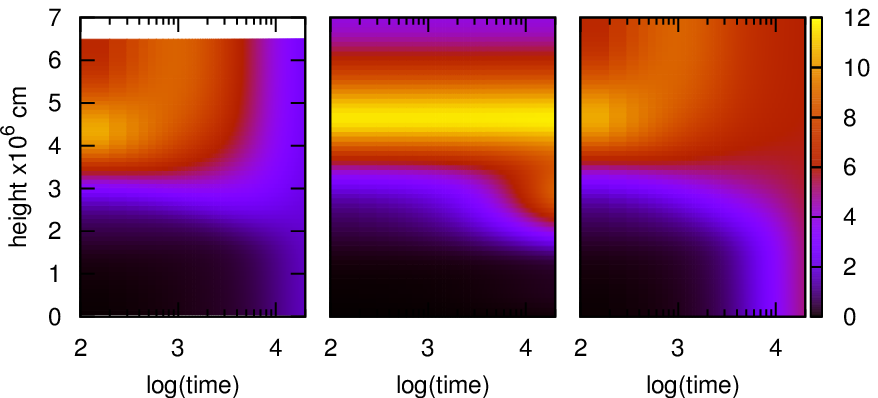}
 \caption{Evolution of the abundances of H$_2$ (top) and CO (bottom) at different heights. Left panel is with diffusion only, middle with chemistry only, and right with both enabled. Note that the colour-scale is in units of $10^{-7}$. Colours online\label{fig:diffuse}}
\end{figure}

We follow the model for $t_\mathrm{end} = 2\times10^{4}$ with (i) diffusion only, (ii) chemistry only, and (iii) chemistry and diffusion together. The results are shown in Fig.\ref{fig:diffuse} for H$_2$ and CO: in the left panel the system follows Eq.(\ref{eqn:diffusion}) only, hence reaches the equilibrium after $t_\mathrm{end}$. With the chemistry only (middle panel) the system acts as 64 one zone models which evolve following the chemical network, with no species exchanged between each cell. Finally, when both are enabled (right panel) the species formed by the chemistry are diffused into the other layers.

To generate the test the short-cut command line is
\begin{verbatim}
 > ./krome -test=atmosphere
\end{verbatim}
which is an alias for  
\begin{verbatim}
 > ./krome -n network/react_kast80 
	-useN
\end{verbatim}

\subsection{Slow-manifold kinetics}\label{sect:test_fake}
We propose here a test based on \citet{Reinhardt2008} to model a simple chemical network in order to show the robustness of the numerical solver \dlsodess. In the phase space of chemical kinetics systems (the space of the species), there are fast and slow dynamics invariant manifolds. For a given set of initial conditions a system quickly converges to a trajectory laying on the slow manifold, where the dynamics spends most of its evolutionary time until the steady-state \citep{Fraser1988,Jones1994,Zagaris2004,Nicolini2013}. In this test we employ a simple H$_2$ combustion model consisting of 12 reactions sketched in Tab.\ref{tab:slowtab} together with the temperature-independent reaction rates, for both the forward ($k_f$) and the reverse process ($k_r$), and involving 6 species (i.e. H$_2$, H, O$_2$, O, H$_2$O, OH). In this test, we demonstrate that the slow-manifold features are well reproduced. 

\begin{figure*}
	\includegraphics[width=0.48\textwidth]{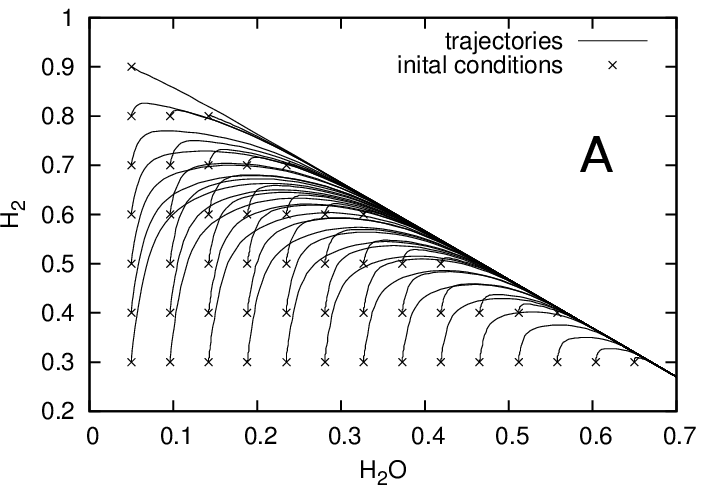}
	\includegraphics[width=0.48\textwidth]{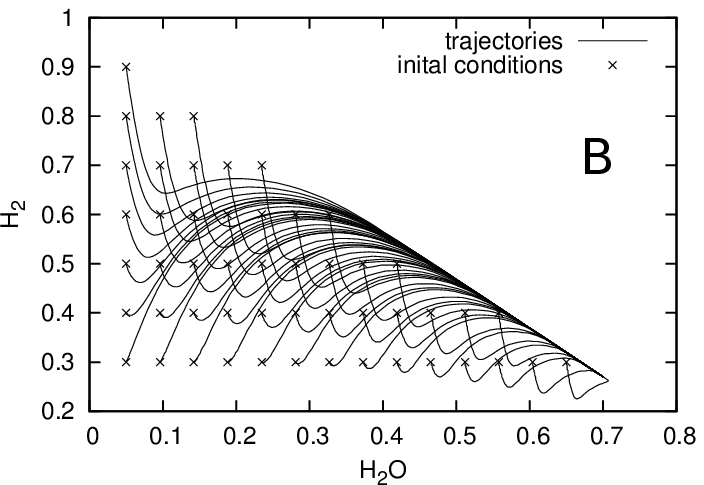}
        \caption{Family of trajectories for the H$_2$ combustion mechanism in \citet{Reinhardt2008}, for two different sets of initial conditions: with a zero (panel A) and a non-zero (panel B) initial oxygen abundance. The slow-manifold is clearly depicted by the transverse line where the trajectories converge.}
\label{fig:slow}
\end{figure*}

The initial conditions are varied over a grid of $n_{\mH_2}$ and $n_{\mH_2\mO}$ according to the constraint $n_{\mH_2}\leq0.96-n_{\mH_2\mO}$ in order to select only the values in the lower left  $n_{\mH_2}$-$n_{\mH_2\mO}$ space as discussed in \citet{Reinhardt2008}. The $n_{\mH_2}$ spans from $0.3$ to $0.95$ spaced by $0.1$, while $n_{\mH_2\mO}$ from $0.05$ to $0.65$ spaced by $0.05$ in arbitrary units. The system must satisfy the two following stoichiometric constraints
\begin{eqnarray}\label{eqn:slowcond}
	2n_{\mH_2} + 2n_{\mH_2\mO} + n_{\mH} + n_{\mO\mH} &=& 2\nonumber\\ 
	2n_{\mO_2} + n_{\mH_2\mO} + n_{\mO} + n_{\mO\mH} &=& 1\,,
\end{eqnarray}
which lead to
\begin{eqnarray}
	n_{\mO} & = & n_{\mO\mH} =  0\nonumber\\
	n_{\mO_2} & = & \left(1-n_{\mH_2\mO}-n_{\mO\mH}-n_{\mO}\right)/2\nonumber\\
	n_{\mH} & = & 2-n_{\mO\mH}-2\,n_{\mH_2\mO}-2\,n_{\mH_2}\,,
\end{eqnarray}
where the first condition is arbitrarily chosen.
We follow the evolution of the system until the steady-state, and we show the results in Fig.\ref{fig:slow} panel A. Our trajectories match the ones reported in Fig.1 of \citet{Reinhardt2008} and the slow-manifold is represented by the transverse line where the system quickly lays before reaching the equilibrium values located at (H$_2$, H, O$_2$, O, H$_2$O, OH) = (0.27, 0.05, 0.135, 0.02, 0.7, 0.01).

As an additional test we perform another calculation changing the initial conditions according to \eqnrefs{eqn:slowcond}, which leads also to 
\begin{eqnarray}
	n_{\mO_2} & = & n_{\mO\mH} =  0\nonumber\\
	n_{\mO} & = & 1-n_{\mH_2\mO}-n_{\mO\mH}-2\,n_{\mO}\nonumber\\
	n_{\mH} & = & 2-n_{\mO\mH}-2\,n_{\mH_2\mO}-2\,n_{\mH_2}\,,
\end{eqnarray}
since the first arbitrarily chosen condition also satisfies Eqs.(\ref{eqn:slowcond}).
In this test $n_{\mO}$ has a non-zero initial value enhancing the efficiency of the reaction $\mH_2+\mO\to\mH_2\mO$, which modifies the trajectories in the very first part of the various evolutions as shown in Fig.\ref{fig:slow} panel B. Also in this test the system converges to the slow-manifold finally reaching the same equilibrium values discussed above, proving the robustness of the solver employed in \kromes. 

The command line to generate the test is
\begin{verbatim}
 > ./krome -test=slowmanifold
\end{verbatim}
which is a short-cut for 
\begin{verbatim}
 > ./krome -n network/react_SM 
    -useN
\end{verbatim}

\begin{table}
	\begin{center}
		\caption{Reactions for the H$_2$ combustion model including the temperature-independent rate coefficients, where $k_f$ represents the forward rate coefficients and $k_r$ the reverse one. Note that the rate coefficient are in arbitrary units of amount/time.} \label{tab:slowtab}
		\begin{tabular}{cccll}
		  \hline
		  \multicolumn{3}{l}{Reaction} & $k_{f}$ & $k_{r}$ \\
		  \hline
		 H$_2$ & ${\overset{k_{\pm1}}{\rightleftharpoons}}$ & 2H & 2.0 & $2.16\times10^2$\\
		 O$_2$ & ${\overset{k_{\pm2}}{\rightleftharpoons}}$ & 2O & 1.0 & $3.375\times10^2$\\
		 H$_2$O & ${\overset{k_{\pm3}}{\rightleftharpoons}}$ & H + OH & 1.0 & $1.4\times10^3$\\
		 H$_2$ + O & ${\overset{k_{\pm4}}{\rightleftharpoons}}$ & H + OH & $10^3$ & $1.08\times10^4$\\
		 O$_2$ + H & ${\overset{k_{\pm5}}{\rightleftharpoons}}$ & O + OH & $10^3$ & $3.375\times10^4$\\
		 H$_2$ + O & ${\overset{k_{\pm6}}{\rightleftharpoons}}$ & H$_2$O & $10^2$ & $7.714\times10^{-1}$\\
                  \hline
		\end{tabular}
	\end{center}
\end{table}

\section{\krome embedded in 3D}\label{sect:3D}
One of the aims of the \krome package is to allow an efficient embedding of the chemistry in 3D hydrodynamical codes. 
It is well-known that an accurate treatment of the microphysical processes is very important in describing 
the dynamics of small scale objects (e.g. star formation processes or the ISM as well as in cosmological calculations). As already explained in section \ref{sect:ODE}, to solve 
a system of ODEs can be very demanding and often approximated approaches are preferred to deal with 3D hydrodynamic complexity.
Hydrodynamical codes as \enzo \citep{Enzo2004,Enzo2013}, and \flash \citep{Fryxell2000} include the chemical and thermal evolution in different ways. Most of them employ simple and very small networks with no more
than 13 species, including H, He, H$^+$, H$_2^+$, H$_2$, He$^+$, He$^{2+}$, D, HD, D$^+$, H$^-$, and e$^-$. 
Even in the latter case, only a few reactions involving those species are considered and often some approximations need to be introduced. 
Most of the problems are related to the time needed in solving the system of ODEs and it became very common to include the following approximations: (i) a decoupling among slower and faster species evolution, and (ii) the decoupling of temperature from the chemical evolution. The latter can generate instabilities at very high density. For instance, a first order Backward Differentiation Formula (BDF) method is typically employed in \enzos. The accuracy of such solvers is well-known and their stability strongly depends on the chosen time-step and on the stiffness of the problem. 
As already discussed in Section \ref{sect:solver}, the \krome package includes more accurate and stable solvers (\textsc{DVODE} and \dlsodess) with a robust internal definition of the time-step aimed at giving more stability and a better solution. It is important to note here that these solvers are also generally slower than a simple first-order BDF and that optimization techniques (e.g. \citealt{Grassi2013}) should be applied to speed-up the calculations. On the contrary, we should consider that an accurate solver is needed for high resolution studies \citep{Bovino2013b} and to solve more stiff problems (e.g. high-density environments, turbulent media). 

In the following, we show how \krome interfaces with complex 3D codes like \enzos, \ramses \citep{Teyssier2002}, and \flashs. We present a typical cosmological problem in \enzos, some standard hydrodynamical tests and cosmological problem in \ramsess, and a 1D and 3D collapse molecular cloud test in \flashs. The \krome patches for these codes are included in the current release\footnote{http://kromechem.org} of \kromes.

\subsection{KROME in RAMSES}\label{sect:ramses}
\ramses is an N-body and hydrodynamical code developed by \citet{Teyssier2002}. It has been designed to study structure formation with high spatial resolution by using an adaptive mesh 
refinement (AMR) technique, with a tree-based data structure. 

The \ramses code solves the hydrodynamical equations in their conservative form with the gravitational term included as a {\it non stiff source term} in the system of equations. Using the conservative form of the system, the discretized equations are solved computing the flux across the cell interfaces. In order to do that the code uses the second-order Godunov method (or Pieceweise Linear Method).

We employed \krome in the current version of the publicly available \ramses code and run a series of hydrodynamical and cosmological tests aimed 
at testing \kromes's flexibility and the possible sources of error when employed in such a complex code. We report results for a 1D test including Sod's shock tube problem and for a 3D cosmological simulations of a collapsing minihalo.
We created the \krome code patch following the \ramses user guide. The chemical species were initialized as passive scalars, and the code has been accordingly modified to include the chemical evolution for each leaf cell. 
An additional flag (chemistry) has been added to switch on/off the chemical evolution. \krome is called if the flag chemistry is TRUE. Note that the standard \ramses flag ``cooling'' can be used to employ the standard cooling already implemented in \ramses. All the tests presented in this work employed a simple primordial network involving H, He, H$^+$, H$_2^+$, H$_2$, He$^+$, He$^{2+}$, H$^-$, and e$^-$  and 20 reactions.  Atomic and H$_2$ cooling are included as well as the H$_2$ three-body formation heating. \\

\textsc{Sod shock-tube:}
\begin{figure}
 \includegraphics[width=.45\textwidth]{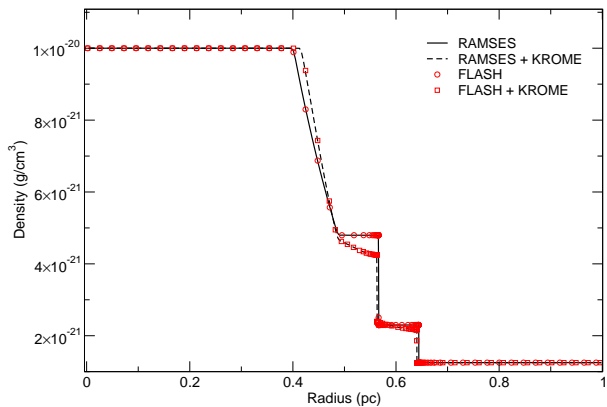}
 \caption{Sod's tube shock density evolution in \ramses, with (black solid curve) and without (black dashed curve) chemistry. The results are given in function of the radius expressed in parsec. The red curves are the results from \flash. See text for details.}
\label{fig:sod}
\end{figure}
the Sod shock is initialized by two separate regions, each initially at rest. The left region has higher pressure and density so that when the system is evolved, the fluid on the left moves to the right supersonically. For instance the density is 10$^{-20}$ g~cm$^{-3}$ and 0.125$\times$10$^{-20}$ g~cm$^{-3}$ on the left and right, respectively. The temperature is about 10$^3$ K on both regions. The solution is simple: a contact discontinuity moves to the right, with a shock wave ahead of it, while a rarefaction wave moves leftwards. All features of the test are well visible: the rarefaction waves, the contact discontinuity and the shocks. The snapshots are taken after 2.45$\times$10$^4$ yr. The effect of the chemistry is due to the cooling and causes a slower evolution of the shock as shown by the results for the density, velocity and temperature profiles reported in Figs. \ref{fig:sod}, \ref{fig:sod1}, and \ref{fig:sod2}, respectively. The behaviour of the hydrodynamical quantities, in particular the lower velocities and temperatures, reflects the chemical evolution reported in Fig.\ref{fig:sod_species}. It is important to discuss the H$_2$ evolution which is the main source of cooling. Indeed where the H$_2$ fraction is smaller the temperature is higher (e.g. around 0.6 pc where the contact discontinuity occurs).  

\begin{figure}
 \includegraphics[width=.45\textwidth]{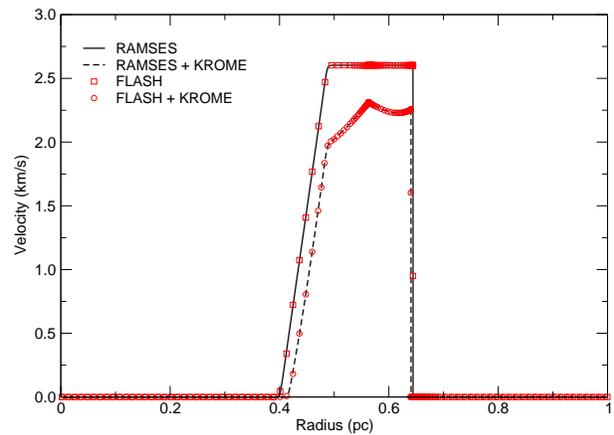}
 \caption{Sod's tube shock velocity evolution in \ramses, with (black solid curve) and without (black dashed curve) chemistry. The results are given in function of the radius expressed in parsec. The red curves are the results from \flash. See text for details.}
\label{fig:sod1}
\end{figure}

\begin{figure}
 \includegraphics[width=.45\textwidth]{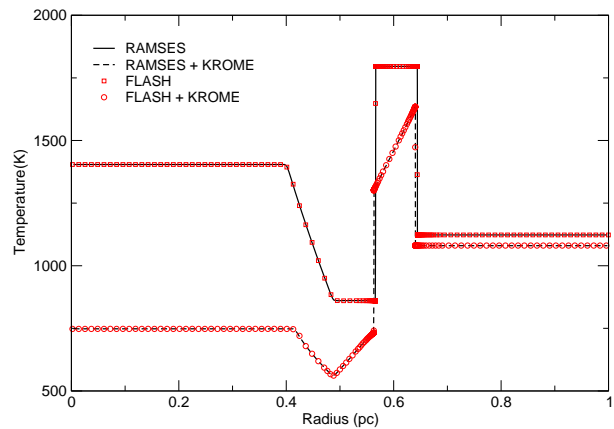}
 \caption{Sod's tube shock temperature evolution in \ramses, with (black solid curve) and without (black dashed curve) chemistry. The results are given in function of the radius expressed in parsec. The red curves are the results from \flash. See text for details.}
\label{fig:sod2}
\end{figure}

\begin{figure}
 \includegraphics[width=.45\textwidth]{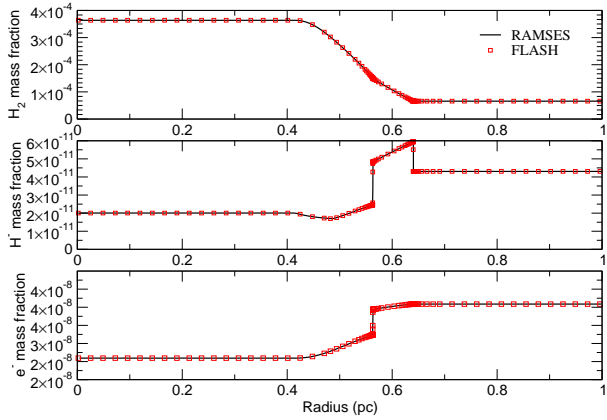}
 \caption{Sod's shock tube chemical species evolution in \ramses. H$_2$ (top), H$^-$ (middle), and e$^-$ (bottom) mass fraction are shown. See text for details.}
\label{fig:sod_species}
\end{figure}

In the same plots we also report a similar test performed with the AMR code \flash, with identical initial conditions (see Sections \ref{sect:flash} and \ref{sect:enzo} for additional details).\\

\textsc{Primordial minihalo evolution:}
To test the code's capability to manage non-equilibrium chemical systems within a cosmological scenario we have simulated the chemo-thermal
evolution of the gas inside a $1.3\times10^6$M$_\odot$ dark matter (DM) halo in the early Universe \citep[see e.g.][]{Abel2002, Bromm02, Yoshida08, Bovino2013b, Latif13b}.

To solve the hydrodynamic equations we used the MUSCL scheme and the HLLC Riemann solver.
The simulation has a starting redshift $z_i\approx 95$. The minihalo was selected from a DM simulation of $256^3$ particles inside a 
$(1\,{\rm Mpc})^3$ volume. After that, we re-simulated the selected halo inside the same box with three nested grids of 512$^3$, 256$^3$ and 128$^3$ 
particles. In this cosmological hydrodynamical simulation we set $l_{min}=7$ and $l_{max}=18$, reaching a resolution of $\Delta x\approx0.3$ proper pc at $z_f\approx11.5$.
At this final redshift the simulation reached the maximum level of refinement.


Our refinement strategy is as follows: (i) each cell is refined if it contains at least 8 DM particles, (ii) each gas cell is refined ensuring that the local
Jeans length is resolved by at least 16 cells, (iii) each cell is refined if the local pressure gradient $\Delta P/P > 2$. The highest resolution region is contained inside a 
spherical volume of radius $r=10^4$ co-moving kpc, which corresponds to three times the halo virial radius at the final redshift $z_f$, approximately.

\begin{figure}
 \includegraphics[width=.45\textwidth]{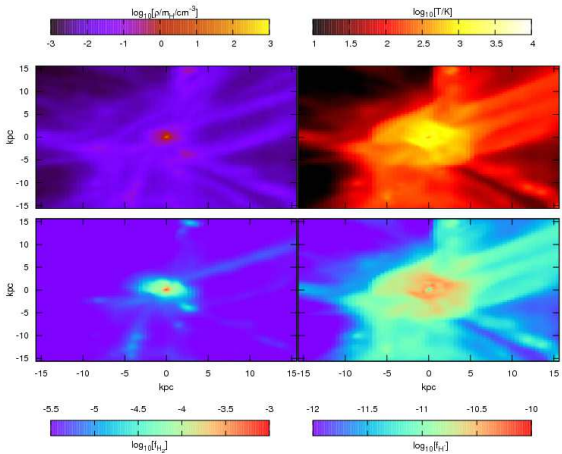}
 \caption{Mass weigthed projections for the gas density (top-left), gas temperature (top-right), H$_2$ mass fraction
(bottom-left) and H$^-$ mass fraction (bottom-right) at a co-moving resolution of $\Delta x\approx61$ pc. The maps correspond to $z_f=11.5$. The maps 
show the common features of a $\sim10^6$M$_\odot$ halo: the high desity central region shows a high 
($\sim10^{-3}$) H$_2$ mass fraction and a low ($\sim200-300$ K) temperature. The distances scales are in co-moving
units.}
\label{fig:mini_halo_large_scale}
\end{figure}
\begin{figure}
 \includegraphics[width=.45\textwidth]{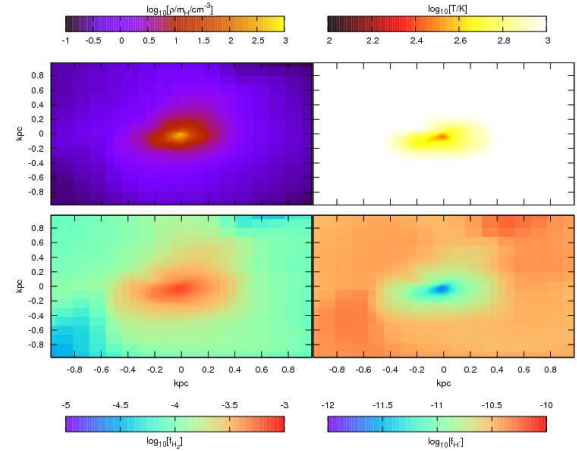}
 \caption{Same as figure \ref{fig:mini_halo_large_scale} but for a co-moving resolution of $\Delta x\approx3.8$ pc.}
\label{fig:mini_halo_small_scale}
\end{figure}

Fig.\ref{fig:mini_halo_large_scale} shows the mass weighted projections for gas density (top-left), temperature (top-right), H$_2$ (bottom-left) and H$^-$ (bottom-right) mass fraction at $z_f=11.5$. 
These maps show the common features for such systems: the central high density region presents the highest H$_2$ mass fraction producing a minimum in the gas temperature. Figure \ref{fig:mini_halo_small_scale} shows the same physical quantities at the same redshift but for the highest level of refinement. 

\begin{figure}
 \includegraphics[width=.45\textwidth]{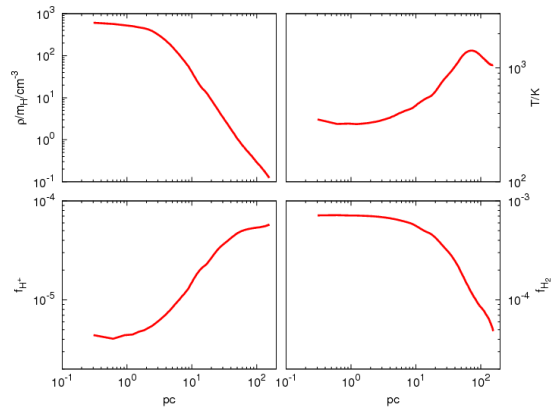}
 \caption{Mass weigthed radial profiles for the gas density (top-left), gas temperature (top-right), H$^+$ mass fraction
and H$_2$ mass mass fraction at $z_f=11.5$. The distance scales are in proper-physical units. The H$_2$ shows a well known behavior reaching a mass fraction $\la10^{-3}$ at scales $\la
10$ pc. The gas temperature also shows the expected features with a maximum inside the virial radius and a minimum at $\approx1$ pc. At smaller scales it rises due to the saturation of the H$_2$ cooling.}
\label{fig:radial_profiles_ramses}
\end{figure}

Fig.\ref{fig:radial_profiles_ramses} shows the mass weighted spherical averaged radial profiles for the gas density 
(top-left), temperature (top-right), H$^+$ (bottom-left) and H$_2$ (bottom-right) at $z_f=11.5$ for the highest resolution central region. The H$_2$ mass fraction in
the halo central region well agrees with the reported values in the literature for such systems (e.g. \citealt{Bovino2013a},\citealt{Bovino2013b}). The gas 
temperature also follows the expected behavior: it reaches a maximum (near the halo virial temperature) inside the virial
radius; it drops because the H$_2$ effect at scales smaller than a few tens of pc and it increases again at scales $\la1$ pc
due to the molecular cooling saturation.

\begin{figure}
 \includegraphics[width=.45\textwidth]{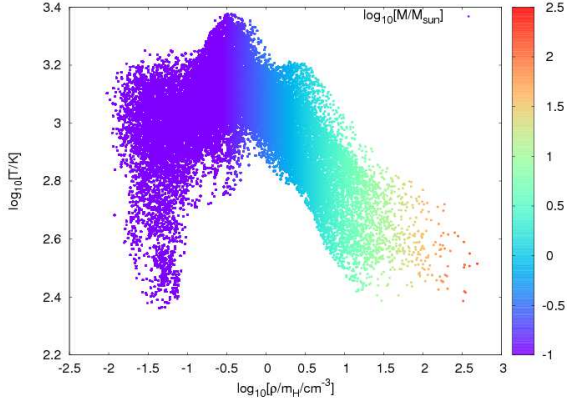}
 \caption{Dentity-temperature phase plane for the simulated mini-halo at $z_f=11.5$. The color bar shows the mass of each
point on the plane. The gas temperature shows the expected behavior rising at low density due to the adiabatic contraction.
Once the gas density reaches the $\approx0.3$cm$^{-3}$ value the H$_2$ abundance is enough to drop the gas temperature 
till ${\rm T}\approx250$ K in the high density regions.}
\label{fig:rhoT_plane_ramses}
\end{figure}

Figure \ref{fig:rhoT_plane_ramses} shows the density-temperature phase plane at the simulation's final redshift. The color
bar shows the mass of each cell on the phase plane. The low density ($\rho/m_H\la1$ cm$^{-3}$) points show the common 
temperature rising due to the gas adiabatic contraction. After reaching a maximum (near the halo virial temperature)
at densities $\rho/m_H\approx0.3$ cm$^{-3}$ the H$_2$ abundance is enough to cool the gas reaching a minimum temperature 
of ${\rm T}\approx250$ K.

\subsection{KROME in ENZO: a first application}\label{sect:enzo}
\enzo is an AMR code which is well-suited to follow gravitational collapse over a large range of scales, which includes various physical modules \citep{Enzo2004}. 
The hydrodynamical equations are solved in the comoving frame which takes into account the cosmological expansion of the Universe employing a split third-order piece-wise parabolic (PPM) solver. Any collisionless components (such as dark matter and stars) are modelled by N-body particles, whose dynamics are governed by Newton's equations. The gravitational potential is taken into account by solving the Poisson equation via multi-grid solver. 


\begin{figure}
	\includegraphics[width=0.45\textwidth]{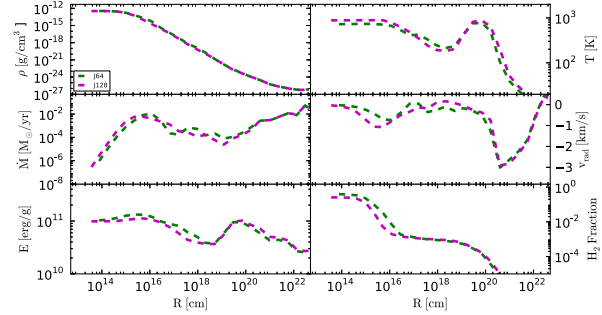}
       \caption{Spherically averaged radial profiles for HaloA runs, taken at the same peak density. Upper left: total density, upper right: temperature, middle left:accretion rate, middle right:radial velocity, bottom left the total energy and bottom right the H$_2$ mass fraction. See text for further details.}\label{fig:enzo_haloA}
\end{figure}

\begin{figure}
	\includegraphics[width=0.45\textwidth]{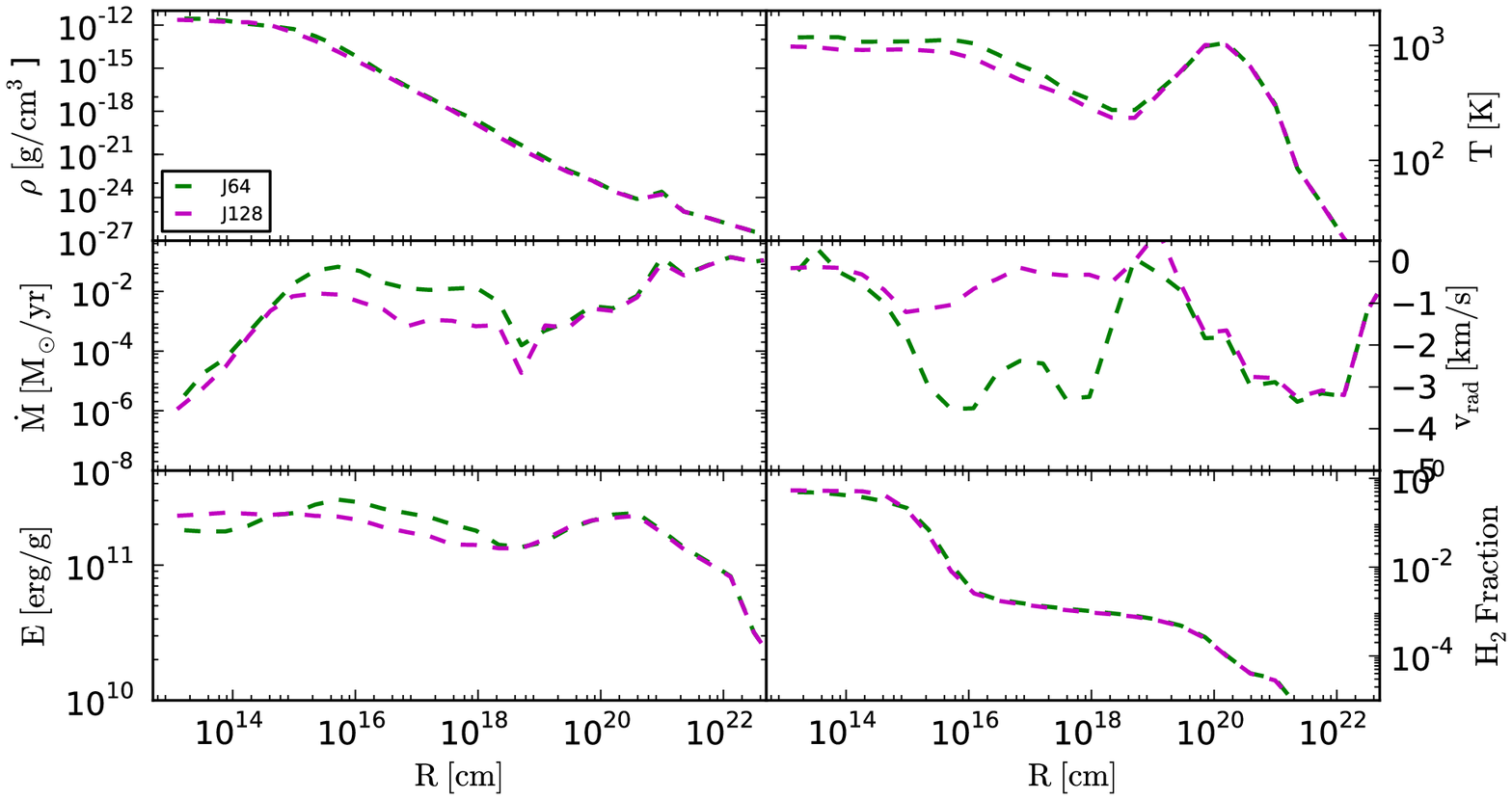}
        \caption{Spherically averaged radial profiles for HaloB runs, taken at the same peak density. Upper left: total density, upper right: temperature, middle left:accretion rate, middle right:radial velocity, bottom left the total energy and bottom right the H$_2$ mass fraction. See text for further details.}\label{fig:enzo_haloB}
\end{figure}

\begin{figure}
	\includegraphics[width=0.45\textwidth]{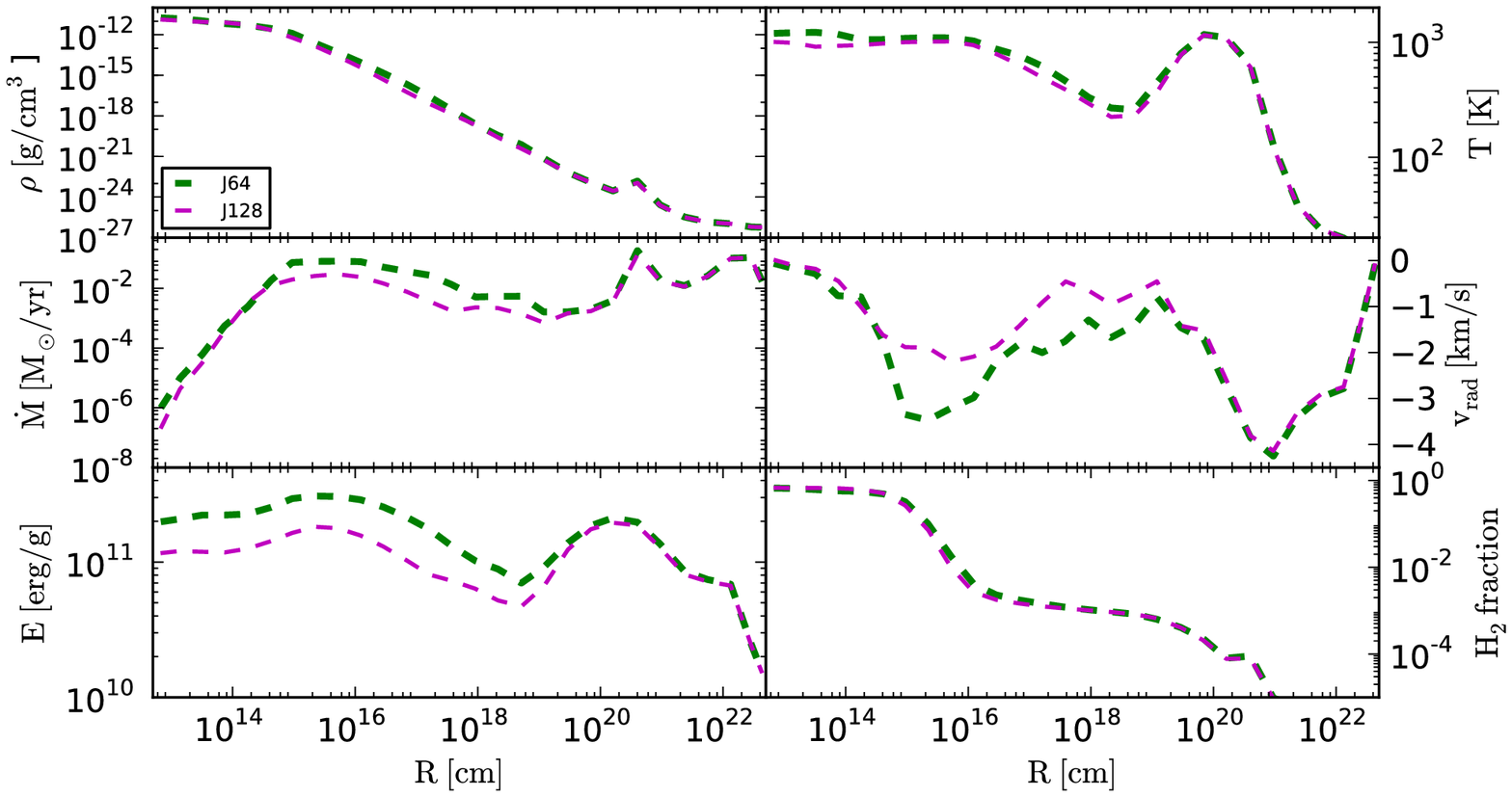}
        \caption{Spherically averaged radial profiles for Halo C runs, taken at the same peak density. Upper left: total density, upper right: temperature, middle left:accretion rate, middle right:radial velocity, bottom left the total energy and bottom right the H$_2$ mass fraction. See text for further details.}\label{fig:enzo_haloC}
\end{figure}

The initial conditions for the three DM primordial minihalos studied here are the same already discussed in \citet{Bovino2013b} and \citet{Bovino2013a}, i.e. three minihalos with masses of 1.3$\times$10$^5$M$_\odot$, 
7$\times$10$^5$M$_\odot$, and 1$\times$10$^6$M$_\odot$, a simulation box of 300 kpc~h$^{-1}$, 27 level of refinement in the central 18 kpc region, with an effective resolution of 0.9 AU in comoving units. The resolution criteria are based on (i) overdensity, (ii) particle mass-resolution, and (iii) Jeans length. In this test we present results from our implementation of \krome in \enzos. 
\begin{figure*}
	\includegraphics[width=0.75\textwidth]{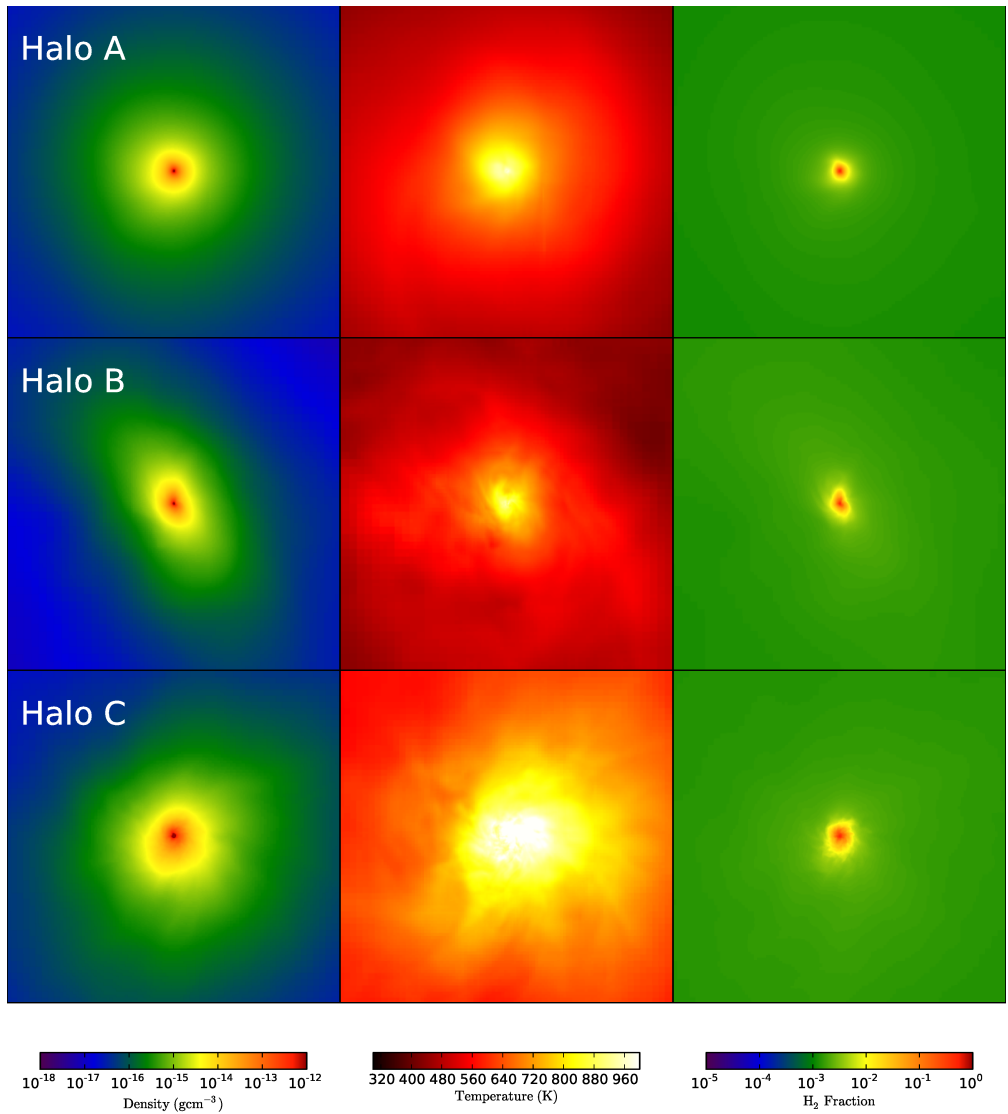}
        \caption{Density, temperature, and H$_2$ fraction projections at scale of 0.03 pc for the \krome runs in \enzo at the highest resolution $J_{128}$ and for the highest refinement level reached. See text for furhter details.}\label{fig:enzo_projection}
\end{figure*}

In Figs. \ref{fig:enzo_haloA}, \ref{fig:enzo_haloB}, and \ref{fig:enzo_haloC}, the results for three different minihalos and for different resolutions per Jeans length (J64, and J128) are shown. It is clear from the figures that the \krome chemistry does not show any dependence on the resolution parameters and the results are always consistent producing a temperature of about 1000 K. In this tests we assume an optical thin medium.

In the same figures other important thermodynamics quantities are presented, in particular accretion rates and radial velocity, both important quantities to handle with possible primordial star formation.  

In Fig. \ref{fig:enzo_projection} we report, for the three minihalos studied, the density, temperature, and H$_2$ fraction projections at scale of 0.03 pc for the last dump of our calculations.

A detailed discussion on the impact of using an accurate solver on high-resolution study, with a comparison with the simple BDF method employed in \enzo is given in \citet{Bovino2013b}. We note that the chemical and thermal structure of the halo is consistent with previous studies by \citet{Abel2002, Bromm02, Yoshida08, Clark11, Greif11, Latif13b, Greif13}, where we refer the reader for a comparison.

It is important to add that in term of computational time the \dlsodes is obviously slower when compared to the simple first-order BDF method employed in the original \enzos. Nevertheless, we expect an additional speed-up when provided the Jacobian and the sparsity structure to the solver or including a linear interpolation for the rates instead of an on the fly evaluation. In addition, our implementation of \krome in \enzo does not provide any interface optimization, it means that an overhead/overwork due to original \enzo routines (now not needed by \krome) exists and should be removed.
Finally, we should consider that a better and converged solution include a computational price that must be payed.

\subsection{KROME in FLASH}\label{sect:flash}

\flash is a versatile AMR code for a wide field of astrophysical applications \citep{Fryxell2000}. It can handle 1D, 2D, and 3D problems and incorporates several MHD solvers. The gravitational potential is calculated by solving the Poisson equation either by making use of a multigrid solver or a Barnes-Hut tree code.

In the following we present results obtained with \flash version 4.0.1 in which the \krome package was implemented. The implementation is straightforward. We handle the chemistry as an additional source term called after the hydro step update. The \krome module is called individually for each (active) cell in each block handing over the particle densities of each species as well as the gas temperature. All routines concerning the chemistry evolution, cooling and heating are provided by the \krome package as described before and do not have to be implemented in \flash itself.\\

\textsc{Sod shock tube}
Here we present the results of the Sod shock tube test described in detail in Section \ref{sect:ramses} for \ramsess. We note that we use identical initial conditions for the test in \flashs. The results are plotted in Figs. \ref{fig:sod}, \ref{fig:sod1}, \ref{fig:sod2}, and \ref{fig:sod_species}. As can be seen, the agreement between \ramses and \flash for the density, temperature and velocity as well as the three species H$_2$, H$^-$ and e$^-$ is almost perfect. We note that also for the other species not shown here the agreement is excellent. It is worth noting that for the test in \flash we use a higher-order solver that is the PPM, compared to the lower-order solver MUSCL used in the \ramses runs. Small differences in the plots can then be attributed to the different solvers employed.\\

\textsc{Cloud collapse}
We have performed simulations of a primordial minihalo in 1D and 3D following the collapse up to densities of the order of 10$^{-10}$ g cm$^{-3}$ corresponding to particle densities of $\sim$ 10$^{13}$ cm$^{-3}$. The grid was adapted dynamically in order to guarantee that the Jeans length is resolved everywhere with at least 32 grid cells using 16 levels of refinement. We used the PPM solver for updating the hydrodynamics and the Barnes-Hut tree code for solving the Poisson equation in 3D and the multipole solver for 1D. The initial conditions are chosen in order to mimic a highly idealized minihalo with a diameter of 0.45 pc consisting of atomic hydrogen and helium (mass fractions of 0.76 and 0.24, respectively) with a uniform number density of 1.67$\times$10$^{-18}$ g cm$^{-3}$. The temperature of the minihalo is set to 1000 K. The core is initially at rest and no magnetic field is present. The ambient medium has a 100 times lower density and is thus dynamically not important for the simulation result. We note that we do not include deuterium in this simulation as we expect no marked differences in the final results. In addition, for the 3D case we performed a run where a transonic turbulent velocity field is included. For all three cases, the (non-turbulent) 1D run, the non-turbulent 3D run, and the turbulent 3D run we performed two simulations, one with and one without the optically thick cooling correction term discussed in Section~\ref{sect:thermal}.

\begin{figure}
 \includegraphics[width=\linewidth]{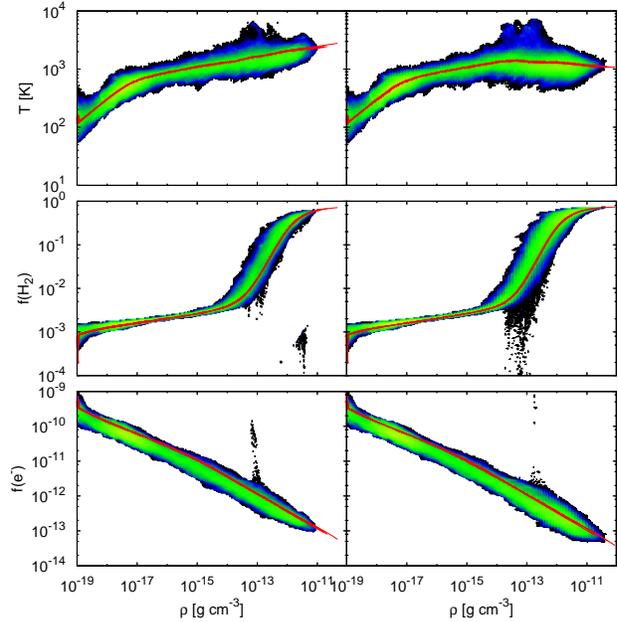}
\caption{Phase diagram of density and temperature (top), mass fraction of H$_2$ (middle) and e$^-$ (bottom) for the FLASH collapse runs at the time the highest refinement level is reached. The results of the non-turbulent 1D and 3D runs are shown by the red, almost completely overlapping lines, results of the turbulence runs by means of the colored areas. Results of the runs with the optically thick correction term are shown in the left panel, results without the correction term in the right panel.}
\label{fig:flash_species}
\end{figure}

In the following we show the results of the simulations at the time the highest refinement level is reached considering the runs with and without the optically thick correction separately. We first show the temperature and the mass fraction of H$_2$ and e$^-$ plotted against the density in Fig.~\ref{fig:flash_species}, with the results including the optically thick correction term shown in the left panel.
First, it can be seen that the results of the 1D run and the non-turbulent 3D run (red lines) are almost indistinguishable for both the case with the optically thick correction (left panel) and the one without it (right panel). For both cases the temperature and mass fractions of H$_2$ and e$^-$ reveal the expected behaviour in particular when comparing with Fig.~\ref{fig:collapsez} for non-metal case. The optically thick correction term becomes recognisable mainly at densities above 10$^{-13}$ g cm$^{-3}$ where the temperature keeps on increasing (top left panel) in contrast to the runs without the correction (top right panel).

Furthermore, for the turbulent run the diminished cooling ability results in a small fraction of gas with a very low H$_2$ fraction at high densities ($\sim$ 10$^{-12}$ -- 10$^{-11}$ g cm$^{-3}$, see middle right panel). This is a consequence of the high gas temperature resulting in the dissociation of molecular hydrogen. In contrast, the high temperature peaks showing up around 10$^{-13}$ g cm$^{-3}$ in the turbulent runs with and without the optically thick correction term (top panel) are most likely due to the presence of shocks created by the turbulence and do not arise in the non-turbulent simulations. In these hot, shocked regions the gas reveals a drop in the H$_2$ fraction and a simultaneous increase in the electron fraction (see also Fig.~\ref{fig:flash_project} below).

\begin{figure}
 \includegraphics[width=\linewidth]{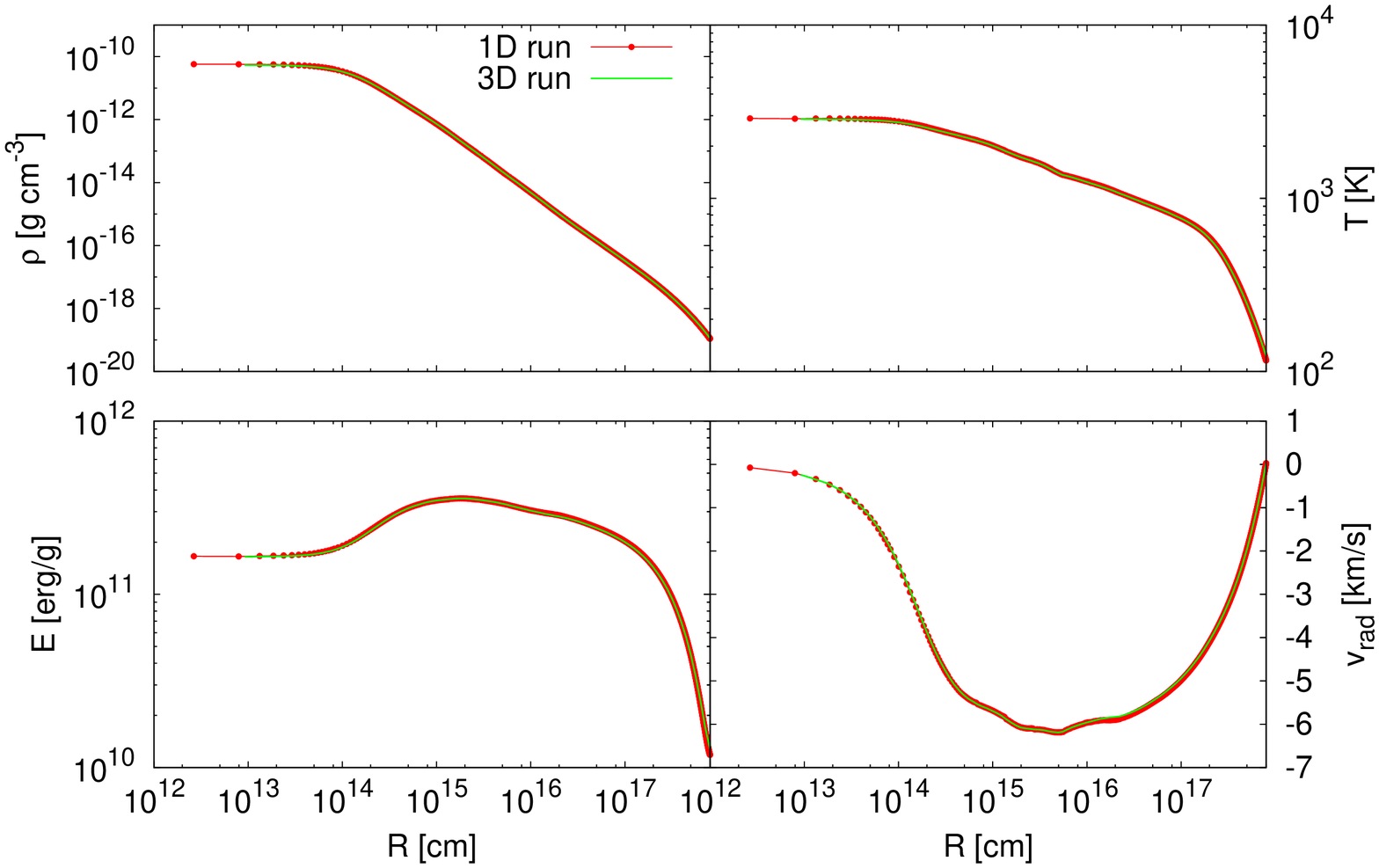}
\caption{Radial dependence of density, temperature, specific total energy and radial velocity for the non-turbulent \flash collapse runs (with the optically thick correction term) at the end of the simulation for the 1D and 3D case.}
\label{fig:flash_radial}
\end{figure}

Next, we plot the radial dependence of selected hydrodynamical variables in Fig.~\ref{fig:flash_radial} for the 1D and non-turbulent 3D run with the optically thick correction\footnote{We do not show the result for the run including turbulence since here the collapse occurs well off from the center.}. Although considering different spatial scales, there is a good qualitative agreement with the results shown in Figs. \ref{fig:enzo_haloA}, \ref{fig:enzo_haloB}, \ref{fig:enzo_haloC}, which confirms that \krome works properly in \flashs.

In order to get a visual impression of the simulation results, in Fig.~\ref{fig:flash_project} we show some projected quantities at the end of the 3D run including turbulence and the optically thick correction term.
\begin{figure*}
 \includegraphics[width=0.3\linewidth]{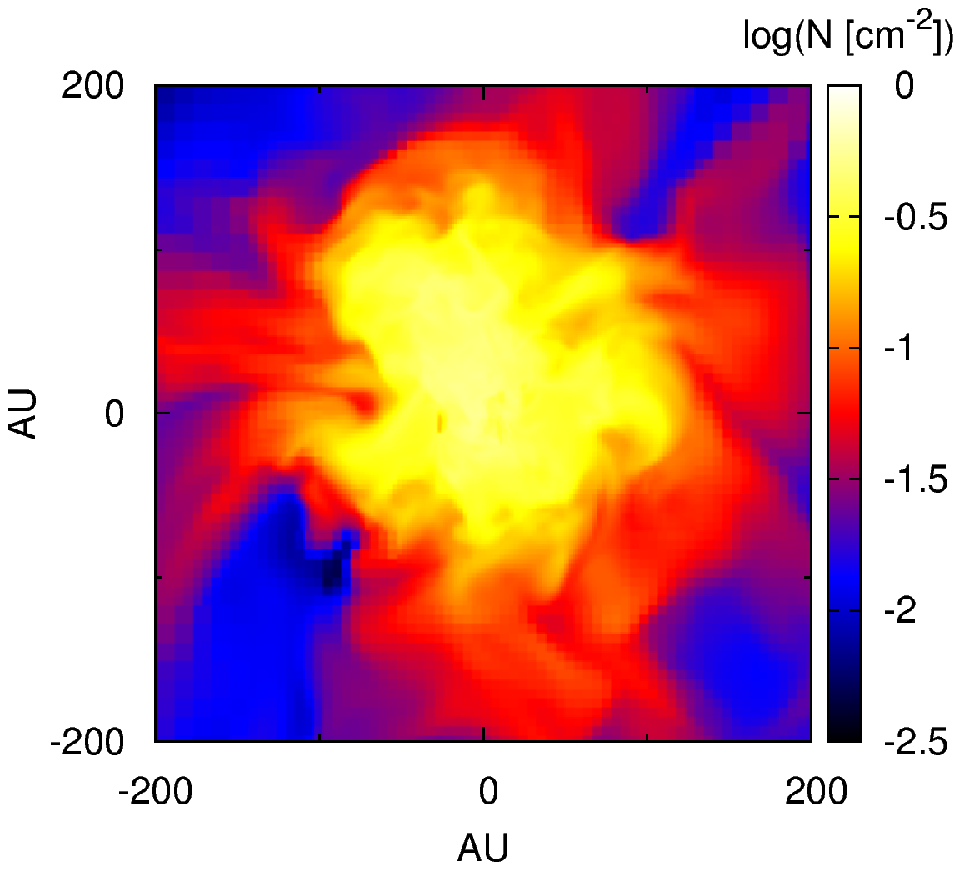}
 \includegraphics[width=0.3\linewidth]{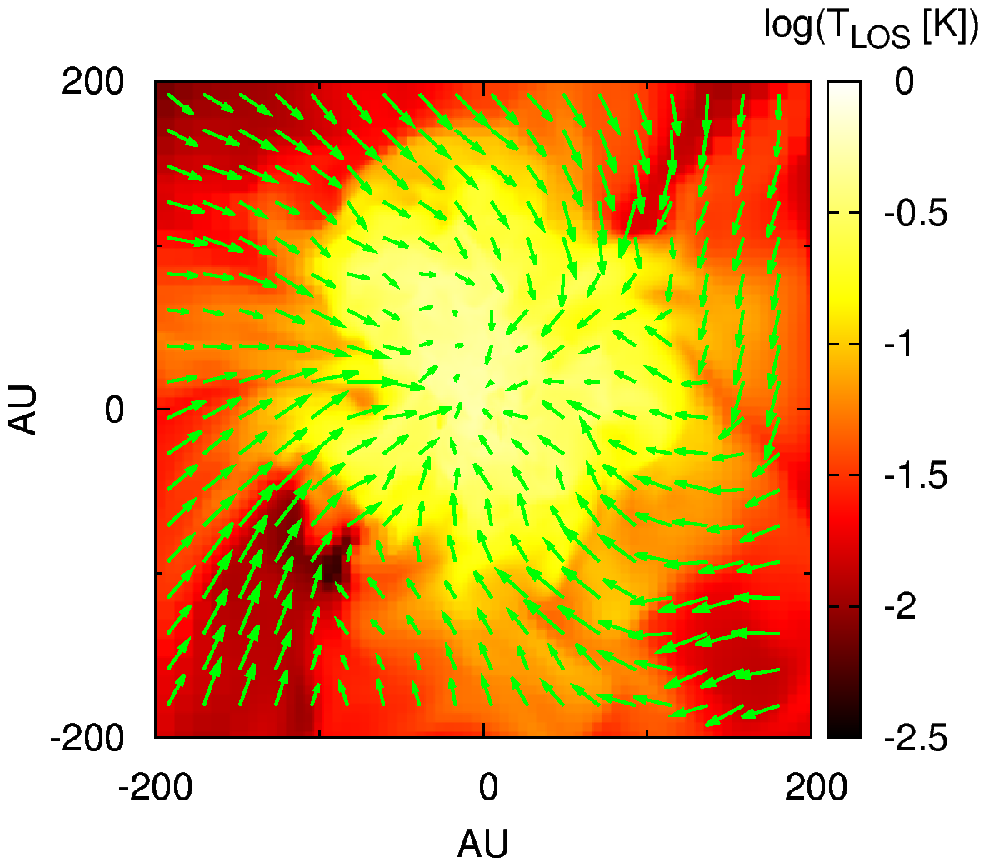}
 \includegraphics[width=0.3\linewidth]{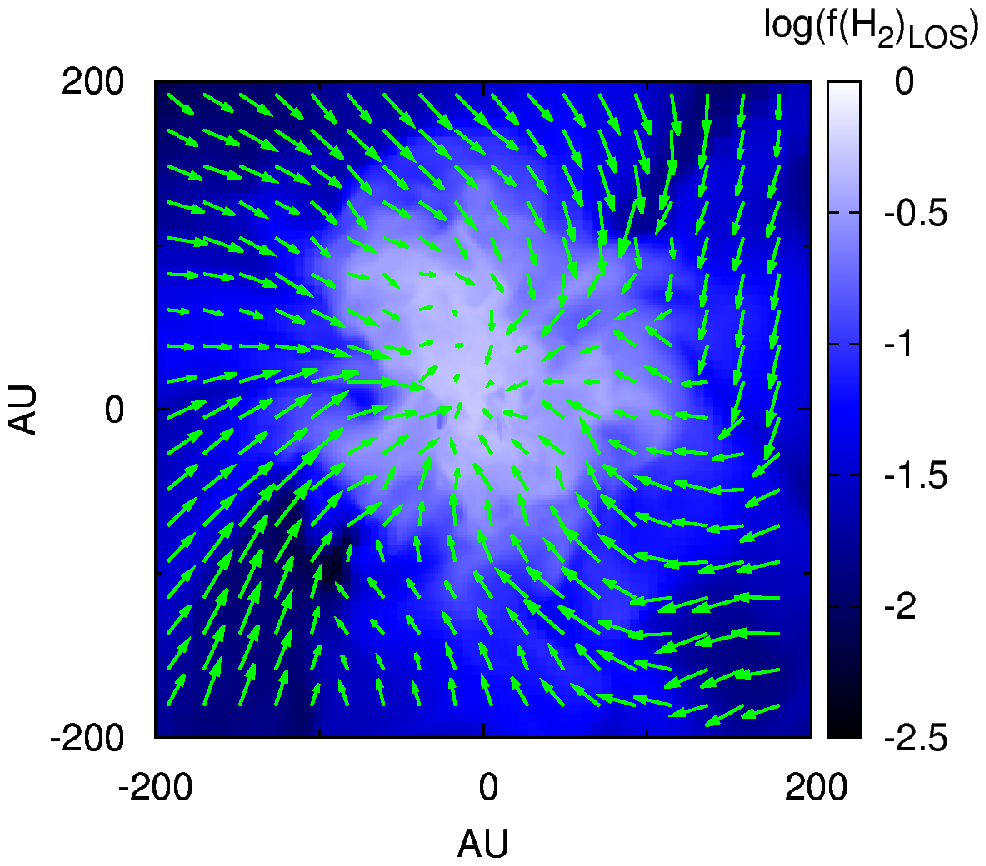}
 \caption{Column density (left), line-of-sight averaged temperature (middle), H$_2$ fraction (left), and velocity (green vectors) of the 3D run including turbulence and the optically thick correction term.}
\label{fig:flash_project}
\end{figure*}
From the middle panel of Fig.~\ref{fig:flash_project} it can be seen that there are regions of hot gas well off from the center, e.g. the hot spots in the lower left and upper right part of the figure. These regions are associated with a significant change in the gas flow indicating the presence of a shock. As stated already before, these hot regions have a relatively low H$_2$ fraction (see right panel).

Finally we state that for the 3D runs performed the amount of computational time for an individual cell is approximately the same independently of whether turbulence is present or not or whether the optically thick cooling term was applied or not. Furthermore, we find that the time required for the chemistry is comparable to the time required for the self-gravity module, i.e. for solving the Poisson equation. Each, the chemistry and the self-gravity module make up about 45\% of the total required computing time. We note here again that we did not apply any optimization procedure (e.g. explicit Jacobian and sparsity) that should improve the global performance.

To conclude, both, the Sod shock tube test and the collapse test show that \krome works nicely within the \flash code and produces reliable results.

\section{Summary and outlook}

We have presented the new chemistry package \krome for the modelling of chemical networks in numerical simulations. The package consists of a \python pre-processor providing \fortran subroutines to solve the rate equations for a given set of chemical reactions provided by the user. Our package provides a set of standard networks as appropriate for primordial chemistry, low metallicity gas, molecular clouds or planetary atmospheres, as well as additional modules for treating photodissociation reactions, the heating and cooling of the gas and the evolution of the dust grain population. In order to solve the chemical rate equations, we employ the high-order solver \dlsodess, which was shown to be both accurate and efficient when applied to sparse chemical networks, which are common in astrophysical applications \citep{Grassi2013, Bovino2013b}.

Our main goals were to create a package which is user-friendly and straightforward to apply for a given set of chemical reactions, and to combine that with an accurate and efficient numerical solver. The latter is particularly important for applications where the package is employed in 3D hydrodynamical simulations, where a substantial amount of time is already spent on solving the equations of hydrodynamics of gravity, and exceeding this time by more than a factor of a few can typically not be afforded. At the same time, it is also important not to choose the simplest numerical technique, which often has the highest efficiency but does not necessarily provide the correct solution. For instance in high-resolution simulations following gravitational collapse, first-order backward differencing techniques fail to converge at high resolution, while the \dlsodes solver provides consistent results as shown in Sect.\ref{sect:enzo} and also discussed in \citet{Bovino2013b}.

In this paper, we have provided an extensive test suite including one-zone models for the chemistry in molecular clouds, gravitational collapse at varying metallicity and radiative backgrounds, 1D shocks, tests of the dust physics, planetary atmospheres and slow-manifold kinetics. In addition, we have applied \krome in the cosmological hydrodynamics codes \enzo and \ramses to explore the chemical evolution in the early Universe and during the formation of the first minihalos, while looking at the evolution of the chemistry during gravitational collapse in the hydrodynamics code \flashs. The results obtained here are consistent with previous studies and show that our package can be efficiently applied in hydrodynamical simulations.

The applications of \krome are not limited to this regime, but the general framework can be employed to investigate the chemistry during a much larger variety of situations, including the formation of molecular clouds \citep{Glover07b, Banerjee09}, the formation of stars \citep{Banerjee07, Hennebelle11, Seifried13}, galactic disks \citep{Tasker09, Shetty12}, the environments of supermassive black holes \citep{Wada09, Hocuk12, Latif13} or planet formation \citep{Lyra09, Johansen11}. 

Of course, pursuing such applications still requires a consideration of the chemical reactions that should be considered. As discussed in the Introduction, large compilations of chemical reactions for different applications are currently available, which can be directly explored in one-zone or 1D models. For three-dimensional simulations, it is however necessary to reduce the chemical complexity and choosing only the main reactions that determine the chemical and in particular the thermal structure. Appropriate reduction techniques have been presented by \citet{Wiebe2003, Semenov2004, Grassi2012, Grassi2013}, which can help to reduce the number of reactions to make more complex simulations computationally feasible.

In a future release of \kromes, we are planning to incorporate such reduction techniques in order to simplify the derivation of new networks. Our current release however contains already a set of different networks which can be employed in astrophysical simulations, and many small astrophysical networks exist in the literature that can be adopted via the \python pre-processor. We therefore expect that our package will make the application of such networks in numerical simulations more straightforward, and encourages new users to include them in their simulations.

The \krome release is available on \mbox{\url{https://bitbucket.org/krome/krome_stable}} and on \mbox{\url{http://kromepackage.org/}} where also a quick guide is included.

\section*{Acknowledgements}
T.G. acknowledges the financial support from the CINECA and S.B. and D.R.G.S. thank for funding through the DFG priority programme `The Physics of the Interstellar Medium' (project SCHL 1964/1-1). D.R.G.S. thanks for funding via the SFB 963/1 on ``Astrophysical Flow Instabilities and Turbulence'' (project A12). D.S. acknowledges funding by the Deutsche Forschungsgemeinschaft via
grant BA 3706/3-1 within the SPP \textit{The Physics of the Interstellar Medium}. E.S. thanks the Agreement ASI/INAF I/025/10/0. F.A.G. thanks the PRIN2009 project by the Ministry of Research (MIUR). We all are grateful to S.~Danielache for the hints about the planetary atmosphere, S.C.O.~Glover and K.~Omukai for useful discussions on the metal cooling, C.~Chiosi, D.~Galli, M.A.~Latif, E.~Merlin, and F.Palla, for the fruitful comments, P.~Nicolini for having indicated us the references of the slow-manifold test, and M.~Satta for the useful comments on thermodynamics. T.G. also thanks J.~R.~Brucato at INAF-Arcetri, and the Institut f\"ur Astrophysik in G\"ottingen for the kind hospitality during the preparation of the final draft. \tgcomment{We thank the anonymous referee for the useful suggestions}.

\bibliographystyle{mn2e}      
\bibliography{mybib} 

\begin{thebibliography}{119}
\expandafter\ifx\csname natexlab\endcsname\relax\def\natexlab#1{#1}\fi

\bibitem[{{Abel} {et~al}\mbox{.}(1997){Abel}, {Anninos}, {Zhang}, \&
  {Norman}}]{Abel97}
{Abel} T., {Anninos} P., {Zhang} Y., {Norman} M.~L., 1997, New Astronomy, 2,
  181

\bibitem[{Abel, Bryan \& Norman(2002)Abel, Bryan, \& Norman}]{Abel2002}
Abel T., Bryan G.~L., Norman M.~L., 2002, Science, 295, 93

\bibitem[{Al-Khateeb {et~al}\mbox{.}(2009)Al-Khateeb, Powers, Paolucci,
  Sommese, Diller, Hauenstein, \& Mengers}]{Ashraf2009}
Al-Khateeb A.~N., Powers J.~M., Paolucci S., Sommese A.~J., Diller J.~A.,
  Hauenstein J.~D., Mengers J.~D., 2009, The Journal of Chemical Physics, 131,
  024118

\bibitem[{Anders \& Grevesse(1989)}]{Anders1989}
Anders E., Grevesse N., 1989, Geochimica et Cosmochimica Acta, 53, 197

\bibitem[{Anderson {et~al}\mbox{.}(1990)Anderson, Bai, Dongarra, Greenbaum,
  McKenney, Du~Croz, Hammerling, Demmel, Bischof, \& Sorensen}]{LAPACK}
Anderson E. {et~al.}, 1990, in Proceedings of the 1990 ACM/IEEE conference on
  Supercomputing, Supercomputing '90, IEEE Computer Society Press, Los
  Alamitos, CA, USA, pp. 2--11

\bibitem[{{Anninos} {et~al}\mbox{.}(1997){Anninos}, {Zhang}, {Abel}, \&
  {Norman}}]{Anninos97}
{Anninos} P., {Zhang} Y., {Abel} T., {Norman} M.~L., 1997, New Astronomy, 2,
  209

\bibitem[{{Bader} \& {Deuflhard}(1983)}]{Bader1983}
{Bader} G., {Deuflhard} P., 1983, Numerische Mathematik, 41, 373

\bibitem[{{Banerjee} \& {Pudritz}(2007)}]{Banerjee07}
{Banerjee} R., {Pudritz} R.~E., 2007, \apj, 660, 479

\bibitem[{{Banerjee} {et~al}\mbox{.}(2009){Banerjee}, {V{\'a}zquez-Semadeni},
  {Hennebelle}, \& {Klessen}}]{Banerjee09}
{Banerjee} R., {V{\'a}zquez-Semadeni} E., {Hennebelle} P., {Klessen} R.~S.,
  2009, \mnras, 398, 1082

\bibitem[{Baulch {et~al}\mbox{.}(1994)Baulch, Cobos, Cox, Frank, Hayman, Just,
  Kerr, Murrells, Pilling, Troe, Walker, \& Warnatz}]{Baulch1994}
Baulch D.~L. {et~al.}, 1994, Journal of Physical and Chemical Reference Data,
  23, 847

\bibitem[{Bodenheimer {et~al}\mbox{.}(2006)Bodenheimer, Laughlin, Rozyczka, \&
  Yorke}]{Bodenheimer2006}
Bodenheimer P., Laughlin G., Rozyczka M., Yorke H., 2006, Numerical Methods in
  Astrophysics: An Introduction, Series in Astronomy and Astrophysics. Taylor
  \& Francis

\bibitem[{{Borysow}(2002)}]{Borysow2002}
{Borysow} A., 2002, \aap, 390, 779

\bibitem[{{Borysow U.~G.~Jorgensen} \& {Fu}(2001)}]{Borysow2001}
{Borysow U.~G.~Jorgensen} A., {Fu} Y., 2001, J. of Quant. Spectr. \& Rad.
  Trans., 68, 235

\bibitem[{{Bovino} {et~al}\mbox{.}(2013){Bovino}, {Grassi}, {Latif}, \&
  {Schleicher}}]{Bovino2013b}
{Bovino} S., {Grassi} T., {Latif} M.~A., {Schleicher} D.~R.~G., 2013, \mnras,
  434, L36

\bibitem[{{Bovino}, {Schleicher} \& {Grassi}(2013){Bovino}, {Schleicher}, \&
  {Grassi}}]{Bovino2013a}
{Bovino} S., {Schleicher} D.~R.~G., {Grassi} T., 2013, ArXiv e-prints

\bibitem[{{Bromm}, {Coppi} \& {Larson}(2002){Bromm}, {Coppi}, \&
  {Larson}}]{Bromm02}
{Bromm} V., {Coppi} P.~S., {Larson} R.~B., 2002, \apj, 564, 23

\bibitem[{Burcat(1984)}]{Burcat1984}
Burcat A., 1984, in Combustion Chemistry, Gardiner WilliamC. J., ed., Springer
  US, pp. 455--473

\bibitem[{{Burrows} \& {Sharp}(1999)}]{Burrows1999}
{Burrows} A., {Sharp} C.~M., 1999, \apj, 512, 843

\bibitem[{{Cazaux} \& {Spaans}(2009)}]{Cazaux2009}
{Cazaux} S., {Spaans} M., 2009, \aap, 496, 365

\bibitem[{{Cen}(1992)}]{Cen1992}
{Cen} R., 1992, ApJS, 78, 341

\bibitem[{Chase(1998)}]{Chase1998}
Chase M. W.~J., 1998, {NIST-JANAF Thermochemical Tables, 4th Edition}. American
  Institute of Physics, New York

\bibitem[{{Clark} {et~al}\mbox{.}(2011){Clark}, {Glover}, {Smith}, {Greif},
  {Klessen}, \& {Bromm}}]{Clark11}
{Clark} P.~C., {Glover} S.~C.~O., {Smith} R.~J., {Greif} T.~H., {Klessen}
  R.~S., {Bromm} V., 2011, Science, 331, 1040

\bibitem[{{Dalgarno} \& {Lepp}(1987)}]{DalgarnoLepp87}
{Dalgarno} A., {Lepp} S., 1987, in IAU Symposium, Vol. 120, Astrochemistry,
  {Vardya} M.~S., {Tarafdar} S.~P., eds., pp. 109--118

\bibitem[{{Draine} \& {Lee}(1984)}]{Draine1984}
{Draine} B.~T., {Lee} H.~M., 1984, \apj, 285, 89

\bibitem[{{Dwek}(1998)}]{Dwek1998}
{Dwek} E., 1998, \apj, 501, 643

\bibitem[{{Efstathiou}(1992)}]{Efstathiu1992}
{Efstathiou} G., 1992, \mnras, 256, 43P

\bibitem[{Eisenstat {et~al}\mbox{.}(1977)Eisenstat, Gursky, Schultz, Sherman,
  \& SCIENCE.}]{Eisenstat1977}
Eisenstat S., Gursky C., Schultz H., Sherman A., SCIENCE. Y. U. N. H. C. D.
  O.~C., 1977, Yale Sparse Matrix Package. II. The Nonsymmetric Codes. Defense
  Technical Information Center

\bibitem[{Eisenstat {et~al}\mbox{.}(1982)Eisenstat, Gursky, Schultz, \&
  Sherman}]{Eisenstat1982}
Eisenstat S.~C., Gursky M.~C., Schultz M.~H., Sherman A.~H., 1982,
  International Journal for Numerical Methods in Engineering, 18, 1145

\bibitem[{Fraser(1988)}]{Fraser1988}
Fraser S.~J., 1988, The Journal of Chemical Physics, 88, 4732

\bibitem[{Fryxell {et~al}\mbox{.}(2000)Fryxell, Olson, Ricker, Timmes, Zingale,
  Lamb, MacNeice, Rosner, Truran, \& Tufo}]{Fryxell2000}
Fryxell B. {et~al.}, 2000, Astrophysical Journal, Supplement, 131, 273

\bibitem[{{Galli} \& {Palla}(1998)}]{Galli1998}
{Galli} D., {Palla} F., 1998, \aap, 335, 403

\bibitem[{{Glover} \& {Abel}(2008)}]{Glover2008}
{Glover} S.~C.~O., {Abel} T., 2008, \mnras, 388, 1627

\bibitem[{{Glover} \& {Jappsen}(2007)}]{Glover2007}
{Glover} S.~C.~O., {Jappsen} A.-K., 2007, \apj, 666, 1

\bibitem[{{Glover} \& {Mac Low}(2007)}]{Glover07b}
{Glover} S.~C.~O., {Mac Low} M.-M., 2007, \apj, 659, 1317

\bibitem[{{Gould} \& {Salpeter}(1963)}]{Gould1963}
{Gould} R.~J., {Salpeter} E.~E., 1963, \apj, 138, 393

\bibitem[{{Grassi} {et~al}\mbox{.}(2012){Grassi}, {Bovino}, {Gianturco},
  {Baiocchi}, \& {Merlin}}]{Grassi2012}
{Grassi} T., {Bovino} S., {Gianturco} F.~A., {Baiocchi} P., {Merlin} E., 2012,
  \mnras, 425, 1332

\bibitem[{{Grassi} {et~al}\mbox{.}(2013){Grassi}, {Bovino}, {Schleicher}, \&
  {Gianturco}}]{Grassi2013}
{Grassi} T., {Bovino} S., {Schleicher} D., {Gianturco} F.~A., 2013, \mnras

\bibitem[{{Grassi} {et~al}\mbox{.}(2011{\natexlab{a}}){Grassi}, {Krstic},
  {Merlin}, {Buonomo}, {Piovan}, \& {Chiosi}}]{Grassi2011}
{Grassi} T., {Krstic} P., {Merlin} E., {Buonomo} U., {Piovan} L., {Chiosi} C.,
  2011{\natexlab{a}}, \aap, 533, A123

\bibitem[{{Grassi} {et~al}\mbox{.}(2011{\natexlab{b}}){Grassi}, {Merlin},
  {Piovan}, {Buonomo}, \& {Chiosi}}]{Grassi11}
{Grassi} T., {Merlin} E., {Piovan} L., {Buonomo} U., {Chiosi} C.,
  2011{\natexlab{b}}, arXiv:1103.0509

\bibitem[{{Greif}, {Springel} \& {Bromm}(2013){Greif}, {Springel}, \&
  {Bromm}}]{Greif13}
{Greif} T.~H., {Springel} V., {Bromm} V., 2013, \mnras, 434, 3408

\bibitem[{{Greif} {et~al}\mbox{.}(2011){Greif}, {Springel}, {White}, {Glover},
  {Clark}, {Smith}, {Klessen}, \& {Bromm}}]{Greif11}
{Greif} T.~H., {Springel} V., {White} S.~D.~M., {Glover} S.~C.~O., {Clark}
  P.~C., {Smith} R.~J., {Klessen} R.~S., {Bromm} V., 2011, \apj, 737, 75

\bibitem[{{Hennebelle} {et~al}\mbox{.}(2011){Hennebelle}, {Commer{\c c}on},
  {Joos}, {Klessen}, {Krumholz}, {Tan}, \& {Teyssier}}]{Hennebelle11}
{Hennebelle} P., {Commer{\c c}on} B., {Joos} M., {Klessen} R.~S., {Krumholz}
  M., {Tan} J.~C., {Teyssier} R., 2011, \aap, 528, A72

\bibitem[{Hindmarsh(1983)}]{Hindmarsh1983}
Hindmarsh A.~C., 1983, IMACS Transactions on Scientific Computation, 1, 55

\bibitem[{Hindmarsh {et~al}\mbox{.}(2005)Hindmarsh, Brown, Grant, Lee, Serban,
  Shumaker, \& Woodward}]{Hindmarsh2005}
Hindmarsh A.~C., Brown P.~N., Grant K.~E., Lee S.~L., Serban R., Shumaker
  D.~E., Woodward C.~S., 2005, ACM Trans. Math. Softw., 31, 363

\bibitem[{{Hirano} \& {Yoshida}(2013)}]{Hirano2013}
{Hirano} S., {Yoshida} N., 2013, \apj, 763, 52

\bibitem[{{Hirashita} \& {Kuo}(2011)}]{Hirashita2011}
{Hirashita} H., {Kuo} T.-M., 2011, \mnras, 416, 1340

\bibitem[{{Hirashita} \& {Yan}(2009)}]{Hirashita2009}
{Hirashita} H., {Yan} H., 2009, \mnras, 394, 1061

\bibitem[{{Hocuk} {et~al}\mbox{.}(2012){Hocuk}, {Schleicher}, {Spaans}, \&
  {Cazaux}}]{Hocuk12}
{Hocuk} S., {Schleicher} D.~R.~G., {Spaans} M., {Cazaux} S., 2012, \aap, 545,
  A46

\bibitem[{{Hollenbach} \& {McKee}(1979)}]{Hollenbach1979}
{Hollenbach} D., {McKee} C.~F., 1979, \apjs, 41, 555

\bibitem[{{Hollenbach} \& {McKee}(1989)}]{Hollenbach1989}
{Hollenbach} D., {McKee} C.~F., 1989, \apj, 342, 306

\bibitem[{{Hu}, {Seager} \& {Bains}(2012){Hu}, {Seager}, \& {Bains}}]{Hu2012}
{Hu} R., {Seager} S., {Bains} W., 2012, \apj, 761, 166

\bibitem[{{Johansen}, {Klahr} \& {Henning}(2011){Johansen}, {Klahr}, \&
  {Henning}}]{Johansen11}
{Johansen} A., {Klahr} H., {Henning} T., 2011, \aap, 529, A62

\bibitem[{Jones(1995)}]{Jones1994}
Jones C., 1995, in Lecture Notes in Mathematics, Vol. 1609, Dynamical Systems,
  Johnson R., ed., Springer Berlin Heidelberg, pp. 44--118

\bibitem[{{Kasting} \& {Donahue}(1980)}]{Kasting1980}
{Kasting} J.~F., {Donahue} T.~M., 1980, \jgr, 85, 3255

\bibitem[{{Katz}, {Weinberg} \& {Hernquist}(1996){Katz}, {Weinberg}, \&
  {Hernquist}}]{Katz96}
{Katz} N., {Weinberg} D.~H., {Hernquist} L., 1996, \apjs, 105, 19

\bibitem[{{Kumar} \& {Fisher}(2013)}]{Kumar2013}
{Kumar} A., {Fisher} R.~T., 2013, arXiv:1302.0330

\bibitem[{{Laor} \& {Draine}(1993)}]{Laor1993}
{Laor} A., {Draine} B.~T., 1993, \apj, 402, 441

\bibitem[{{Latif} {et~al}\mbox{.}(2013{\natexlab{a}}){Latif}, {Schleicher},
  {Schmidt}, \& {Niemeyer}}]{Latif13}
{Latif} M.~A., {Schleicher} D.~R.~G., {Schmidt} W., {Niemeyer} J.,
  2013{\natexlab{a}}, \mnras, 433, 1607

\bibitem[{{Latif} {et~al}\mbox{.}(2013{\natexlab{b}}){Latif}, {Schleicher},
  {Schmidt}, \& {Niemeyer}}]{Latif13b}
{Latif} M.~A., {Schleicher} D.~R.~G., {Schmidt} W., {Niemeyer} J.,
  2013{\natexlab{b}}, \apjl, 772, L3

\bibitem[{{Le Petit} {et~al}\mbox{.}(2006){Le Petit}, {Nehm{\'e}}, {Le
  Bourlot}, \& {Roueff}}]{LePetit2006}
{Le Petit} F., {Nehm{\'e}} C., {Le Bourlot} J., {Roueff} E., 2006, \apjs, 164,
  506

\bibitem[{{Leitch-Devlin} \& {Williams}(1985)}]{LeitchDevlin1985}
{Leitch-Devlin} M.~A., {Williams} D.~A., 1985, \mnras, 213, 295

\bibitem[{{Lenzuni}, {Chernoff} \& {Salpeter}(1991){Lenzuni}, {Chernoff}, \&
  {Salpeter}}]{Lenzuni1991}
{Lenzuni} P., {Chernoff} D.~F., {Salpeter} E.~E., 1991, \apjs, 76, 759

\bibitem[{{Leung}, {Herbst} \& {Huebner}(1984){Leung}, {Herbst}, \&
  {Huebner}}]{Leung1984}
{Leung} C.~M., {Herbst} E., {Huebner} W.~F., 1984, \apjs, 56, 231

\bibitem[{{Lewis}(1997)}]{lewis_book_1997}
{Lewis} J.~S., 1997, {Physics and chemistry of the solar system}

\bibitem[{{Li} \& {Draine}(2001)}]{LiDraine2001b}
{Li} A., {Draine} B.~T., 2001, \apj, 554, 778

\bibitem[{{Li}, {Klessen} \& {Mac Low}(2003){Li}, {Klessen}, \& {Mac
  Low}}]{Li2003}
{Li} Y., {Klessen} R.~S., {Mac Low} M.-M., 2003, \apj, 592, 975

\bibitem[{{Lipovka}, {N{\'u}{\~n}ez-L{\'o}pez} \& {Avila-Reese}(2005){Lipovka},
  {N{\'u}{\~n}ez-L{\'o}pez}, \& {Avila-Reese}}]{Lipovka2005}
{Lipovka} A., {N{\'u}{\~n}ez-L{\'o}pez} R., {Avila-Reese} V., 2005, \mnras,
  361, 850

\bibitem[{{Lodato}(2007)}]{Lodato2007}
{Lodato} G., 2007, Nuovo Cimento Rivista Serie, 30, 293

\bibitem[{{Lodders} \& {Fegley}(2002)}]{Lodders2002}
{Lodders} K., {Fegley} B., 2002, \icarus, 155, 393

\bibitem[{{Lyra} {et~al}\mbox{.}(2009){Lyra}, {Johansen}, {Zsom}, {Klahr}, \&
  {Piskunov}}]{Lyra09}
{Lyra} W., {Johansen} A., {Zsom} A., {Klahr} H., {Piskunov} N., 2009, \aap,
  497, 869

\bibitem[{{Maio} {et~al}\mbox{.}(2007){Maio}, {Dolag}, {Ciardi}, \&
  {Tornatore}}]{Maio2007}
{Maio} U., {Dolag} K., {Ciardi} B., {Tornatore} L., 2007, \mnras, 379, 963

\bibitem[{{Maloney}, {Hollenbach} \& {Tielens}(1996){Maloney}, {Hollenbach}, \&
  {Tielens}}]{Maloney1996}
{Maloney} P.~R., {Hollenbach} D.~J., {Tielens} A.~G.~G.~M., 1996, \apj, 466,
  561

\bibitem[{{Maret}, {Bergin} \& {Tafalla}(2013){Maret}, {Bergin}, \&
  {Tafalla}}]{Maret2013}
{Maret} S., {Bergin} E.~A., {Tafalla} M., 2013, ArXiv e-prints

\bibitem[{{Martin}, {Keogh} \& {Mandy}(1998){Martin}, {Keogh}, \&
  {Mandy}}]{Martin1998}
{Martin} P.~G., {Keogh} W.~J., {Mandy} M.~E., 1998, \apj, 499, 793

\bibitem[{{Mathis}, {Rumpl} \& {Nordsieck}(1977){Mathis}, {Rumpl}, \&
  {Nordsieck}}]{Mathis1977}
{Mathis} J.~S., {Rumpl} W., {Nordsieck} K.~H., 1977, \apj, 217, 425

\bibitem[{{Meijerink} \& {Spaans}(2005)}]{Meijerink2005}
{Meijerink} R., {Spaans} M., 2005, \aap, 436, 397

\bibitem[{{Navarro} \& {Steinmetz}(1997)}]{Navarro1997}
{Navarro} J.~F., {Steinmetz} M., 1997, \apj, 478, 13

\bibitem[{{Nejad}(2005)}]{Nejad2005}
{Nejad} L.~A.~M., 2005, \apss, 299, 1

\bibitem[{Nicolini \& Frezzato(2013)}]{Nicolini2013}
Nicolini P., Frezzato D., 2013, The Journal of Chemical Physics, 138, 234102

\bibitem[{{Nozawa}, {Kozasa} \& {Habe}(2006){Nozawa}, {Kozasa}, \&
  {Habe}}]{Nozawa2006}
{Nozawa} T., {Kozasa} T., {Habe} A., 2006, \apj, 648, 435

\bibitem[{{Omukai}(2000)}]{Omukai2000}
{Omukai} K., 2000, ApJ, 534, 809

\bibitem[{{Omukai}(2001)}]{Omukai2001}
{Omukai} K., 2001, \apj, 546, 635

\bibitem[{{Omukai} {et~al}\mbox{.}(2005){Omukai}, {Tsuribe}, {Schneider}, \&
  {Ferrara}}]{Omukai2005}
{Omukai} K., {Tsuribe} T., {Schneider} R., {Ferrara} A., 2005, \apj, 626, 627

\bibitem[{{O'Shea} {et~al}\mbox{.}(2004){O'Shea}, {Bryan}, {Bordner}, {Norman},
  {Abel}, {Harkness}, \& {Kritsuk}}]{Enzo2004}
{O'Shea} B.~W., {Bryan} G., {Bordner} J., {Norman} M.~L., {Abel} T., {Harkness}
  R., {Kritsuk} A., 2004, arXiv:0403044

\bibitem[{{Peters} {et~al}\mbox{.}(2012){Peters}, {Schleicher}, {Klessen},
  {Banerjee}, {Federrath}, {Smith}, \& {Sur}}]{Peters2012}
{Peters} T., {Schleicher} D.~R.~G., {Klessen} R.~S., {Banerjee} R., {Federrath}
  C., {Smith} R.~J., {Sur} S., 2012, \apjl, 760, L28

\bibitem[{{Press} {et~al}\mbox{.}(1992){Press}, {Teukolsky}, {Vetterling}, \&
  {Flannery}}]{Press1992}
{Press} W.~H., {Teukolsky} S.~A., {Vetterling} W.~T., {Flannery} B.~P., 1992,
  {Numerical recipes in FORTRAN. The art of scientific computing}. Cambridge
  University Press

\bibitem[{Reinhardt, Winckler \& Lebiedz(2008)Reinhardt, Winckler, \&
  Lebiedz}]{Reinhardt2008}
Reinhardt V., Winckler M., Lebiedz D., 2008, J.Phys.Chem.A, 112, 1712

\bibitem[{{Rentrop} \& {Kaps}(1979)}]{Rentrop1979}
{Rentrop} P., {Kaps} P., 1979, Numerische Mathematik, 33, 55

\bibitem[{{Ripamonti} \& {Abel}(2004)}]{Ripamonti2004}
{Ripamonti} E., {Abel} T., 2004, \mnras, 348, 1019

\bibitem[{{Santoro} \& {Shull}(2006)}]{Santoro2006}
{Santoro} F., {Shull} J.~M., 2006, \apj, 643, 26

\bibitem[{{Schleicher} {et~al}\mbox{.}(2008){Schleicher}, {Galli}, {Palla},
  {Camenzind}, {Klessen}, {Bartelmann}, \& {Glover}}]{Schleicher2008}
{Schleicher} D.~R.~G., {Galli} D., {Palla} F., {Camenzind} M., {Klessen} R.~S.,
  {Bartelmann} M., {Glover} S.~C.~O., 2008, \aap, 490, 521

\bibitem[{{Schleicher}, {Spaans} \& {Glover}(2010){Schleicher}, {Spaans}, \&
  {Glover}}]{Schleicher2010}
{Schleicher} D.~R.~G., {Spaans} M., {Glover} S.~C.~O., 2010, \apjl, 712, L69

\bibitem[{{Schneider} {et~al}\mbox{.}(2006){Schneider}, {Omukai}, {Inoue}, \&
  {Ferrara}}]{Schneider2006}
{Schneider} R., {Omukai} K., {Inoue} A.~K., {Ferrara} A., 2006, \mnras, 369,
  1437

\bibitem[{{Segura} {et~al}\mbox{.}(2003){Segura}, {Krelove}, {Kasting},
  {Sommerlatt}, {Meadows}, {Crisp}, {Cohen}, \& {Mlawer}}]{Segura2003}
{Segura} A., {Krelove} K., {Kasting} J.~F., {Sommerlatt} D., {Meadows} V.,
  {Crisp} D., {Cohen} M., {Mlawer} E., 2003, Astrobiology, 3, 689

\bibitem[{{Seifried} {et~al}\mbox{.}(2013){Seifried}, {Banerjee}, {Pudritz}, \&
  {Klessen}}]{Seifried13}
{Seifried} D., {Banerjee} R., {Pudritz} R.~E., {Klessen} R.~S., 2013, \mnras,
  432, 3320

\bibitem[{{Semenov} {et~al}\mbox{.}(2010){Semenov}, {Hersant}, {Wakelam},
  {Dutrey}, {Chapillon}, {Guilloteau}, {Henning}, {Launhardt}, {Pi{\'e}tu}, \&
  {Schreyer}}]{Semenov2010}
{Semenov} D. {et~al.}, 2010, \aap, 522, A42

\bibitem[{{Semenov}, {Wiebe} \& {Henning}(2004){Semenov}, {Wiebe}, \&
  {Henning}}]{Semenov2004}
{Semenov} D., {Wiebe} D., {Henning} T., 2004, \aap, 417, 93

\bibitem[{{Shang}, {Bryan} \& {Haiman}(2010){Shang}, {Bryan}, \&
  {Haiman}}]{Shang2010}
{Shang} C., {Bryan} G.~L., {Haiman} Z., 2010, \mnras, 402, 1249

\bibitem[{{Shetty} \& {Ostriker}(2012)}]{Shetty12}
{Shetty} R., {Ostriker} E.~C., 2012, \apj, 754, 2

\bibitem[{{Stancil}, {Lepp} \& {Dalgarno}(1998){Stancil}, {Lepp}, \&
  {Dalgarno}}]{Stancil1998}
{Stancil} P.~C., {Lepp} S., {Dalgarno} A., 1998, \apj, 509, 1

\bibitem[{{Tasker} \& {Tan}(2009)}]{Tasker09}
{Tasker} E.~J., {Tan} J.~C., 2009, \apj, 700, 358

\bibitem[{{Teyssier}(2002)}]{Teyssier2002}
{Teyssier} R., 2002, \aap, 385, 337

\bibitem[{{The Enzo Collaboration}(2013)}]{Enzo2013}
{The Enzo Collaboration}, 2013, arXiv:1307.2265

\bibitem[{{Tielens} {et~al}\mbox{.}(1994){Tielens}, {McKee}, {Seab}, \&
  {Hollenbach}}]{Tielens1994}
{Tielens} A.~G.~G.~M., {McKee} C.~F., {Seab} C.~G., {Hollenbach} D.~J., 1994,
  \apj, 431, 321

\bibitem[{{Vedel}, {Hellsten} \& {Sommer-Larsen}(1994){Vedel}, {Hellsten}, \&
  {Sommer-Larsen}}]{Vedel1994}
{Vedel} H., {Hellsten} U., {Sommer-Larsen} J., 1994, \mnras, 271, 743

\bibitem[{{Verner} \& {Ferland}(1996)}]{Verner1996}
{Verner} D.~A., {Ferland} G.~J., 1996, \apjs, 103, 467

\bibitem[{{Visscher} \& {Moses}(2011)}]{visschermoses2011}
{Visscher} C., {Moses} J.~I., 2011, \apj, 738, 72

\bibitem[{{Wada}, {Papadopoulos} \& {Spaans}(2009){Wada}, {Papadopoulos}, \&
  {Spaans}}]{Wada09}
{Wada} K., {Papadopoulos} P.~P., {Spaans} M., 2009, \apj, 702, 63

\bibitem[{{Wakelam} \& {Herbst}(2008)}]{Wakelam2008}
{Wakelam} V., {Herbst} E., 2008, \apj, 680, 371

\bibitem[{{Wakelam} {et~al}\mbox{.}(2012){Wakelam}, {Herbst}, {Loison},
  {Smith}, {Chandrasekaran}, {Pavone}, {Adams}, {Bacchus-Montabonel},
  {Bergeat}, {B{\'e}roff}, {Bierbaum}, {Chabot}, {Dalgarno}, {van Dishoeck},
  {Faure}, {Geppert}, {Gerlich}, {Galli}, {H{\'e}brard}, {Hersant}, {Hickson},
  {Honvault}, {Klippenstein}, {Le Picard}, {Nyman}, {Pernot}, {Schlemmer},
  {Selsis}, {Sims}, {Talbi}, {Tennyson}, {Troe}, {Wester}, \&
  {Wiesenfeld}}]{Wakelam2012}
{Wakelam} V. {et~al.}, 2012, \apjs, 199, 21

\bibitem[{{Wakelam} {et~al}\mbox{.}(2010){Wakelam}, {Smith}, {Herbst}, {Troe},
  {Geppert}, {Linnartz}, {{\"O}berg}, {Roueff}, {Ag{\'u}ndez}, {Pernot},
  {Cuppen}, {Loison}, \& {Talbi}}]{Wakelam2010}
{Wakelam} V. {et~al.}, 2010, \ssr, 156, 13

\bibitem[{{Walsh} {et~al}\mbox{.}(2009){Walsh}, {Harada}, {Herbst}, \&
  {Millar}}]{Walsh2009}
{Walsh} C., {Harada} N., {Herbst} E., {Millar} T.~J., 2009, \apj, 700, 752

\bibitem[{{Weingartner} \& {Draine}(2001)}]{Weingartner2001}
{Weingartner} J.~C., {Draine} B.~T., 2001, \apj, 548, 296

\bibitem[{{Wiebe}, {Semenov} \& {Henning}(2003){Wiebe}, {Semenov}, \&
  {Henning}}]{Wiebe2003}
{Wiebe} D., {Semenov} D., {Henning} T., 2003, \aap, 399, 197

\bibitem[{{Woitke}, {Kamp} \& {Thi}(2009){Woitke}, {Kamp}, \&
  {Thi}}]{Woitke2009}
{Woitke} P., {Kamp} I., {Thi} W.-F., 2009, \aap, 501, 383

\bibitem[{{Woodward} \& {Colella}(1984)}]{Woodward1984}
{Woodward} P.~R., {Colella} P., 1984, Journal of Computational Physics, 54, 115

\bibitem[{{Yahil}(1983)}]{Yahil1983}
{Yahil} A., 1983, \apj, 265, 1047

\bibitem[{{Yoshida}, {Omukai} \& {Hernquist}(2008){Yoshida}, {Omukai}, \&
  {Hernquist}}]{Yoshida08}
{Yoshida} N., {Omukai} K., {Hernquist} L., 2008, Science, 321, 669

\bibitem[{Zagaris, Kaper \& Kaper(2004)Zagaris, Kaper, \& Kaper}]{Zagaris2004}
Zagaris A., Kaper H.~G., Kaper T.~J., 2004, Journal of Nonlinear Science, 14,
  59

\end{thebibliography}

\appendix

\section{Values adopted for H$_2$ cooling}
Here we present the fitting coefficients employed for the H$_2$ cooling function as discussed in \ref{sect:cooling} and adopted by \citet{Glover2008}. The list of fitting coefficients for H$_2$ cooling rates is reported in Tab.\ref{tab:total-coeffs}.

\begin{table*}
\caption{Fitting coefficients for H$_2$ cooling rates, for a 3:1 ortho-para ratio}\label{tab:total-coeffs}
\begin{tabular}{cclcclccl}
\hline
Species & Temperature range (K) & Coefficients & Species & Temperature range (K) & Coefficients\\
\hline
H & $10 < T \le 100$ & $a_{0} = -16.818342$ & H & $100 < T \le 1000$ & $a_{0} = -24.311209$ \\
 & &  $a_{1} =  37.383713 $ & & & $a_{1} = 3.5692468$ \\
 & &   $a_{2} = 58.145166 $ & & & $a_{2} =  -11.332860$ \\
 & &   $a_{3} = 48.656103 $ & & & $a_{3} = -27.850082$ \\
 & &   $a_{4} = 20.159831 $ & & & $a_{4} = -21.328264$ \\
  & &   $a_{5} = 3.8479610 $ & & & $a_{5} = -4.2519023$ \\
  & &  & & \\
H & $1000 < T \le 6000$ &  $a_{0} = -24.311209$ & H$_2$ & $100 < T \le 6000$ & $ a_{0} =  -23.962112$ \\
& & $a_{1} = 4.6450521$   &  & & $a_{1} = 2.09433740$ \\
& & $a_{2} =  -3.7209846$ &  & & $a_{2} =  -0.77151436$ \\
& & $a_{3} = 5.9369081$   &  & & $a_{3} = 0.43693353$ \\
& & $a_{4} = -5.5108047$  &  & & $a_{4} = -0.14913216$ \\
& & $a_{5} =  1.5538288$  &  & & $a_{5} = -0.033638326$ \\
& &  & & \\
He & $10 < T \le 6000$ & $a_{0} = -23.689237$ &  H$^+$ & $10 < T \le 10000$ & $a_{0} = -21.716699$ \\
& & $a_{1} =   2.1892372$  &  & & $a_{1} =  1.3865783$ \\
& & $a_{2} =  -0.81520438$ &  & & $a_{2} =  -0.37915285$ \\
& & $a_{3} =  0.29036281$  &  & & $a_{3} =  0.11453688$ \\
& & $a_{4} =  -0.16596184$ &  & & $a_{4} = -0.23214154$ \\
& & $a_{5} =  0.19191375$  &  & & $a_{5} = 0.058538864$ \\
& & & & \\
e$^-$ & $10 < T \le 200$ & $a_{0} =  -34.286155$ &  e$^-$ & $200 < T \le 10000$ & $a_{0} = -22.190316$ \\
& & $a_{1} = -48.537163 $  & & & $a_{1} = 1.5728955$ \\
& & $a_{2} =  -77.121176$  & & & $a_{2} = -0.21335100$ \\
& & $a_{3} =  -51.352459$  & & & $a_{3} = 0.96149759$ \\
& & $a_{4} =  -15.169160 $ & & & $a_{4} = -0.91023195$ \\
& & $a_{5} = -0.98120322$  & & & $a_{5} = 0.13749749$ \\
& & \\
\hline
\end{tabular}
\end{table*}

\section{Values adopted for metal cooling}
\begin{figure}
	\includegraphics[width=0.45\textwidth]{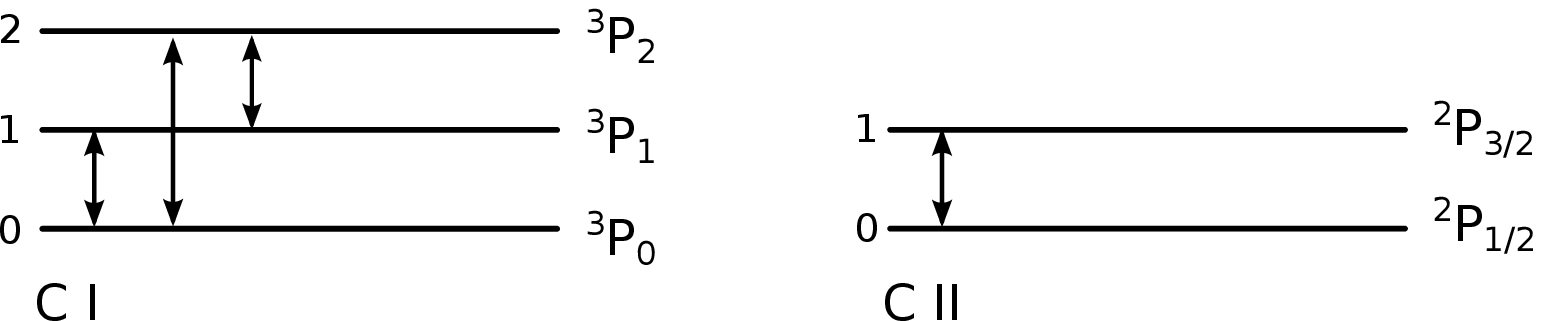}
        \caption{Pictorial view for the line transitions of carbon.}\label{fig:metalsline}
\end{figure}

\begin{figure}
	\includegraphics[width=0.45\textwidth]{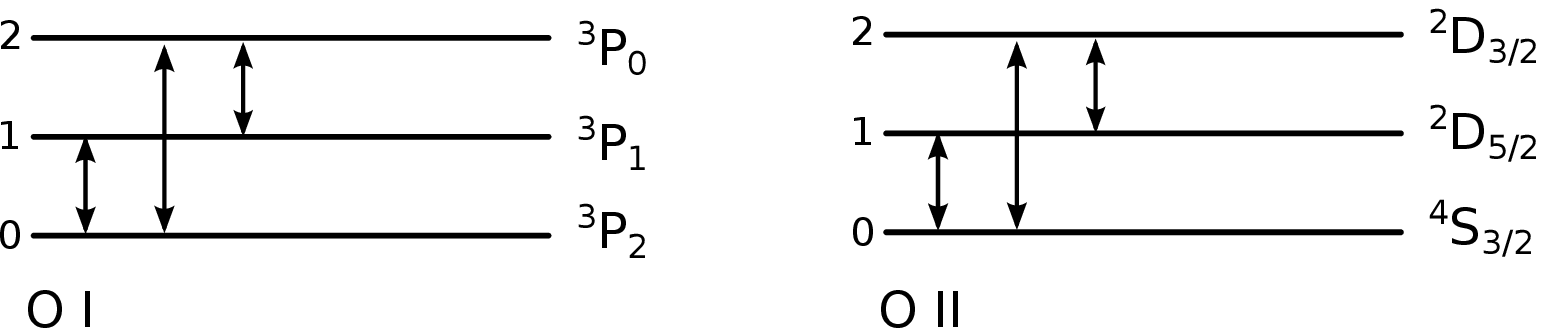}
        \caption{Pictorial view for the line transitions of oxygen.}\label{fig:metalsline1}
\end{figure}

\begin{figure}
	\includegraphics[width=0.45\textwidth]{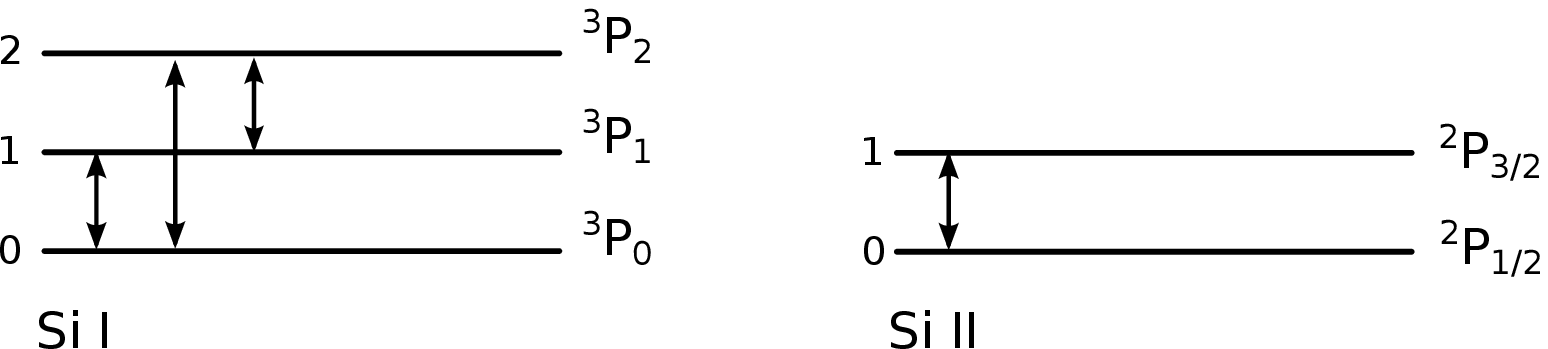}
        \caption{Pictorial view for the line transitions of silicon.}\label{fig:metalsline2}
\end{figure}

\begin{figure}
	\includegraphics[width=0.45\textwidth]{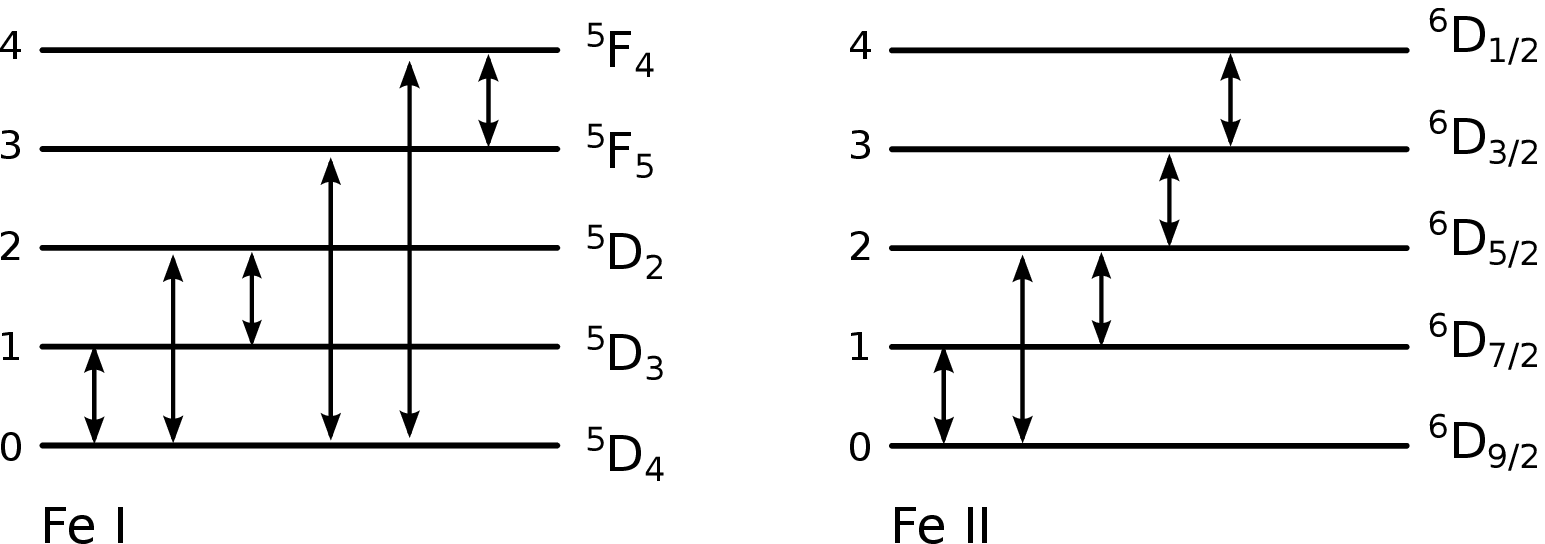}
        \caption{Pictorial view for the line transitions of iron.}\label{fig:metalsline3}
\end{figure}

In this Section we report the data used for the fine-structure metal cooling. The data are taken from \citet{Hollenbach1989,Maio2007,Glover2007,Grassi2012}.
The transitions included in \krome implementation are sketched in Figs.~\ref{fig:metalsline}, \ref{fig:metalsline1}, \ref{fig:metalsline2}, and \ref{fig:metalsline3}. The rate coefficients with several colliders and the atomic data are reported in Tabs.\ref{tab:metal_coeffs2}, \ref{tab:metal_coeffs3}, \ref{tab:metal_coeffs4}, and \ref{tab:metal_coeffs}.

\begin{table*}
\caption{De-excitation rate coeffients for neutral metals, and atomic data for each transition. Note that $a(b)=a\times10^b$.}\label{tab:metal_coeffs2}
\begin{tabular}{lllllllll}
\hline
Coolant & j$\to$i & $\gamma_{ji}^\mH$ (cm$^3$s$^{-1}$) & $\gamma_{ji}^{\mH^+}$ (cm$^3$s$^{-1}$)& A$_{ji}$(s$^{-1})$ & $\Delta E_{ji}$ (K) & $g_i$ & $g_j$\\
\hline
CI & 1$\to$0 & $1.6(-10) T_2^{0.14}$ & $[9.6(-11) -1.8(-14)T +1.9(-18)T^2]T^{0.45}$ & $7.9(-8)$ & $24$ & $3$ & $1$\\
   &         & 			      & if $(T>5\times10^3)$ $8.9(-10)T^{0.117}$ &&&&  \\
CI & 2$\to$0 & $9.2(-11) T_2^{0.26}$ & $[3.1(-12) -6(-16)T +3.9(-20)T^2]T$ &$2.7(-7)$ & $63$ & $5$ & $1$\\ 
   &         & 			      & if $(T>5\times10^3)$ $2.3(-9)T^{0.0965}$ &&&&  \\
CI & 2$\to$1 & $2.9(-10) T_2^{0.26}$ & $[1(-10) -2.2(-14)T +1.7(-18)T^2]T^{0.7}$ &$2.1(-14)$ & $39$ & $5$ & $3$\\ 
   &         & 			      & if $(T>5\times10^3)$ $9.2(-9)T^{0.0535}$ &&&&  \\

OI & 1$\to$0 & $9.2(-11)(T_2)^{0.67}$ & $6.38(-11)T^{0.4}$ & $8.9(-5)$ & $230$ &  $3$ & $5$ \\ 
 & &  & if $(T>194)$ $7.75(-12)T^{0.8}$ & & & & \\
 & &  & if $(T>3686)$ $2.65(-10)T^{0.37}$ & & & & \\
OI & 2$\to$0 & $4.3(-11)(T_2)^{0.80}$ & $6.1(-13)T^{1.1}$ & $1.8(-5)$ & $330$ & $1$  & $5$ \\
 & &  & if $(T>511)$ $2.12(-12)T^{0.9}$ & & & & \\
 & &  & if $(T>7510)$ $4.49(-10)T^{0.3}$ & & & & \\
OI & 2$\to$1 & $1.1(-10)(T_2)^{0.44}$ & $2.03(-11)T^{0.56}$ & $1.3(-10)$ & $98$ & $1$ & $3$ \\
 & &  & if $(T>2090)$ $3.43(-10)T^{0.19}$ & & & & \\

SiI & 1$\to$0 & $3.5(-10) T_2^{-0.03}$ & $7.2(-9)$ & $8.4(-6)$ & $110$ & $3$ & $1$ \\  
SiI & 2$\to$0 & $1.7(-11) T_2^{0.17}$ & $7.2(-9)$ & $2.4(-10)$ & $320$ & $5$ & $1$ \\  
SiI & 2$\to$1 & $5(-10) T_2^{0.17}$ & $2.2(-8)$ & $1.0(-9)$ & $210$ & $5$ & $3$ \\  

FeI & 1$\to$0 & $8(-10)T_2^{0.17}$ & - & $2.5(-3)$& $594.43$ & $9$ & $7$ \\ 
FeI & 2$\to$0 & $6.9(-10)T_2^{0.17}$ & - & $1.0(-9)$& $1012.9$ & $9$ & $5$ \\ 
FeI & 2$\to$1 & $5.3(-10)T_2^{0.17}$ & - & $1.6(-3)$& $414.47$ & $7$ & $5$ \\ 
FeI & 3$\to$0 & - & - & $2.0(-3)$& $594.43$ & $9$ & $11$ \\ 
FeI & 4$\to$0 & - & - & $1.5(-3)$& $1012.9$ & $9$ & $8$ \\ 
FeI & 4$\to$3 & - & - & $3.6(-3)$& $414.47$ & $11$ & $8$ \\ 
\hline
\end{tabular}
\end{table*}

\begin{table*}
\caption{De-excitation rate coeffients for neutral metals with H$_2$ ortho and para, and electrons. See Tab.\ref{tab:metal_coeffs2} for additional values, while other de-excitation rates are listed in Tab.\ref{tab:metal_coeffs4}. Note that $a(b)=a\times10^b$ and $T_2=T/(100$~K).}\label{tab:metal_coeffs3}
\begin{tabular}{lllll}
\hline
Coolant & j$\to$i & $\gamma_{ji}^{\mH_2^o}$ (cm$^3$s$^{-1}$) & $\gamma_{ji}^{\mH_2^p}$ (cm$^3$s$^{-1}$)& $\gamma_{ji}^{\me^-}$ (cm$^3$s$^{-1}$)\\
\hline
CI & 1$\to$0 & $8.7(-11) -6.6(-11)\exp(-T/218.3)$ & $7.9(-11) -8.7(-11)\exp(-T/126.4)$ & see Tab.\ref{tab:metal_coeffs4}\\
CI & 2$\to$0 & $1.2(-11) -6.1(-11)\exp(-T/387.3)$ & $1.1(-10) -8.6(-11)\exp(-T/223)$ & see Tab.\ref{tab:metal_coeffs4}\\ 
  &  & & $+8.7(-11)\exp(-2T/223) $\\
CI & 2$\to$1 & $2.9(-10) -1.9(-10)\exp(-T/348.9)$& $2.7(-10) -1.9(-10)\exp(-T/348.9)$ & see Tab.\ref{tab:metal_coeffs4}\\ 
  &  & & $+1.8(-10)\exp(2T/250.7) $\\

OI & 1$\to$0 & $2.7(-11)T^{0.362}$ & $3.46(-11)T^{0.316}$ & $5.12(-10)T^{-0.075}$ \\ 
OI & 2$\to$0 & $5.49(-11)T^{0.317}$ & $7.07(-11)T^{0.268}$ & $4.86(-10)T^{-0.026}$\\ 
OI & 2$\to$1 & $2.74(-14)T^{1.06}$ & $3.33(-15)T^{1.36}$ & $1.08(-14)T^{0.926}$\\ 

FeI & 1$\to$0 & - & - & $1.2(-7)$ \\ 
FeI & 2$\to$0 & - & - & $1.2(-7)$ \\ 
FeI & 2$\to$1 & - & - & $9.3(-8)$ \\
FeI & 3$\to$0 & - & - & $2(-7)(T/10^4)^{0.57}$ \\
FeI & 4$\to$0 & - & - & $1(-7)(T/10^4)^{0.57}$ \\
FeI & 4$\to$3 & - & - & $1.5(-7)$ \\
\hline
\end{tabular}
\end{table*}

\begin{table*}
\caption{De-excitation rate coeffients for neutral carbon, Einstein's coefficients, energy level differences, and level multeplicities. See Tab.\ref{tab:metal_coeffs2} for additional atomic values. Note that $a(b)=a\times10^b$.}\label{tab:metal_coeffs4}
\begin{tabular}{lll}
\hline
Coolant & j$\to$i & $\gamma_{ji}^{\me^-}$ (cm$^3$s$^{-1}$)\\
\hline
CI & 1$\to$0 & $2.88(6)T^{-0.5}\exp(-9.25141 -7.73782(-1)\ln(T) +3.61184(-1)\ln(T)^2$\\
  & & $ -1.50892(-2)\ln(T)^3 -6.56325(-4)\ln(T)^4)$\\
  & & if$(T>10^3)$ $2.88(-6)T^{-0.5} \exp(-4.446(2) -2.27913(2)\ln(T)$\\
  & & $+4.2595(1)\ln(T)^2 -3.4762\ln(T)^3 +1.0508(-1)\ln(T)^4)$\\

CI & 2$\to$0 & $1.73(-6)T^{-0.5}\exp(-7.69735 1.30743\ln(T) -0.111338\ln(T)^3$\\
  & & $+0.705277(-2)\ln(T)^4)$\\

  & & if$(T>10^3)$ $1.73(-6)T^{-0.5} \exp(3.50609(2) -1.87474(2)\ln(T)$\\
  & & $+3.61803(1)\ln(T)^2 -3.03283\ln(T)^3 +9.38138(-2)\ln(T)^4)$\\

CI & 2$\to$1 & $2 1.73(-6)T^{-0.5}\exp(-7.4387 -0.57443\ln(T) +0.358264\ln(T)^2$\\
  & & $-4.18166(-2)\ln(T)^3 +2.35272(-3)\ln(T)^4)$\\
  & & if$(T>10^3)$ $1.73(-6)T^{-0.5} \exp(3.86186(2) -2.02192(2)\ln(T)$\\
  & & $+3.85049(1)\ln(T)^2 -3.19268\ln(T)^3 +9.78573(-2)\ln(T)^4)$\\

\hline
\end{tabular}
\end{table*}

\begin{table*}
\caption{De-excitation rate coeffients for ions, Einstein's coefficients, energy level differences, and level multeplicities. Note that $a(b)=a\times10^b$, $T_2=T/(100$~K), and $T_4=T/(10^4$~K)}\label{tab:metal_coeffs}
\begin{tabular}{lllllllll}
\hline
Coolant & j$\to$i & $\gamma_{ji}^\mH$ (cm$^3$s$^{-1}$) & $\gamma_{ji}^{\me^-}$ (cm$^3$s$^{-1}$)& A$_{ji}$(s$^{-1})$ & $\Delta E_{ji}$ (K) & $g_i$ & $g_j$\\
\hline
CII & 1$\to$0 & $8\cdot 10^{-10} T_2^{0.07}$ & $2.8\cdot 10^{-7} T_2^{-0.5}$ &$2.4\cdot 10^{-6}$ & $91.2$ & $4$ & $2$\\ 
OII & 1$\to$0 &  & $1.3\cdot 10^{-8}(T_4)^{-0.5}$ & $5.1 \cdot 10^{-5}$ & $38574.4$ & $11$ & $7$ \\   
OII & 2$\to$0 &  & $1.3\cdot 10^{-8}(T_4)^{-0.5}$ & $1.7 \cdot 10^{-4}$ & $38603.2$ & $7$  & $7$ \\   
OII & 2$\to$1 &  & $2.5\cdot 10^{-8}(T_4)^{-0.5}$ & $1.3 \cdot 10^{-7}$ & $28.8$ $7$ & $11$ \\   
SiII & 1$\to$0 & $8\cdot 10^{-10} T_2^{-0.07}$ & $1.7\cdot 10^{-6} T_2^{-0.5}$ & $2.1\cdot 10^{-4}$ & $413.6$ & $4$ & $2$ \\  
FeII & 1$\to$0 & $9.5 \cdot 10^{-10}$ & $1.8\cdot 10^{-6} T_2^{-0.5}$ & $2.13 \cdot 10^{-3}$& $553.58$ & $10$ & $8$ \\ 
FeII & 2$\to$1 & $4.7 \cdot 10^{-10}$ & $8.7\cdot 10^{-7}T_2^{-0.5}$ & $1.57 \cdot 10^{-3}$& $407.01$  & $6$ & $8$ \\ 
FeII & 3$\to$2 & $5.0 \cdot 10^{-10}$ & $10^{-5}T^{-0.5}$ & $ 1.50 \cdot 10^{-9}$ & $280.57$ & $6$ & $4$\\  
FeII & 4$\to$3 & $5.0 \cdot 10^{-10}$ & $10^{-5}T^{-0.5}$ &  & $164.60$ & $4$ & $2$ \\
FeII & 2$\to$0 & $5.7 \cdot 10^{-10}$ & $1.8\cdot 10^{-6}T_2^{-0.5}$ & &  $960.59$ & $10$ & $6$\\ 
\hline
\end{tabular}
\end{table*}

\section{Reaction rates employed in the primordial network}\label{appx:primordial}
The reaction rates employed in the primordial network are listed in Tab.\ref{tab:rates}.
\begin{table*}
	\caption{List of reactions and rates included in our chemical network. See \citet{Abel97} for further details.}\label{tab:rates}
        \begin{tabular}{@{}lllc}
		\hline\hline
		Reaction & Rate coefficient (cm$^3$ s$^{-1}$) & &\\	
		\hline
       	H + e$^-$ $\rightarrow$ H$^+$ + 2e$^-$  & $k_1$ = exp[-32.71396786+13.5365560 ln $T_e$ & &  \\
	& - 5.73932875 (ln $T_e$)$^2$+1.56315498 (ln $T_e$)$^3$ & &  \\
	& - 0.28770560 (ln $T_e$)$^4$+3.48255977 $\times$ 10$^{-2}$(ln $T_e$)$^5$ & & \\
	& - 2.63197617 $\times$ 10$^{-3}$(ln $T_e$)$^6$+1.11954395 $\times$ 10$^{-4}$(ln $T_e$)$^7$ & \\
	& - 2.03914985 $\times$ 10$^{-6}$(ln $T_e$)$^8$] & \\
       	H$^+$ + e$^-$ $\rightarrow$ H  + $\gamma$ & $k_2$ = 3.92 $\times$ 10$^{-13}$ $T_e$ $^{-0.6353}$ & $T \le 5500$ K & \\
	& $k_2$ = $\exp$[-28.61303380689232 & $T > 5500$ K & \\
& - 7.241 125 657 826 851 $\times$ 10$^{-1}$ ln $T_e$\\ 
& - 2.026 044 731 984 691 $\times$ 10$^{-2}$ (ln $T_e$)$^2$\\
& - 2.380 861 877 349 834 $\times$ 10$^{-3}$ (ln $T_e$)$^3$\\
& - 3.212 605 213 188 796 $\times$ 10$^{-4}$ (ln $T_e$)$^4$\\
& - 1.421 502 914 054 107 $\times$ 10$^{-5}$ (ln $T_e$)$^5$\\
& + 4.989 108 920 299 510  $\times$ 10$^{-6}$ (ln $T_e$)$^6$\\
& + 5.755 614 137 575 750  $\times$ 10$^{-7}$ (ln $T_e$)$^7$\\
& - 1.856 767 039 775 260  $\times$ 10$^{-8}$  (ln $T_e$)$^8$\\
& - 3.071 135 243 196 590  $\times$ 10$^{-9}$  (ln $T_e$)$^9$]  \\
	He + e$^-$ $\rightarrow$ He$^+$ + 2e$^-$  & $k_3$ = $\exp$[-44.09864886 & $T_e > 0.8$ eV \\
	& + 23.915 965 63 ln$T_e$ \\
	& - 10.753 230 2 (ln $T_e$)$^2$\\
	& + 3.058 038 75 (ln $T_e$)$^3$\\
	& - 5.685 118 9 $\times$ 10$^{-1}$ (ln $T_e$)$^4$\\
	& + 6.795 391 23 $\times$ 10$^{-2}$ (ln $T_e$)$^5$\\
	& - 5.009 056 10 $\times$ 10$^{-3}$ (ln $T_e$)$^6$\\
	& + 2.067 236 16 $\times$ 10$^{-4}$ (ln$T_e$)$^7$\\
	& - 3.649 161 41 $\times$ 10$^{-6}$ (ln $T_e$)$^8$]  & \\
	He$^+$ + e$^-$ $\rightarrow$ He + $\gamma$ & $k_4$ =  3.92 $\times$ 10$^{-13}$ $T_e$ $^{-0.6353}$ &  $T_e \le 0.8$ eV \\
	& $k_4 = $ + 3.92 $\times$ 10$^{-13}$ $T_e^{-0.6353}$ & $T > 0.8$ eV \\
	& + 1.54 $\times$ 10$^{-9}$ $T_e^{-1.5}$ [1.0 + 0.3 / $\exp$(8.099 328 789 667/$T_e$)]  \\
	& /[$\exp$(40.496 643 948 336 62/$T_e$)]\\
	&  & \\
	He$^+$ + e$^-$ $\rightarrow$ He$^{2+}$ + 2e$^-$ & $k_5$ = $\exp$[-68.710 409 902 120 01 & $T_e > 0.8 $ eV \\
	& + 43.933 476 326 35 ln$T_e$ \\
	& - 18.480 669 935 68 (ln $T_e$)$^2$ \\
	& + 4.701 626 486 759 002 (ln $T_e$)$^3$ \\
	& - 7.692 466 334 492 $\times$ 10$^{-1}$ (ln $T_e$)$^4$\\
	& + 8.113 042 097 303 $\times$ 10$^{-2}$ (ln $T_e$)$^5$\\
	& - 5.324 020 628 287 001 $\times$ 10$^{-3}$ (ln $T_e$)$^6$\\
	& + 1.975 705 312 221 $\times$ 10$^{-4}$ (ln $T_e$)$^7$ \\
	& - 3.165581065665 $\times$ 10$^{-6}$ (ln $T_e$)$^8$] & \\ 
	He$^{2+}$ + e$^-$ $\rightarrow$ He$^+$ + $\gamma$ & $k_6$ = 3.36 $\times$ 10$^{-10}$ $T^{-1/2} (T/1000)^{-0.2} (1+(T/10^6)^{0.7})^{-1}$ & \\
	H + e $\rightarrow$ H$^-$ + $\gamma$ & $k_7$ = 6.77 $\times$ 10$^{-15}$ $T_e^{0.8779}$ & \\
	H$^-$ + H $\rightarrow$ H$_2$ + e$^-$ & $k_8$ = 1.43 $\times$ 10$^{-9}$& $T \le 1160 K$ \\
	& $k_8$ = $\exp$[-20.069 138 975 870 03 & $T > 1160$ K \\
	& + 2.289 800 603 272 916 $\times$ 10$^{-1}$ ln $T_e$\\ 
	& + 3.599 837 721 023 835 $\times$ 10$^{-2}$ (ln $T_e$)$^2$\\
	& - 4.555 120 027 032 095 $\times$ 10$^{-3}$ (ln $T_e$)$^3$\\
	& - 3.105 115 447 124 016 $\times$ 10$^{-4}$ (ln $T_e$)$^4$\\
	& + 1.073 294 010 367 247 $\times$ 10$^{-4}$ (ln $T_e$)$^5$\\
	& - 8.366 719 604 678 64 $\times$ 10$^{-6}$ (ln $T_e$)$^6$\\
	& + 2.238 306 228 891 639 $\times$ 10$^{-7}$ (ln $T_e$)$^7$]&\\
	H + H$^+$ $\rightarrow$ H$_2^+$ + $\gamma$ & $k_9$ = 1.85 $\times$ 10$^{-23}$ $T^{1.8}$ & $T \le 6700$ K\\
	& $k_9$ = 5.81 $\times$ 10$^{-16}$ ($T/56200$)$^{(-0.6657\log_{10}(T/56200))}$ &  $T > 6700$ K\\
	\hline
\end{tabular}
\end{table*}

\begin{table*}
	\contcaption{List of reactions and rates included in our chemical network. See \citet{Abel97} for further details.}
        \begin{tabular}{@{}lllc}
		\hline\hline
		Reaction & Rate coefficient (cm$^3$ s$^{-1}$) & &\\	
		\hline
	H$_2^+$ + H $\rightarrow$ H$_2$ + H$^+$ & $k_{10}$ = 6.0 $\times$ 10$^{-10}$ & \\
	H$_2$ + H$^+$ $\rightarrow$ H$_2^+$ + H & $k_{11}$ = $\exp$[-24.249 146 877 315 36 & $T_e >  0.3$ eV \\
	& + 3.400 824 447 095 291 ln $T_e$ \\
	& - 3.898 003 964 650 152  (ln$T_e$)$^2$\\
	& + 2.045 587 822 403 071 (ln $T_e$)$^3$\\	
	& - 5.416 182 856 220 388 $\times$ 10$^{-1}$ (ln $T_e$)$^4$\\
	& + 8.410 775 037 634 12 $\times$ 10$^{-2}$ (ln $T_e$)$^5$\\
	& - 7.879 026 154 483 455 $\times$ 10$^{-3}$ (ln $T_e$)$^6$\\
	& + 4.138 398 421 504 563 $\times$ 10$^{-4}$ (ln $T_e$)$^7$\\
	& - 9.363 458 889 286 11 $\times$ 10$^{-6}$ (ln $T_e$)$^8$] & \\
	H$_2$ + e$^-$ $\rightarrow$ 2H + e$^-$ & $k_{12}$ = 5.6 $\times$ 10$^{-11} T^{0.5} \exp(-102124.0/T)$ & $T_e >  0.3$ eV\\
	H$^-$ + e$^-$ $\rightarrow$ H + 2e$^-$  & $k_{13}$ = $\exp$(-18.018 493 342 73 & $T_e > 0.04$ eV \\
	& + 2.360 852 208 681 ln$T_e$\\
	& - 2.827 443 061 704 $\times$ 10$^{-1}$ (ln $T_e$)$^2$\\
	& + 1.623 316 639 567 $\times$ 10$^{-2}$ (ln $T_e$)$^3$\\
	& - 3.365 012 031 362 999 $\times$ 10$^{-2}$ (ln $T_e$)$^4$\\
	& + 1.178 329 782 711  $\times$ 10$^{-2}$ (ln $T_e$)$^5$\\
	& - 1.656 194 699 504  $\times$ 10$^{-3}$ (ln $T_e$)$^6$\\
	& + 1.068 275 202 678  $\times$ 10$^{-4}$ (ln $T_e$)$^7$\\
	& - 2.631 285 809 207  $\times$ 10$^{-6}$ (ln $T_e$)$^8$& \\
	H$^-$ + H $\rightarrow$ 2H +  e$^-$ & $k_{14}$ = 2.56 $\times$ 10$^{-9}$ $T_e^{1.78186}$ & $T_e \le  0.04$ eV \\
	& $k_{14}$ =  $\exp$[-20.372 608 965 333 24 & $T_e >  0.04$ eV \\
	& + 1.139 449 335 841 631 ln $T_e$ \\
	& - 1.421 013 521 554 148 $\times$ 10$^{-1}$ (ln $T_e$)$^2$\\
	& + 8.464 455 386 63 $\times$ 10$^{-3}$ (ln $T_e$)$^3$\\
	& - 1.432 764 121 299 2 $\times$ 10$^{-3}$ (ln $T_e$)$^4$\\
	& +2.012 250 284 791 $\times$ 10$^{-4}$ (ln $T_e$)$^5$\\
	& + 8.663 963 243 09 $\times$ 10$^{-5}$ (ln $T_e$)$^6$\\
	& - 2.585 009 680 264 $\times$ 10$^{-5}$ (ln $T_e$)$^7$\\
	& + 2.455 501 197 039 2 $\times$ 10$^{-6}$ (ln $T_e$)$^8$\\
	& - 8.068 382 461 18 $\times$ 10$^{-8}$ (ln $T_e$)$^9$]\\
	H$^-$ + H$^+$ $\rightarrow$ 2H + $\gamma$ & $k_{15}$ = 6.5 $\times$ 10$^{-9}$ $T_e^{-0.5}$ & \\
	H$^-$ + H$^+$ $\rightarrow$ H$_2^+$ + e$^-$ & $k_{16}$ = 10$^{-8} \times T^{-0.4}$& \\
	H$_2^+$ + e $\rightarrow$ 2H + $\gamma$ & $ k_{17}$ = 10$^{-8}$ & $T \le 617$ K  \\
	& $k_{17}$ =  1.32 $\times$ 10$^{-6}$ $T^{-0.76}$ & $T > 617$ K \\
	H$_2^+$ + H$^-$ $\rightarrow$ H + H$_2$ & $k_{18}$ = 5.0 $\times$ 10$^{-7}$ (10$^2 \times T$)$^{-0.5}$& \\
	3H $\rightarrow$ H$_2$ + H & $k_{19}$ = 1.3$\times$10$^{-32}(T/300)^{-0.38}$ & $T \le 300 K$\\
	& $k_{19}$ = 1.3$\times$10$^{-32}(T/300)^{-1.00}$ & $T > 300 K$\\
	H$_2$ + H $\rightarrow$ 3H & $k_{20}$ = $(1.0670825\times10^{-10}\times T_e^{2.012})/(\exp(4.463/T_e)\times(1 + 0.2472 T_e)^{3.512})$\\
	\hline
\end{tabular}
\end{table*}

\section{Reaction file format}\label{appx:file_format}
The reaction file has a default template, namely
\begin{verbatim}
index, R1, R2, R3, P1, P2, P3, P4, Tmin, Tmax, rate
\end{verbatim}
that can be modified according to other formats. In particular one can indicate the number of items simply using the following case non-sensitive tokens: \verb+R+ for reactants, \verb+P+ for products, \verb+Tmin+ and \verb+Tmax+ for the temperature limits, \verb+rate+ for the reaction rate, and \verb+idx+ for the index. The format string must begin with \verb+@format:+ to be recognized by the file parser of the \krome package. The user can employ several format strings within the same reaction file, e.g.
\begin{verbatim}
@format:R, R, P, Tmin, Tmax, rate
H, H, H2, 1e1, 1e4, 1d-9
C, O, CO, 1e1, 1e4, 3d-9
@format:Tmin, Tmax, R, R, P, P rate
1e1, 1e4, H+, C, H, C+, 1d-10
\end{verbatim}
where the first line indicates a block of reaction with two reactants, one product, the temperature limits, and the rate coefficient, while the fouth line is employed to use a format that starts with the temperature limits, than two reactants and two products, and the rate coefficient. Note that the coefficient rates in this example have only an illustrative purpose, and do not represent the real coefficients for the reactions reported here. 

Finally, \krome allows in the file to use comments and variables, as already discussed, and also individual temperature operators, e.g.
\begin{verbatim}
@format:Tmin, Tmax, R, R, P, P rate
>1e1, 1e4, H+, C, H, C+, 1d-10
\end{verbatim}
where the $>$ sign is employed for this reaction only.

\bsp

\label{lastpage}

\end{document}